\documentclass[aps,reprint,superscriptaddress,nofootinbib]{revtex4-2}
\usepackage[table]{xcolor}
\usepackage{tabularx}
\usepackage{enumitem}

\setlist{nolistsep}
\definecolor{green}{HTML}{66FF66}
\definecolor{myGreen}{HTML}{009900}

\usepackage{quantikz}
\usepackage[T1]{fontenc}
\usepackage[utf8]{inputenc}
\usepackage{cmap}
\usepackage{amsmath}
\usepackage{amssymb}
\usepackage{amstext}
\usepackage{amsfonts}
\usepackage{mathtools}
\usepackage{hyperref}
\usepackage{todonotes}
\usepackage{physics}
\usepackage{amsthm}
\usepackage{enumitem}
\usepackage{dcolumn}
\usepackage{bm}
\usepackage{hyperref}

\usepackage[capitalise]{cleveref}
\hypersetup{
 colorlinks=true,
 linkcolor=blue,
 citecolor=blue,
 filecolor=blue,
 urlcolor=blue,
 anchorcolor=blue,
 bookmarksopen=true,
 bookmarksopenlevel=1,
 pdfpagemode=UseOutlines,
 pdfstartview={XYZ null null 1},
 linktocpage=true,
}

\usepackage{times}

\usepackage{graphicx}
\usepackage[caption = false, font=small]{subfig}

\usepackage{bm}
\usepackage{tikz}
\usetikzlibrary{shapes.geometric}
\usetikzlibrary{shapes,arrows}
\usetikzlibrary{positioning}
\usetikzlibrary{shadows}

\tikzstyle{line} = [draw, -latex']
\tikzstyle{image} = [rectangle, draw, fill= white, 
text width=0.3em, minimum height = 2.3em, align= center,
copy shadow={draw, shadow xshift=1mm, shadow yshift=-1mm}]

\usepackage{setspace}
\usepackage{makecell}
\usepackage[export]{adjustbox}
\usepackage{multirow}
\usepackage{booktabs}

\definecolor{LightCyan}{rgb}{0.88,1,1}

\newcommand{\orcidicon}[1]{\href{https://orcid.org/#1}{\includegraphics[height=\fontcharht\font`\B]{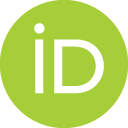}}}

\newcommand{\figref}[1]{Figure~\ref{#1}}
\newcommand{\eqnref}[1]{Eq.~\eqref{#1}}
\newcommand{\tabref}[1]{Table~\ref{#1}}
\newcommand{\secref}[1]{Section~\ref{#1}}
\newcommand{\adxref}[1]{Appendix~\ref{#1}}

\newcommand{\model}{LaSt-QGAN }
\newcommand{\etal}{\textit{et al.}}
\newcommand{\E}{\mathbb{E}}
\newcommand{\Var}{\text{Var}}
\newcommand{\img}{\mathcal{I}}
\newcommand{\bfx}{\mathbf{x}} 
\newcommand{\tbfx}{\tilde{\mathbf{x}}} 
\newcommand{\mcl}{\mathcal{L}}
\newcommand{\Dz}{\mathcal{D}_{\mathbf{z}}}
\newcommand{\Dl}{\mathcal{D}_{\ell}}
\newcommand{\Np}{N_{\bm{\Theta}}}
\newcommand{\G}{G_{\bm{\theta}}}
\newcommand{\D}{D_{\bm{\phi}}}

\renewcommand{\tabcolsep}{0pt}

\begin{document}

\title{Latent Style-based Quantum GAN for high-quality Image Generation}

\author{Su Yeon Chang\orcidicon{0000-0001-5768-2434}}
\email{su.yeon.chang@cern.ch}
\affiliation{European Organization for Nuclear Research (CERN), Geneva, Switzerland}
\affiliation{Laboratory of Theoretical Physics of Nanosystems (LTPN), Ecole Polytechnique F\'ed\'erale de Lausanne (EPFL), Lausanne, Switzerland}

\author{Supanut Thanasilp\orcidicon{0000-0002-8252-4056}} 
\affiliation{Institute of Physics, Ecole Polytechnique F\'{e}d\'{e}rale de Lausanne (EPFL), CH-1015 Lausanne, Switzerland}
\affiliation{Chula Intelligent and Complex Systems, Department of Physics, Faculty of Science, Chulalongkorn University, Bangkok, Thailand, 10330}

\author{Bertrand Le Saux}
\affiliation{\textit{$\Phi$-lab}, European Space Agency, Frscati (RM), Italy  }

\author{Sofia Vallecorsa}
\affiliation{European Organization for Nuclear Research (CERN), Geneva, Switzerland}

\author{Michele Grossi\orcidicon{0000-0003-1718-1314}}
\email{michele.grossi@cern.ch}
\affiliation{European Organization for Nuclear Research (CERN), Geneva, Switzerland}

\begin{abstract}
Quantum generative modeling is among the promising candidates for achieving a practical advantage in data analysis. Nevertheless, one key challenge is to generate large-size images comparable to those generated by their classical counterparts. In this work, we take an initial step in this direction and introduce the \textit{Latent Style-based Quantum GAN (LaSt-QGAN)}, which employs a hybrid classical-quantum approach in training Generative Adversarial Networks (GANs) for arbitrary complex data generation. This novel approach relies on powerful classical auto-encoders to map a high-dimensional original image dataset into a latent representation. The hybrid classical-quantum GAN operates in this latent space to generate an arbitrary number of fake features, which are then passed back to the auto-encoder to reconstruct the original data. 
Our \model can be successfully trained on realistic computer vision datasets beyond the standard MNIST, namely Fashion MNIST (fashion products) and SAT4 (Earth Observation images) with 10 qubits, resulting in a comparable performance (and even better in some metrics) with the classical GANs.
Moreover, we analyze the barren plateau phenomena within this context of the continuous quantum generative model using a polynomial depth circuit and propose a method to mitigate the detrimental effect during the training of deep-depth networks. 
Through empirical experiments and theoretical analysis, we demonstrate the potential of \model for the practical usage in the context of image generation and open the possibility of applying it to a larger dataset in the future.
\end{abstract}
\maketitle

\section{Introduction}
\label{sec:introduction}

Over the past few decades, generative modeling has stood as one of the main pillars in machine learning (ML), revolutionizing not only academia but also industries and everyday life~\cite{RAY2023ChatGPT, Daras2022DALLE, Isola2017pix, Paganini2018CaloGAN}. These models aim to generate synthetic data that closely resembles the original data by learning the underlying probability distribution. While operating on high-dimensional data manifolds posts some key challenges, it also inspires researchers to propose diverse architectures~\cite{Kingma2022AE, Goodfellow2014GAN, Dickstein2015DM, Du2019EM, Papamakarios2021Normalizing} and training strategies~\cite{Gulrajani2017wasserstein, Arjovsky2017stability}. 

Among those various architectures, generative adversarial networks (GANs)~\cite{Goodfellow2014GAN, Mirza2014cGAN, Gulrajani2017wasserstein} and diffusion models (DMs)~\cite{Dickstein2015DM, Jonathan2020DMs, Dhariwal2021DM} have emerged as two of the most developed and widely used. On one hand, GANs learn the implicit data distribution of an arbitrary dataset by simultaneously training two distinct neural networks in an adversarial minimax game, successfully being used for a wide range of applications such as image generation~\cite{Radford2016unsupervised, Karras2018StyleGAN}, text-to-image synthesis~\cite{Scott2016, Kang2023scaling}, image-to-image translation~\cite{Zhu2017CycleGAN, Yi2017DualGAN, Isola2017pix} and high-energy physics particle shower simulation~\cite{Paganini2018CaloGAN, DeOliveira2017LAGAN}. 
On the other hand, DMs rely on iteratively learning to reconstruct data which are intentionally perturbed by noise. 
Of particular interest, one DM variant known as a latent diffusion model (LDM)~\cite{Rombach_2022_LDM_CVPR} incorporates a strength of pre-trained autocoders to embed original data into a low-dimensional latent space and learns data generation at this level, directly circumventing the issue of operating in a high dimensional space. This approach significantly reduces the computational resources while retaining high fidelity of generated data~\cite{Rombach_2022_LDM_CVPR}.

\begin{figure*}
\subfloat[Discrete Quantum GAN]{
    \includegraphics[width = 0.495\textwidth]{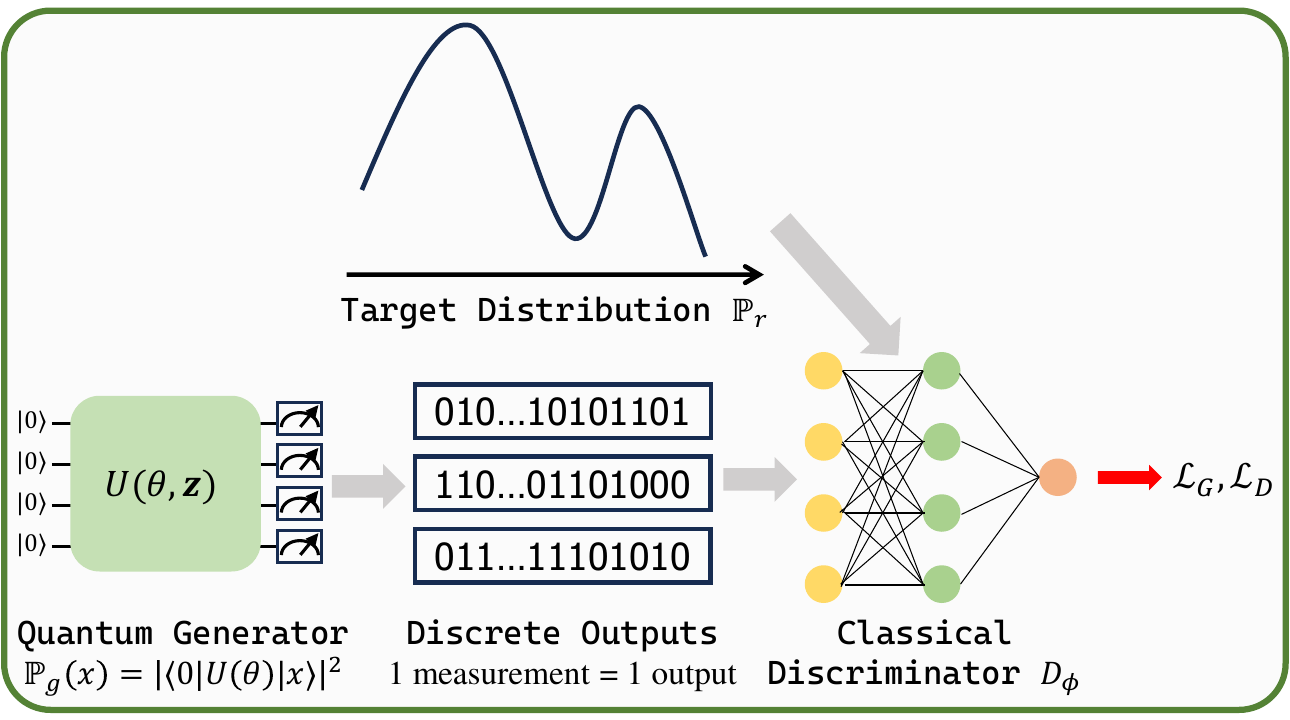}
} 
\hspace*{\fill}
\subfloat[Continuous Quantum GAN]{
    \includegraphics[width = 0.495\textwidth]{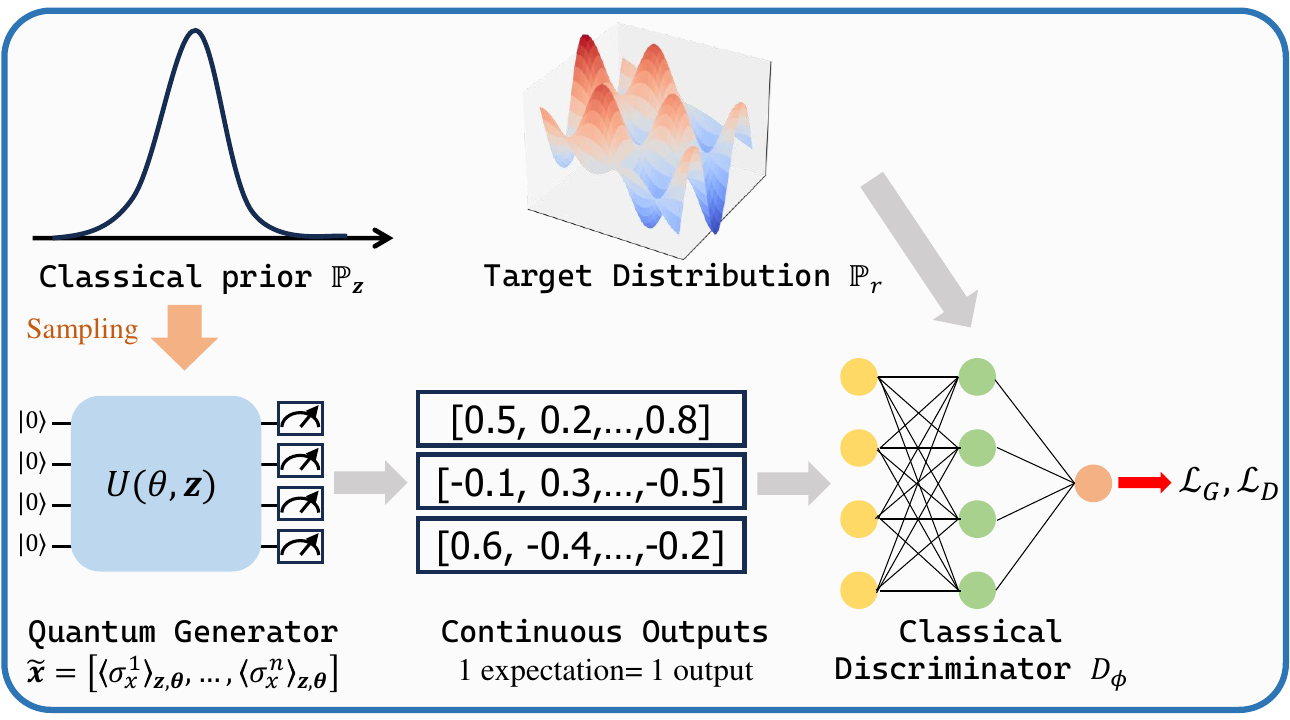}
} 
\caption{\label{fig:discrete_continuous} \textbf{Schematic diagram summarizing the general training framework of discrete and continuous quantum GANs}. 
Frequently, we use a hybrid approach with a quantum generator and classical discriminator~\cite{Zoufal2019, BravoPrieto2022}, although an alternative option exists where a quantum discriminator is employed~~\cite{Romero2019}. }
\end{figure*}

Meanwhile, due to the rise of quantum computers, quantum machine learning (QML) has emerged as a new paradigm for data analysis, harnessing the power of quantum mechanics in the hope of achieving a practical advantage over conventional classical ML~\cite{biamonte2017quantum, cerezo2020variationalreview, huang2021quantum, Liu2021Speedup, aharonov2021quantum, gao2022enhancing, huang2021power, wu2023quantum,nietner2023average, tangpanitanon2020expressibility}. 
Such growing interest has also spurred efforts to extend QML to the context of generative models by employing parametrized quantum circuits to learn either discrete or continuous distributions (see \figref{fig:discrete_continuous} for a visual summary). Discrete generative models (including quantum Born machines~\cite{Benedetti2019QCBM, nietner2023average, rudolph2023trainability, coyle2020born, liu2018differentiable, Rudolph2022, Kiss2022QCBM, kyriienko2022protocols}, quantum GANs~\cite{Zoufal2019, chang2021cvqgan, chang2021dual, Huang2021qGAN, letcher2023qgan} and quantum Boltzmann machines~\cite{Amin2018QBM,coopmans2023sample}) employ a parametrized $n$-qubit quantum state to represent a discrete distribution of $2^n$ bit-strings with generated samples efficiently obtained as measurements in a computational basis. On the other hand, in the case of continuous models such as variational quantum generator~\cite{Romero2019} or style-based quantum GANs~\cite{BravoPrieto2022, barthe2024expressivity}, a quantum circuit acts as a feature map and takes classical random input to produce expectation values as new samples. Despite less sampling efficiency, this approach by design naturally handles continuous data generation and is expected to have a wide range of applications, such as image synthesis, where each pixel takes a continuous value.

Compared to discrete models where there exists a relatively larger body of literature~\cite{Benedetti2019QCBM, nietner2023average, rudolph2023trainability, coyle2020born, liu2018differentiable, Rudolph2022, Kiss2022QCBM, kyriienko2022protocols, Zoufal2019, Huang2021qGAN, letcher2023qgan, Amin2018QBM,coopmans2023sample, Gili2023,tangpanitanon2020expressibility}, studies of the continuous quantum generative modeling are much less explored (see the \secref{sec:related} for details of the recent research advancements). The proof of concept on small-size data generation was demonstrated in the context of 3-dimensional Monte Carlo event generation~\cite{BravoPrieto2022}. Recently, an expressivity of the continuous model has been investigated and universality is shown to be achievable under some sufficient conditions~\cite{barthe2024expressivity}. Yet, there remain many open questions, both fundamentally and practically. One of which is \textit{how to achieve the capability of producing an arbitrary number of high-quality images with large sizes}. Given large-size images being generated by classical ML today, resolving this particular problem is one of the key pieces for a practical quantum advantage or utility. 

In this work, we propose a hybrid quantum-classical GAN approach, which we call \textit{Latent Style-based Quantum GAN (LaSt-QGAN)}, capable of generating large-size images. We leverage the idea of LDMs by first embedding the complex high-dimensional data into a lower-dimensional latent space using a pre-trained classical autoencoder and then training a style-based quantum GAN directly on this compressed latent representation. After training, expectation values produced by the quantum generator are considered as new features in the latent space, which are then mapped back to the original data space by the autoencoder leading to a new set of large-size images. Compared with standard style-based quantum GANs, our method allows us to push a limit to generate much larger size images despite having the same quantum resources.

Our study focuses on two main objectives. First, we conduct empirical research on \model in order to understand its potential in practical applications. In comparison to existing frameworks, we empirically showcase the model's capacity to generate diverse images by testing it on the standard MNIST, the Fashion MNIST dataset, and on Earth observation image dataset, known as SAT4~\cite{SAT4}. Second, we perform further analysis on the model to assess its robustness against statistical fluctuations caused by shot noises and evaluate its trainability at the initial step, a pivotal factor in QML. Crucially, we investigate the barren plateau phenomena of continuous quantum GANs (using both analytical and numerical tools) for the first time, as a complementary contribution to ~\cite{rudolph2023trainability, letcher2023qgan} who pioneered this work in the case of the discrete quantum generative models. In particular, this study suggests a possible method to trigger the training of \model at the initial step with a small angle initialization around the identity for a polynomial depth quantum circuit, whose loss landscape is exponentially flat on average. As the training of our model happens at the level of latent space, the barren plateau results here are directly applied to continuous generative models based on expectation values in general, including the standard style-based quantum GANs.

The paper is organized as follows. \secref{sec:theory} introduces the general training framework of our novel \model approach. We apply the proposed model to three different datasets and summarize the training results in \secref{sec:results}. In addition, we compare the performance of \model with a classical GAN which has the same training schema but with a classical generator. 
The results empirically demonstrate that \model outperforms the classical generator with a similar model size on these particular tasks. In \secref{sec:statistics}, we study the impact of the finite number of measurements and argue the robustness of the model against the statistical fluctuations. We confirm that the errors due to the shot noise cannot be detected by the standard methods for image evaluation. \secref{sec:bp} investigates the trainability of \model in the case of the shallow-depth circuit with numerical simulations and extends the study to the deep-depth circuit case. Finally, in \secref{sec:conclusion}, we summarize our study with proposals for future research.

\section{\label{sec:theory} General Framework}
\subsection{Overall training schema}

Our work proposes a hybrid classical-quantum GAN approach, so-called, \textit{Latent Style-based Quantum GAN (LaSt-QGAN)}, which integrates two distinct components: a classical autoencoder and a quantum GAN. The autoencoder is an unsupervised neural network used for dimensionality reduction and data compression. It consists of an \textit{encoder}, which embeds the high dimensional data into a low dimensional latent space, and a \textit{decoder}, which reconstructs the data from these latent features. 
In our approach, the autoencoder functions as an invertible image preprocessing tool, efficiently reducing the dimensionality of the complex images and reconstructing fake images from the generated latent features.
On the other hand, the quantum GAN serves as a generative model for producing fake features, employing a quantum \textit{generator} and a classical \textit{discriminator}. The quantum generator is responsible for generating fake features from a randomly sampled noise, while the classical discriminator differentiates between real and fake features. 

\begin{figure}[h]
    \centering
    \includegraphics[width = 0.49\textwidth, trim = {3.7cm, 4.8cm, 7.25cm, 1.75cm}]{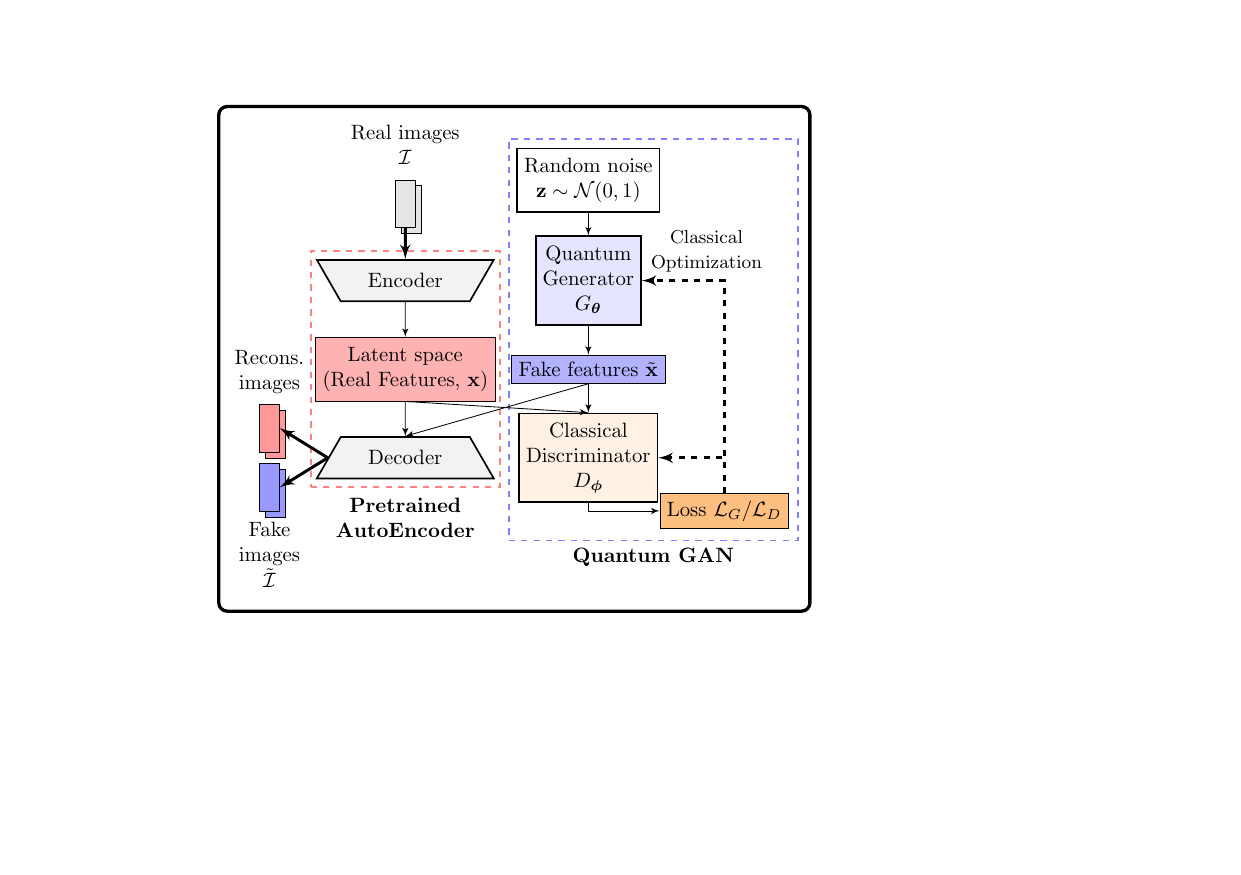}
\caption{\textbf{Schematic diagram for \model training.} The model consists of a convolutional auto-encoder that embeds the original images into a low-dimensional latent space and a quantum GAN with a quantum generator $\G$ and a classical discriminator $\D$. The features extracted with the autoencoder are used as the training set of the GAN. At the end of the training, images are reconstructed by inversely transforming the features generated by the quantum generator using the pre-trained convolutional auto-encoder.}
    \label{fig:GAN}
\end{figure}

The overall training schema of \model is illustrated on \figref{fig:GAN}. First of all, we extract the essential features, denoted as $\mathbf{x}\sim\mathbb{P}_r$, in the latent space of dimension $\Dl$  from real images $\img$ via a classical convolutional auto-encoder. The auto-encoder is pre-trained on the original image dataset, thus used as an invertible dimensionality reduction technique.
Those extracted features are utilized as the real training dataset $\mathcal{X}_{train} \subset \mathbb{R}^{\Dl}$ for the quantum GAN training. At each step, $G_{\bm{\theta}}$ reproduces fake data $G_{\bm{\theta}}(\mathbf{z}) = \mathbf{\tilde{x}}\in \tilde{\mathcal{X}} \sim\mathbb{P}_g,\tilde{\mathcal{X}} \subset \mathbb{R}^{\Dl}  $ from a latent noise $\mathbf{z} \in \mathbb{R}^{\Dz}$ sampled randomly from a prior $\mathbb{P}_\mathbf{z}$. Then, the fake and the real features are given as input to the discriminator $D_{\bm{\phi}}$, which returns a scalar value measuring the \textit{realness} of the samples (i.e., a larger value implies an image is more likely to be real). We note that for the following of the paper, we will keep the tilde mark $\tilde{\circ}$ to denote the generated samples. 

Additionally, Wasserstein loss with gradient penalty~\cite{Arjovsky2017wGAN,Gulrajani2017wasserstein} is used for better convergence in the model. The gradient penalty corresponds to a regularization term to enforce Lipschitz constraint on the gradients of the discriminator (often called as \textit{critic} in Wasserstein GAN). This helps to avoid the vanishing gradients in the generator observed in the classical GAN, by excluding the sigmoid functions in the discriminator activations~\cite{Arjovsky2017wGAN}.  In this setup, the generator and the discriminator loss functions measure the Wasserstein distance or the \textit{Earth Mover} distance between the output distributions of the real and the fake samples, with the following expression:
\begin{align}
    \mcl_G(\bm{\theta}, \bm{\phi}) = \;& - \mathop{\mathbb{E}}_{\mathbf{z}\sim \mathbb{P}_\mathbf{z}}[D_{\bm{\phi}}\left(G_{\bm{\theta}}(\mathbf{z})\right)] 
    \label{eq:loss_G} \\
\mcl_D(\bm{\theta}, \bm{\phi}) =\; &  - \mathop{\mathbb{E}}_{\mathbf{x}\sim \mathbb{P}_r}[D_{\bm{\phi}}(\mathbf{x})] \nonumber +  \mathop{\mathbb{E}}_{\mathbf{z}\sim \mathbb{P}_\mathbf{z}}[D_{\bm{\phi}}(G_{\bm{\theta}}(\mathbf{z}))]   \\ 
            & + \lambda \mathop{\mathbb{E}}_{\hat{\mathbf{x}}\sim\mathbb{P}_{\hat{\mathbf{x}}}}[(||\nabla_{\hat{\mathbf{x}}} D_{\bm{\phi}}(\mathbf{\tilde{x}})||_2 -1)^2 ]
\label{eq:wasserstein}
\end{align}
where the last term corresponds to the gradient penalty of the discriminator. In this term, $\hat{\mathbf{x}} = \epsilon \mathbf{x} + (1-\epsilon)\tilde{\mathbf{x}}$ correspond to points interpolated between real and generated samples with a random value $\epsilon$ sampled from a uniform distribution and $\lambda$ the penalty coefficient. These formulas imply that the discriminator aims to maximize the distance, while the generator aims to minimize it. 

At the end of the training, the generated data distribution $\mathbb{P}_g$ should approach as close as possible to the real data distribution $\mathbb{P}_r$. The generated features are then passed back to the and inversely transformed into images $\tilde{\img}$. Thanks to continuity in the latent space, the inverse transform of the generated features leads to the reconstruction of the correct images in the image space. 

\begin{figure*}[t]
\subfloat[Circuit1.]{
\includegraphics[width = 0.48\textwidth,  trim = {2.6cm, 17cm, 1.8cm, 2cm }]{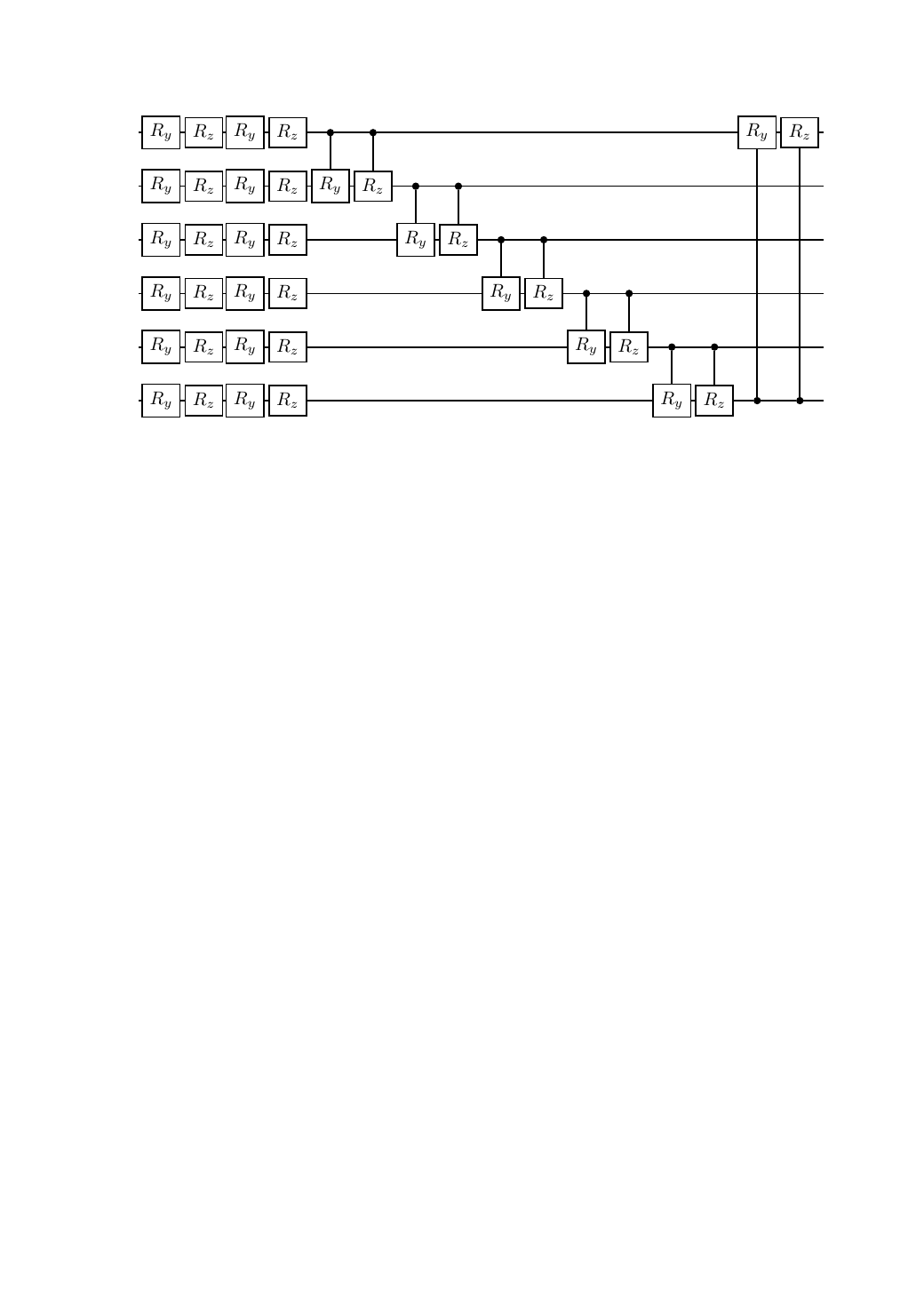} 
}
\hspace*{\fill}
\subfloat[Circuit2.]{
\includegraphics[width = 0.18\textwidth,  trim = {2.6cm, 17cm, 10cm, 2cm }]{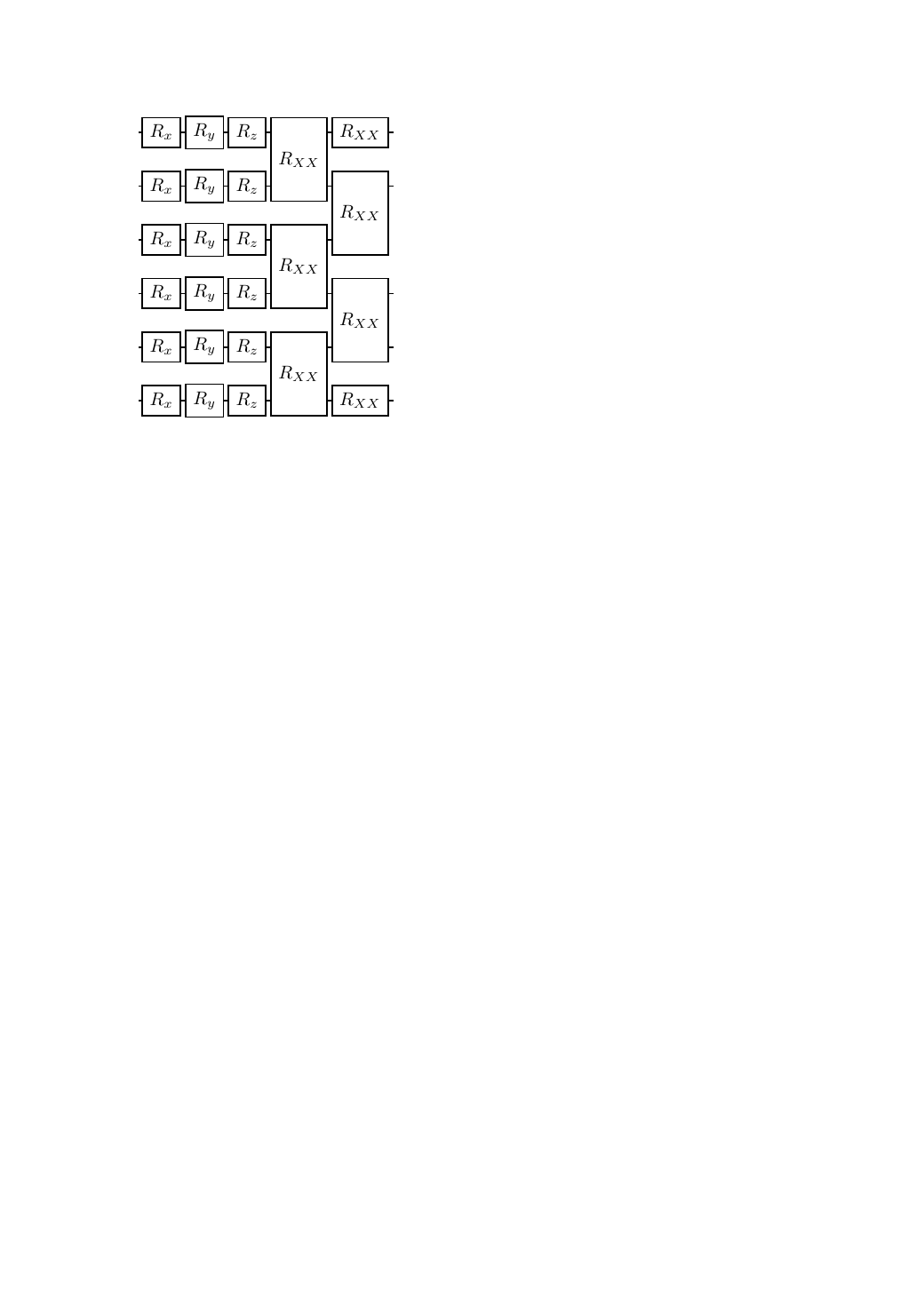} 
}
\hspace*{\fill}
\subfloat[Circuit3.]{
\includegraphics[width = 0.33\textwidth,  trim = {2.2cm, 16.5cm, 5.9cm, 2cm }]{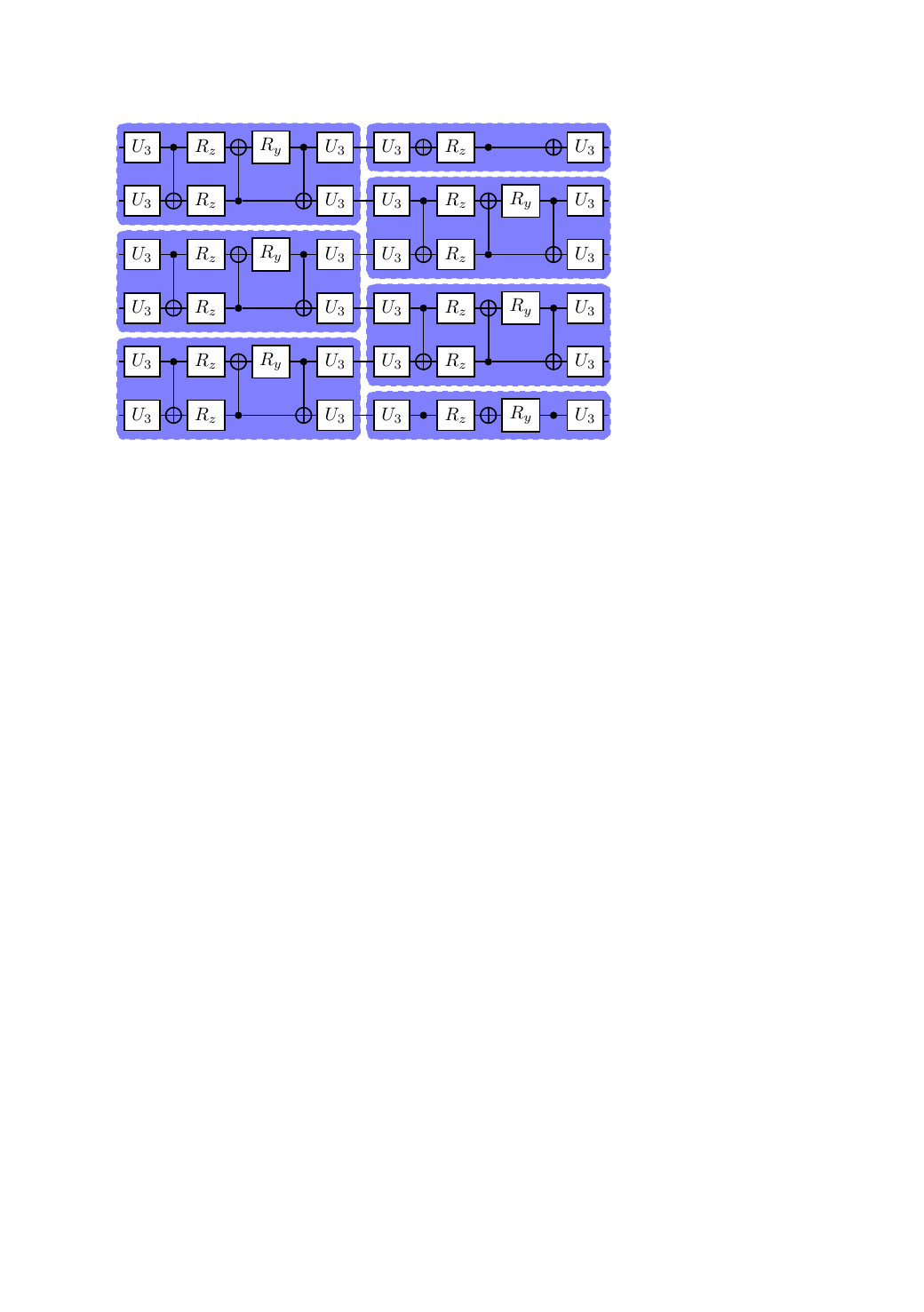} 
}
 \caption{\textbf{Different circuit architecture used for learning layers, $U^\ell_{\bm{\theta}} $, in the quantum generator.} (a) Circuit1 and (b) Circuit2 are taken from two different quantum GAN papers for continuous data generation by C. Bravo Prieto \etal~\cite{BravoPrieto2022} and J. Romero \etal~\cite{Romero2019}, respectively. (c) Circuit3 is composed of repeated two-qubit quantum circuits (blue square), responsible for an arbitrary $SU(4)$ state generation~\cite{MacCormack2020SU4}.
}
 \label{fig:circuits}
\end{figure*}

\subsection{Style-based quantum generator}
The quantum generator takes the form of a parameterized quantum circuit, also known as a quantum neural network (QNN). The $n$-qubit unitary quantum circuit $\mathcal{U}_{\bm{\theta}}(\mathbf{z})$ with parameters $\bm{\theta}$ transforms the classical latent noise $\mathbf{z}$ into an encoded quantum state $\ket{\Psi_{\bm{\theta},\mathbf{z}}}\in \mathcal{H}$ with $\mathcal{H}$ as a $2^n$ dimensional Hilbert space. In other words, the parameterized circuit acts as a feature map that maps for the classical input.

Unlike the architecture firstly introduced in Ref~\cite{Romero2019} where the classical noise embedding layer and trainable layers are separated, the particularity of the style-based architecture is that the rotation angles in the learning layers are also parameterized by the latent noises. Mathematically, the unitary transformation $\mathcal{U}_{\bm{\theta}}(\mathbf{z})$ can be written as a $L$-repetition of learning layers $U^\ell_{\bm{\theta}_\ell}(\mathbf{z})$ parameterized by the set of parameters $\bm{\theta}_\ell$ for each layer $\ell = 1,...,L$  : 
\begin{equation}
    \mathcal{U}_{\bm{\theta}}(\mathbf{z}) = U^L_{\bm{\theta_L}}(\mathbf{z})\cdots U^1_{\bm{\theta_1}}(\mathbf{z}). 
\end{equation}
Then, the latent vectors, $\mathbf{z}$ are embedded into the angles of qubit rotation action, $\bm{\theta}_\ell$ for each layer $\ell$ by an affine transformation : 
\begin{equation}
    \bm{\theta}_\ell = W_\ell\mathbf{z} + \mathbf{b}_\ell
\end{equation}
where $W_\ell$ is the weight matrix of size $N_{\bm{\theta}} \times \Dz$ with $N_{\bm{\theta}}$ the number of rotation angles in QNNs and $\mathbf{b}_\ell \in \mathbb{R}^{N_{\bm{\theta}}}$ the bias. During the training, the model will be trained by varying $ \bm{\Theta} = \{W_{\ell}, \mathbf{b}_{\ell}\}_{\ell = 1,..., L}$ with $N_{\bm{\Theta}}$ the total number of trainable parameters. 
This can also be regarded as an equivalence of data reuploading technique, where the input data are embedded into the rotation angles in the learning layers for classification task~\cite{PerezSalinas2020datareuploading, Easom2021datareuploding}, in the context of generative models. 

\figref{fig:circuits} illustrates three different circuit types for a single parameterized layer, $U^\ell_{\bm{\theta}}(\mathbf{z})$,  used in this paper for numerical simulations. Circuit1 is the quantum circuit architecture employed in style-based quantum GAN for Monte Carlo event generation~\cite{BravoPrieto2022}, and Circuit2 is inspired by the quantum circuit presented in Ref.~\cite{Romero2019}, used as a variational quantum generator (VQG) for continuous distribution. Additionally, we also consider Circuit3, which consists of repeated two-qubit quantum filters (blue square). These filters are responsible for generating an arbitrary $SU(4)$ state~\cite{MacCormack2020SU4}, thus serving as a universal quantum state generator at least at the level of two-qubits. 

After transforming the classical input through the QNN, we measure the expectation values of some observables at the end of the generator to extract some information from the encoded state. Unlike the original architecture~\cite{BravoPrieto2022}, which performs only the measurement of the Pauli Z operator, $\sigma_z$, our architecture uses expectation values of both Pauli X and Z operators, $\sigma_x$ and $\sigma_z$. The measured values are then concatenated into a single vector, also called a latent feature, which will be given as input to the discriminator:  
\begin{equation}
\mathbf{x} = \{\langle\sigma_{x}^1\rangle_{\mathbf{z}, \bm{\theta}},...,\langle\sigma_{x}^{n}\rangle_{\mathbf{z}, \bm{\theta}}, \langle\sigma_{z}^1\rangle_{\mathbf{z}, \bm{\theta}},...,\langle\sigma_{z}^{n} \rangle\}_{\mathbf{z}, \bm{\theta}} \in \mathbb{R}^{2n} ,  
\end{equation}
where $\langle\sigma_{x}^i\rangle_{\mathbf{z}, \bm{\theta}} ,\langle\sigma_{z}^i\rangle_{\mathbf{z}, \bm{\theta}}$ denote the expectation values of $\sigma_x$ and $\sigma_z$ on $i$-th qubit for an input latent noise $\mathbf{z}$ and the generator angle $\bm{\theta}$, i.e., : 
\begin{equation}
    \langle\sigma_\mu^i\rangle_{\mathbf{z}, \bm{\theta}} = \langle 0 | \mathcal{U}_{\bm{\theta}}(\mathbf{z})^\dagger \sigma_\mu^i \mathcal{U}_{\bm{\theta}}(\mathbf{z}) |0\rangle
\end{equation}
with $\mu \in \{x, z\}$. 
We note that this strategy does not satisfy the sufficient conditions specified in Ref.~\cite{barthe2024expressivity} for universality. Nevertheless, the numerical results in the following sections demonstrate the model can be adequately used to generate the samples in the training set for the given tasks. More generally, one can employ a polynomial number of expectation values to construct a latent feature with a larger dimension.

We note that the multi-observable or multi-basis strategy has also been employed in the previous study for multi-classification task~\cite{Zeng2021} or probability learning task~\cite{Rudolph2022} to capture the hidden information of the quantum circuit adequately.
This way of interpreting the quantum output state allows using only $n$ qubits for $\Dl = 2n$ values, also bringing an advantage in terms of quantum resources. 

\subsection{\label{sec:metrics}Evaluation metrics}
\renewcommand{\tabcolsep}{2.5pt}
\begin{figure*}[t]
    \centering
    \begin{tabular}{|c||c|c|c|}
    \hline
         & MNIST & FashionMNIST & SAT4 \\ 
         \hline
         \rotatebox{90}{\model} &  
         \includegraphics[width = 0.3\textwidth]{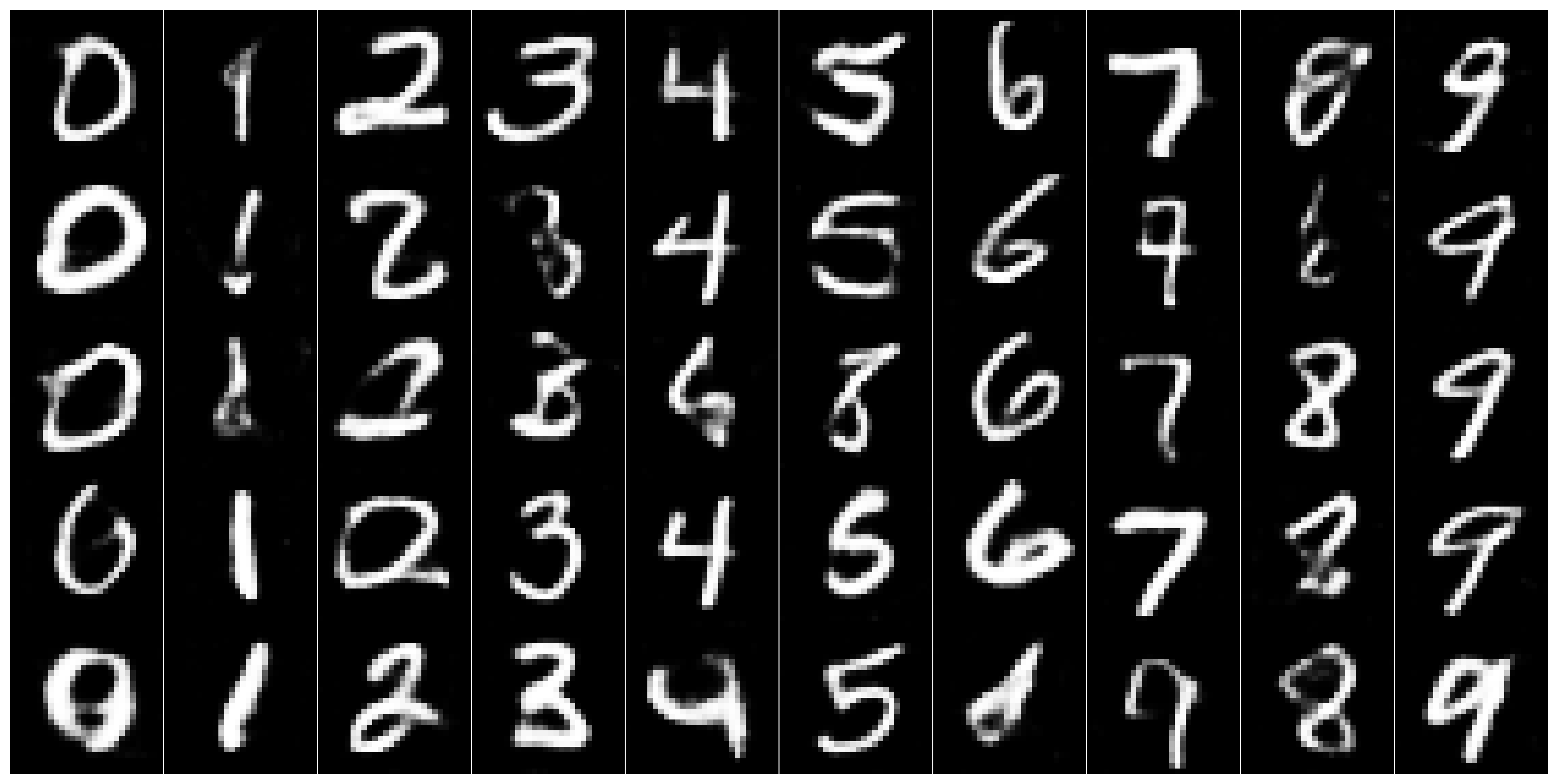} & 
         \includegraphics[width = 0.3\textwidth]{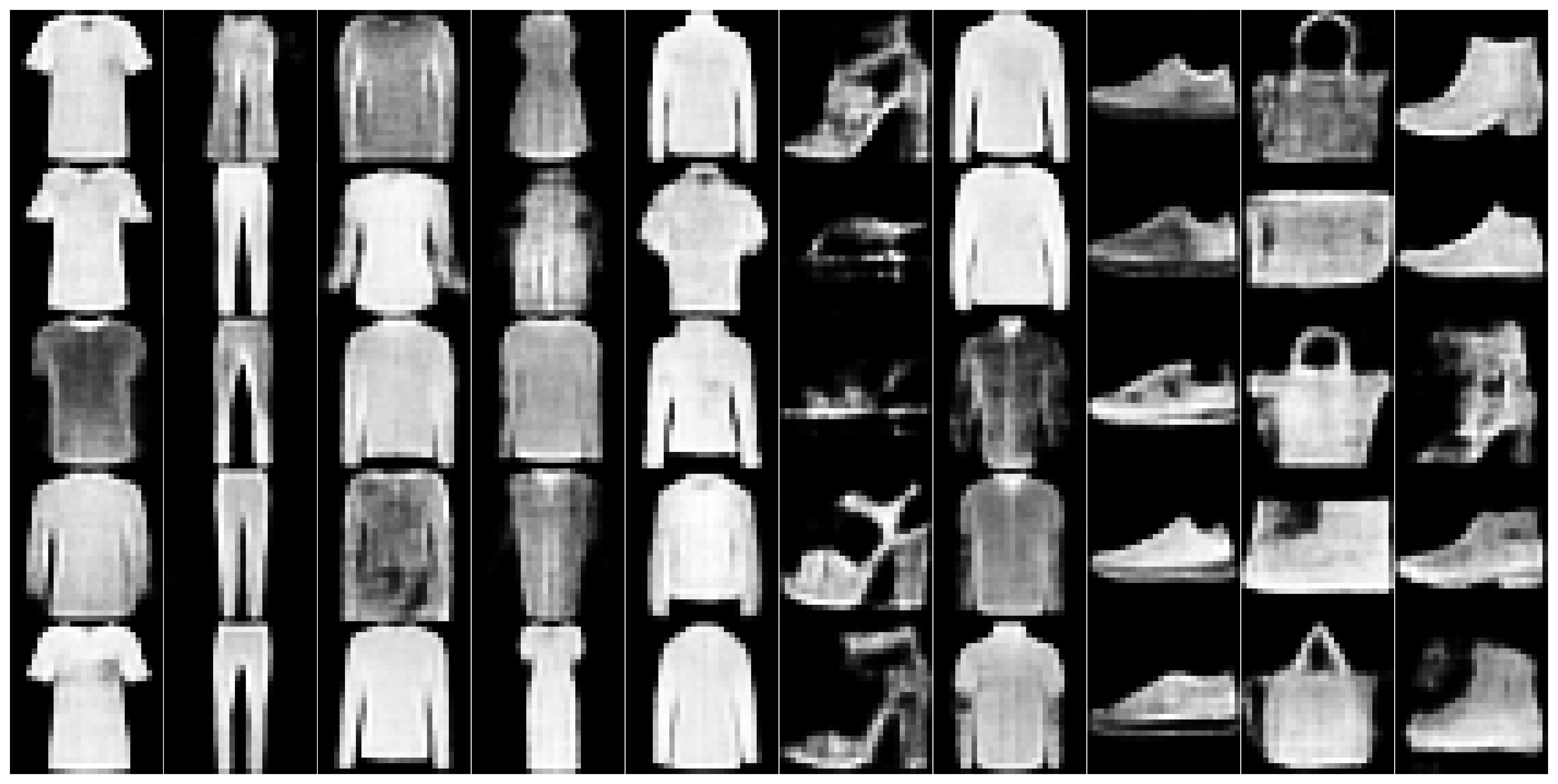} & 
         \includegraphics[width = 0.3\textwidth]{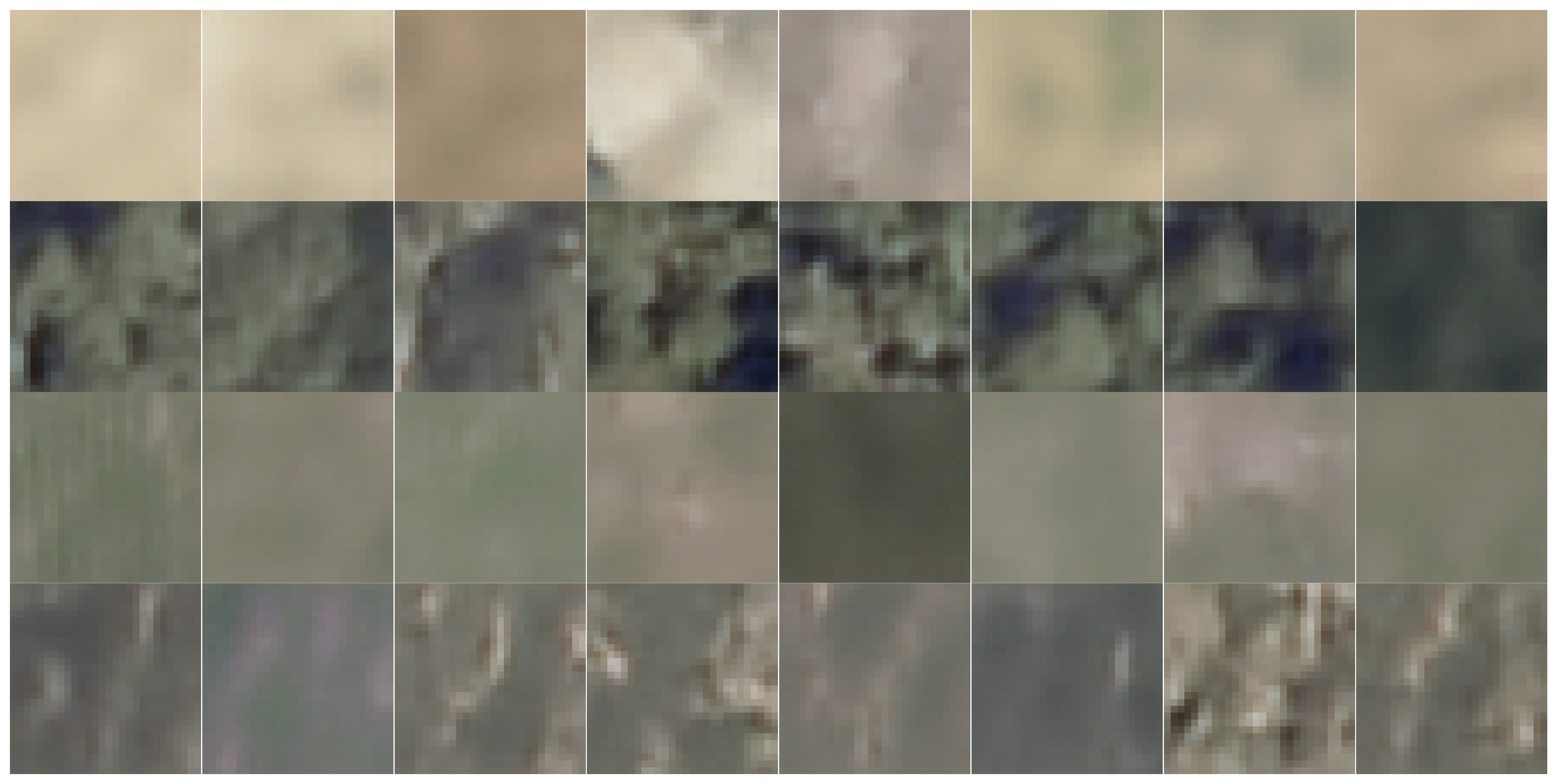} \\ \hline
         \rotatebox{90}{Classical GAN}& 
         \includegraphics[width = 0.3\textwidth]{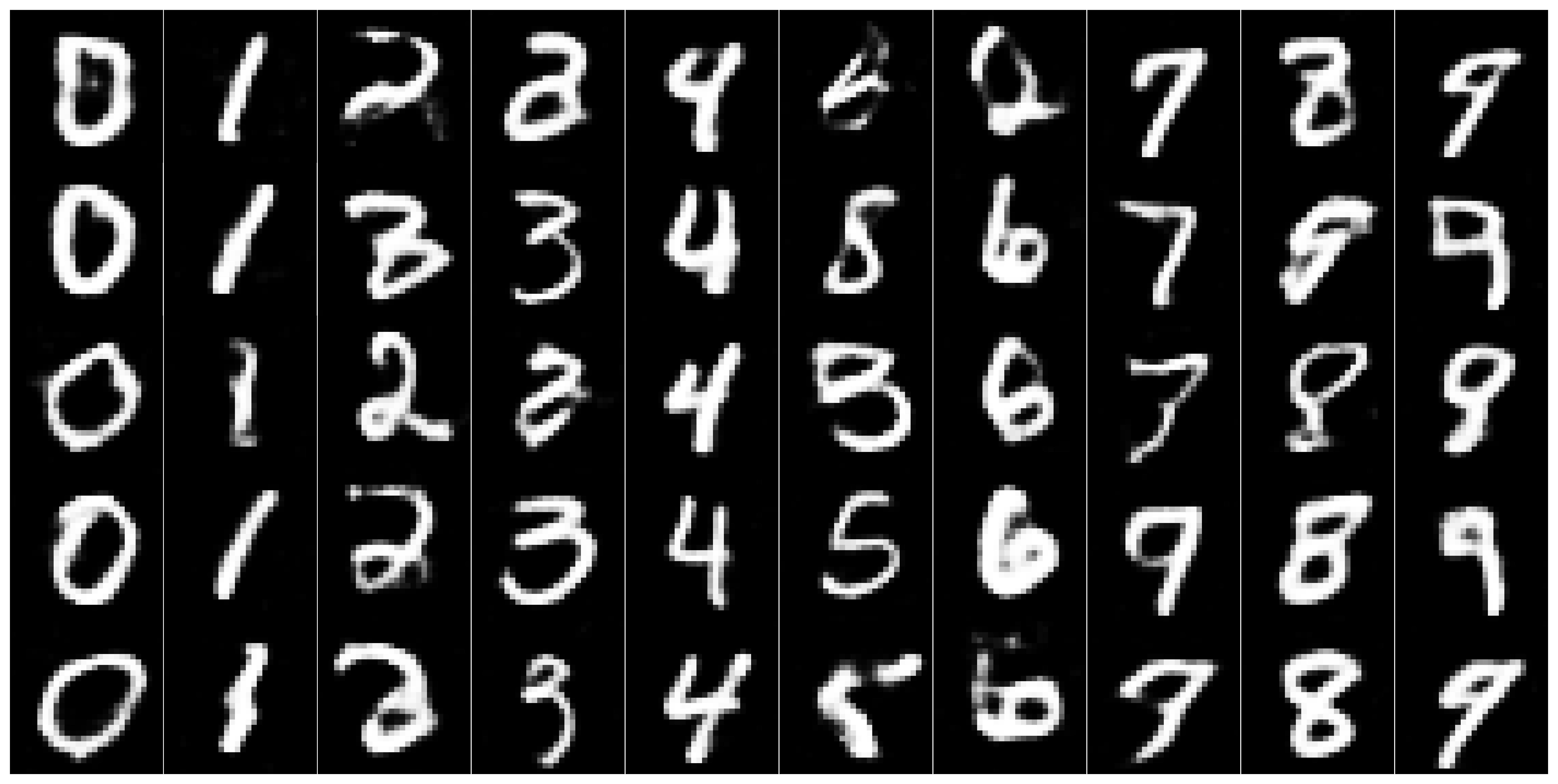} & 
         \includegraphics[width = 0.3\textwidth]{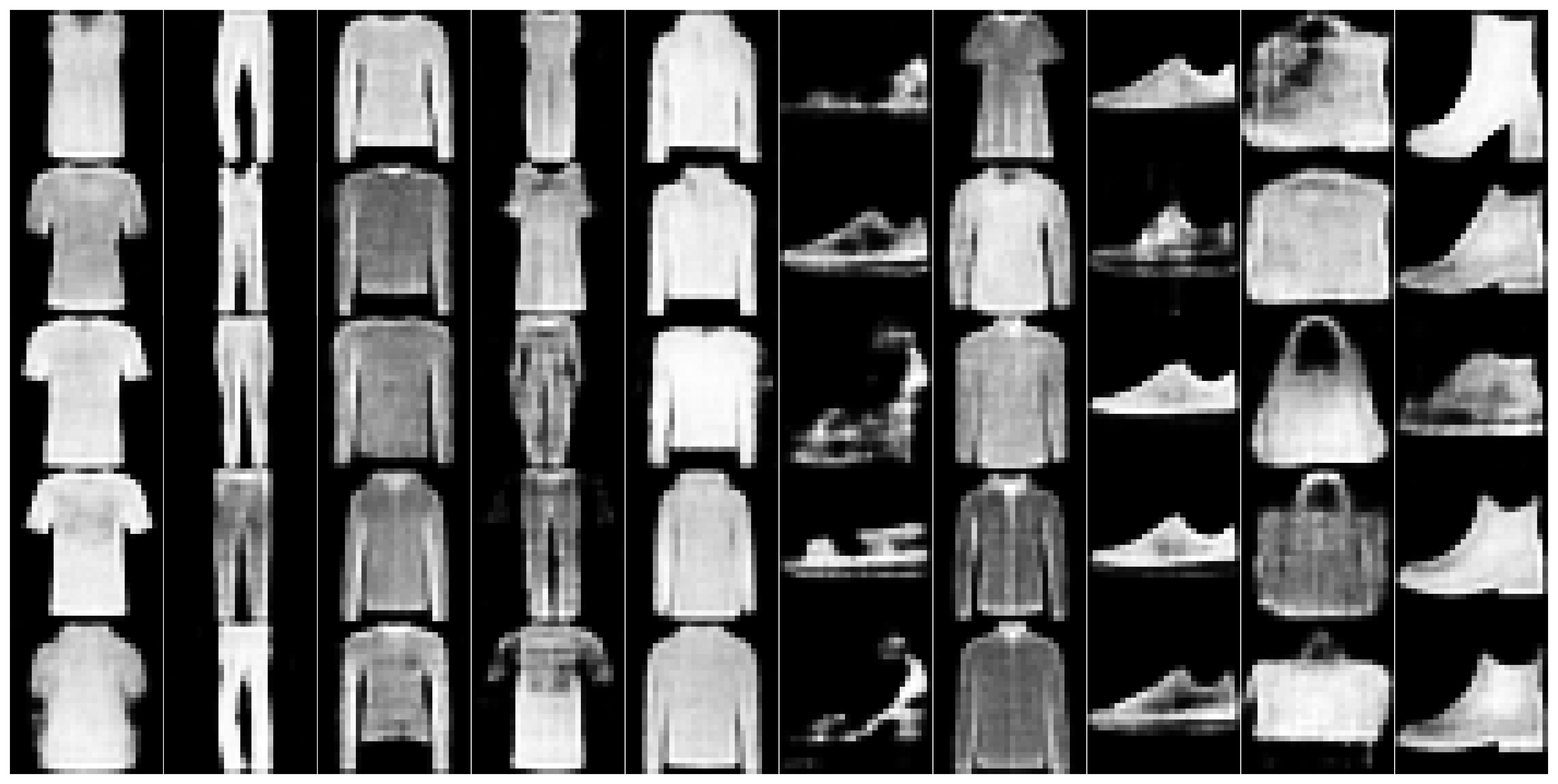} &
         \includegraphics[width = 0.3\textwidth]{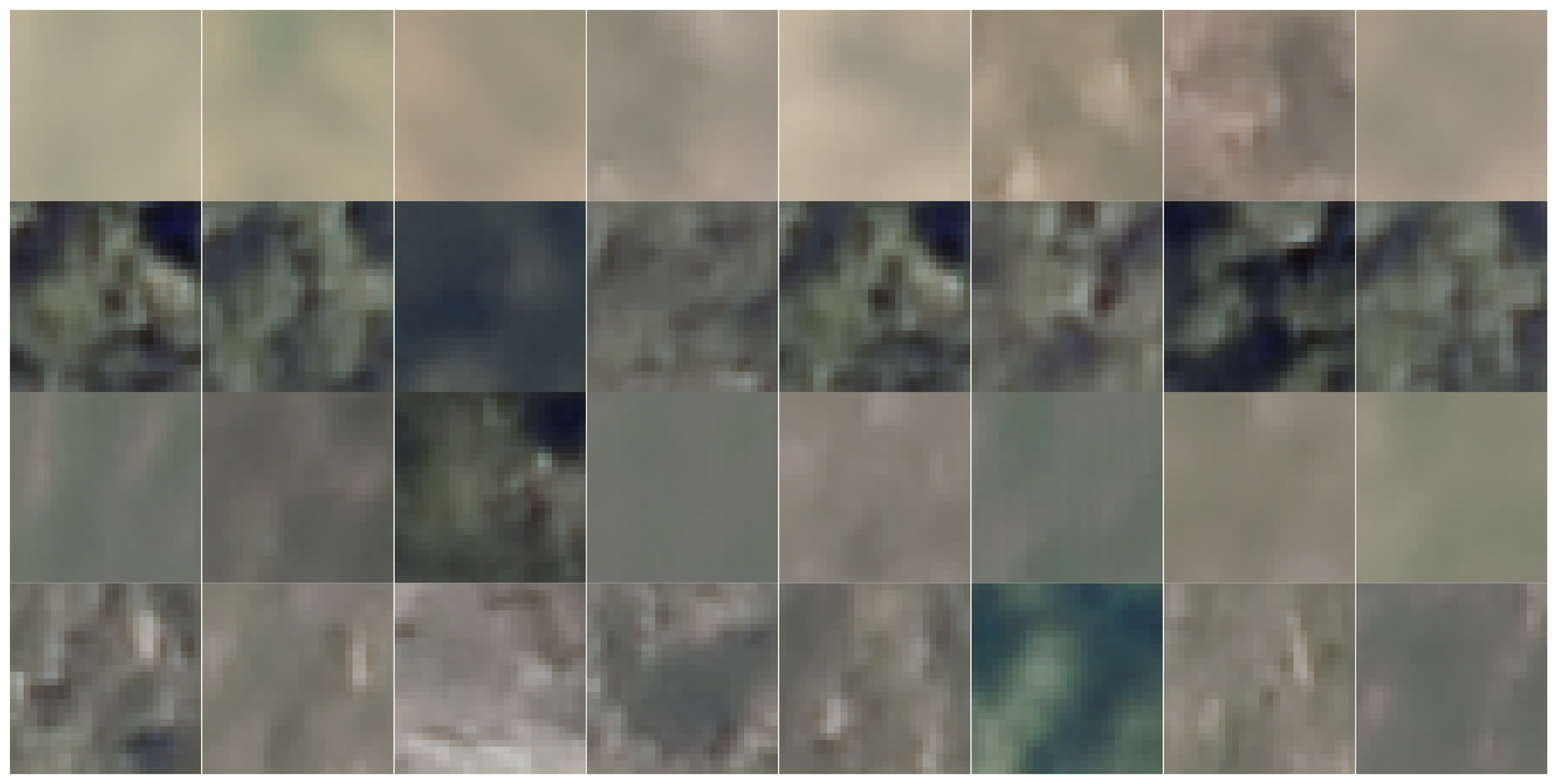}\\ \hline
    \end{tabular}
    \caption{\textbf{Examples of images generated via \model (Circuit1, depth 2) and a classical GAN ([50, 30]) for different datasets: MNIST, FashionMNIST and SAT4.} The fake features are obtained using $\Dz = 10$ and $\Dl = 20$ and the images are reconstructed using a pre-trained convolutional auto-encoder from the features obtained by the GAN in the latent space. The images are presented in columns classified using a pre-trained ResNet50~\cite{He2015resnet} for MNIST and FashionMNIST, and in rows for SAT4. } 
    \label{fig:generated_images}
\end{figure*}

Unlike the classification task, where the evaluation methods are quite straightforward for the final test accuracy, it is less clear how to evaluate the generative models.
There have been efforts to define the appropriate metrics to evaluate the performance of Quantum GAN in previous studies. However, the proposed metrics are limited for discrete generative models~\cite{gili2023generalization} as they require one-to-one comparisons of the dataset, or for continuous data with small dimensions~\cite{riofrío2023performance} where the direct comparison of the probability distribution is available. Therefore, it is important to choose the appropriate metrics to compare the performance between models for image generation. In this paper, we evaluate the performance of GAN with three different metrics: Inception Score (IS)~\cite{Salimans2016gans, barratt2018IS}, Fr\'echet Inception Distance (FID)~\cite{Heusel2017gans, Karras2018StyleGAN} and Jensen-Shannon divergence (JSD)~\cite{metrics_gan}.

IS evaluates the quality and the diversity of generated images by calculating the Kullback-Leibler (KL) divergence between marginal distributions obtained by summing up the outputs of the Inception V3 Network~\cite{Szegedy2016inceptionv3} applied on real and generated images. Inception V3 Network is a convolutional neural network widely used in classical ML for image recognition task, pretraind on ImageNet dataset (only used for metrics calculation in this paper). Note that the KL divergence for discrete distribution is computed with the following formula : 
\begin{equation}
    D_{KL}(P||Q) = \sum_{\mathbf{x}\in\mathcal{X}} P(\mathbf{x})\log(\frac{P(\mathbf{x})}{Q(\mathbf{x})})
\end{equation}
for the empirical distribution $P$ and target distribution $Q$ defined on discrete space $\mathcal{X}$.  The maximum value of IS is the number of classes in the dataset, and the higher the IS value, the better the result.

FID also measures the quality of the images using the output of the Inception V3 model, but calculates the Frechet distance between the real and fake embedding from the model given by the expression: 
\begin{equation}
    d(X, Y) = ||\mu_X -  \mu_Y||^2 - \Tr(\Sigma_X + \Sigma_Y - 2\sqrt{\Sigma_X\Sigma_Y})
\end{equation}
where $\mu_X, \mu_Y$ are the mean vector of multi-dimensional data $X$ and $Y$ (in this case, the real and fake embedding from the Inception V3) and $\Sigma_X, \Sigma_Y$ their covariance matrices. This distance measures the similarity between the distribution of the real and the generated images in the feature space obtained using the Inception V3. The lower the FID value, the better the quality of the images. 

Finally, the Jensen-Shannon divergence measures the distance between two discrete probability distributions, similar to the Kullback-Leibler divergence but symmetric and smoother with the following formula : 
\begin{align}
    D_{JS}(P||Q)  =\;& \frac{1}{2}D_{KL}\Big(P\big\|\frac{1}{2}(P+Q)\Big) \nonumber  \\ 
    &+ \frac{1}{2}D_{KL}\Big(Q\big\|\frac{1}{2}(P+Q)\Big).
\end{align}
As JSD requires discrete distributions, in the case of the unlabelled continuous dataset, we first classify the train samples into $K$ bins using K-mean clustering to generate a discrete target distribution, $Q$. The generated samples are also classified according to the lowest distance from the centers of the train set clusters, returning the generated distribution $P$, on which we compute the JSD value. The lower the JSD, the more diverse the images.

\section{\label{sec:results} Main results}

\subsection{Experimental setup}

This section presents the results of \model trained on MNIST~\cite{lecun2010mnist} ($28\times 28 \times 1$ pixels), Fashion MNIST~\cite{xiao2017fashionMNIST} ($28\times 28 \times 1$ pixels) and SAT4~\cite{SAT4} ($28\times 28 \times 4$ pixels), which contains 4 classes of Earth Observation images with RGB and Near Infrared channels. We then compare them with the results of their classical counterpart. To keep the latent space embedding model comparable, we used only the RGB channels in the SAT4 dataset. Unless specified, the same model architecture and hyperparameters are used for the following simulations. 
\renewcommand{\tabcolsep}{2.5pt}

\begin{table*}[t]
    \centering
    \begin{tabular}{c|c|c|c|c|c|c}
          & $G_{\bm{\theta}}$ config. & $\Np$ & FID $\downarrow$  & IS $\uparrow$   & JSD (features/$10^{-2}$)  $\downarrow$ & JSD (images/$10^{-2}$) $\downarrow$\\
         \hline
         \hline
          \multirow{7}{*}{LaSt-QGAN}   & Circ.~1 ($d = 2$) & 1360 & $17.2 \pm 0.35$ &  $8.29 \pm 0.02$ &  $0.79 \pm 0.05$ &  $1.63 \pm 0.09$ \\
         & Circ.~1 ($d = 4$) & 2280 & $14.85 \pm 0.34$ &  $8.49 \pm 0.04$&  $0.75 \pm 0.07$&  $1.49 \pm 0.18$ \\
         & Circ.~1 ($d = 6$) &  3200 &$14.13 \pm 0.73$ &   $8.53 \pm 0.05$ &  $\mathbf{0.71 \pm 0.07}$ &  $1.29 \pm 0.1$ \\ 
         \cline{2-7}
         & Circ.~2 ($d = 2$) &  1010 &$19.13 \pm 0.54$ &  $8.10 \pm 0.06$ &  $1.22 \pm 0.19$ &  $2.08 \pm 0.17$\\
         & Circ.~2 ($d = 4$) &  1690 &$16.2 \pm 0.32$ &  $8.34 \pm 0.03$ &  $0.94 \pm 0.09 $ &  $1.66 \pm 0.17$\\
         & Circ.~2 ($d = 6$) &  2370 &$14.85 \pm 0.61$ &  $8.47 \pm 0.06$ &  $0.85 \pm 0.05$ &  $1.39 \pm 0.11$\\
         
         \cline{2-7}
         & Circ.~3 ($d = 2$) &  3300 &$14.29 \pm 0.38$ &  $8.5 \pm 0.04$ &  $0.76 \pm 0.06$ &  $1.5 \pm 0.12$\\
         & Circ.~3 ($d = 4$) &  6600 &$12.72 \pm 0.4$ &  $8.65 \pm 0.05$ &  $0.71 \pm 0.07$ &  $1.14 \pm 0.12$\\
         & Circ.~3 ($d = 6$) &  9900 &$\mathbf{11.99 \pm 0.56}$ &  $\mathbf{8.71 \pm 0.04}$ &  $0.72 \pm 0.09$ &  $\mathbf{1.13 \pm 0.12}$\\
         \hline
         \hline
         \multirow{2}{*}{Classical}  & $[50, 30]$ & 2960 &$18.24 \pm 3.6$ & $8.24 \pm 0.28$ & $3.74 \pm 1.64$ & $4.51 \pm 2.0$  \\
         & $[100, 50]$ & 7660 & $12.56 \pm 0.91$ & $8.8 \pm 0.06$ & $1.18 \pm 0.17$ & $1.56 \pm 0.13$ \\ \hline
    \end{tabular} 
    \caption{\textbf{Training results of \model and the classical GAN for MNIST dataset.} Number of parameters used in the generator and different metrics (averaged over 10 runs) to compare the performance of \model and the corresponding classical GAN using 10,000 generated images (best results highlighted in bold). For FID and JSD, the lower, the better and for IS, the higher, the better. We observe that with a similar model size ($\approx 3k $ parameters), \model outperforms the classical GAN for all metrics. Note that our results are close to the result of SoTA vanilla GAN models, which have FID of 7.87~\cite{Wei2021duelGAN} and 12.88~\cite{Lazcano2021hypGAN}.}
    \label{tab:metrics_MNIST}
\end{table*}

The images are embedded into the latent space of dimension $\Dl = 20$ with a convolutional auto-encoder. The detailed architecture of the auto-encoder is given in Appendix.~\ref{adx:autoencoder}. As the output of quantum generator $\langle\sigma_x\rangle$ and $\langle\sigma_z \rangle$ are defined in $[-1, 1]$, the latent space should also be constrained in the same interval. The quantum generator takes $n = 10$ qubits with the latent noises of $\Dz = 10$, each component sampled independently from a normal distribution, $\mathcal{N}(0,1)$. To guarantee the convergence of the model, the initial quantum generator parameters, $W_\ell$ and $\mathbf{b}_\ell$, are chosen randomly from a uniform distribution between $[-0.01, 0.01]$. 
The classical discriminator consists of two hidden dense layers with 100 and 50 nodes for MNIST and FashionMNIST and 200 and 100 nodes for the SAT4 dataset, followed by leaky Relu activation functions and an output node of size one. 

To assess the performance of \model against classical models, we construct a classical GAN that follows the identical training framework as LaSt-QGAN, as depicted in \figref{fig:GAN}, but employing a classical linear generator instead of a quantum one. 
The classical generator consists of an input layer with $\mathcal{D}_{\mathbf{z}}$ nodes, two hidden layers with $[h_1, h_2]$ nodes followed by a leaky Relu activation function, and the output layer with $\Dl$ nodes attached to a Tanh function to constrain the generator output between -1 and 1.  
For the following simulations, we consider two different classical generators: 1) $[h_1, h_2],  = [50, 30]$,  2) $[h_1, h_2] = [100, 50]$. The first one is chosen to have a similar number of parameters as the quantum generators, while the second one is constructed to have the same hidden layers as the discriminator for a balanced GAN architecture. In order to also guarantee faster convergence for classical neural networks, we use LeCun normal initialization~\cite{lecun2010mnist} for the parameters. 

In all cases, the model parameters are updated with an Adam optimizer using a learning rate of 0.001 for both discriminator and generator with $\beta_1 = 0.5$ and $\beta_2 = 0.999$. Those hyperparameters are chosen empirically to assure the fastest convergence and stability of the model.  For loss calculation, $\lambda = 10$ is chosen as the penalty coefficient (c.f. \eqnref{eq:wasserstein}). 

Our codes used Jax~\cite{jax2018github} and Flax~\cite{flax2020github} packages for training algorithms implementation, and Pennylane~\cite{pennylane2018} for quantum circuits construction and optimization. 

\subsection{\label{sec:generic_results}Generic results}

We display in \figref{fig:generated_images} the images of different datasets generated by \model 
 and the corresponding classical counterpart using the features extracted by the pre-trained convolutional auto-encoder. The results prove that the model can reproduce images correctly, although further improvements are required for a higher quality of the results. 

 \begin{figure}[!h]
\centering
\hspace*{\fill}\subfloat[\label{fig:tsne_MNIST}MNIST]{\includegraphics[width = 0.49\textwidth]{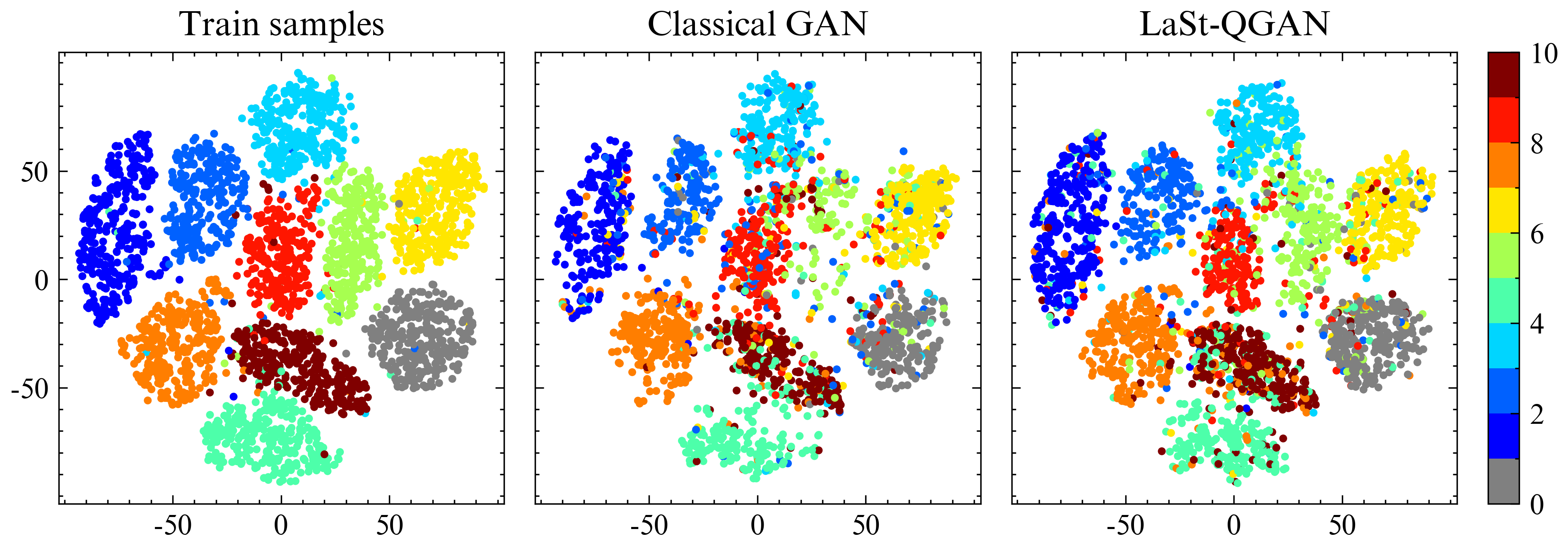}}
\hspace*{\fill}
\newline
\hspace*{\fill}\subfloat[\label{fig:tsne_Fashion}FashionMNIST]{\includegraphics[width = 0.49\textwidth]{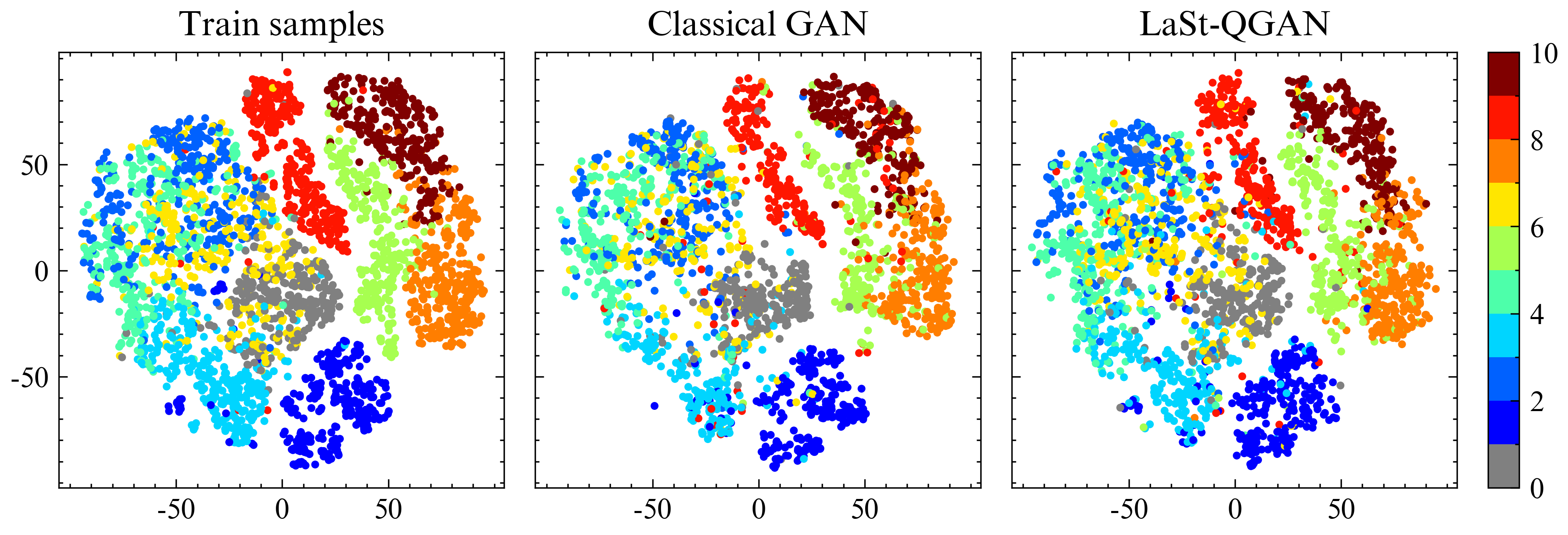}}
\hspace*{\fill}
\caption{\label{fig:tsne} \textbf{Visualization of generated features embedded into two dimensions using t-SNE for MNIST and FashionMNIST dataset.} The labels of generated samples are obtained via classification with pre-trained ResNet50. The clustering of features reveals that the underlying similarity in each class is preserved in the latent space with the proposed models. }
\end{figure}

\begin{table*}[t]
    \centering
    \begin{tabular}{c|c|c|c|c|c|c}
          & $G_{\bm{\theta}}$ config. & $\Np$. & FID $\downarrow$  & IS $\uparrow$   & JSD (features/$10^{-2}$) $\downarrow$ & JSD (images/$10^{-2}$) $\downarrow$\\
         \hline
            \multirow{7}{*}{LaSt-QGAN}  & Circ.~1 ($d = 2$) & 1360 & $29.42 \pm 0.59$ &  $8.27 \pm 0.04$ &  $1.01 \pm 0.095$ &  $1.61 \pm 0.2$ \\
         
         & Circ.~1 ($d = 4$) &  2280 &$27.59 \pm 0.56$ &   $8.37 \pm 0.02$ &  $0.85 \pm 0.05$ & $1.42 \pm 0.1$ \\
         & Circ.~1 ($d = 6$) & 3200 & $26.89 \pm 0.57$ &  $8.44 \pm 0.02$&  $0.76 \pm 0.08$&  $\mathbf{1.28 \pm 0.11}$ \\
         \cline{2-7}
         & Circ.~2 ($d = 2$) &  1010 & $32.26 \pm 0.43$ &  $8.12 \pm 0.05$ &  $1.28 \pm 0.12$ &  $2.06 \pm 0.14$\\
         & Circ.~2 ($d = 4$) &  1690 & $29.2 \pm 0.3$  &  $8.34 \pm 0.03$ &  $0.94 \pm 0.09 $ &  $1.69 \pm 0.12$\\
         & Circ.~2 ($d = 6$) &  2370 & $28.1 \pm 0.77$ &  $8.47 \pm 0.06$ &  $0.85 \pm 0.05$ &  $1.40 \pm 0.16$\\
         \cline{2-7}
          & Circ.~3 ($d = 2$) &  3300 & $27.8 \pm 0.88$ &  $8.34 \pm 0.03$ &  $0.81 \pm 0.08$ &  $1.35 \pm 0.12$\\
          & Circ.~3 ($d = 4$) &  6600 & $25.96 \pm 0.52$ &  $8.5 \pm 0.05$ &  $\mathbf{0.75 \pm 0.09}$ &  $1.50 \pm 0.23$ \\
          & Circ.~3 ($d = 6$) &  9900 & $\mathbf{25.43 \pm 0.4}$ &  $\mathbf{8.56 \pm 0.04}$ &  $1.08 \pm 0.2$ &  $1.50 \pm 0.23$ \\
         \hline
         \hline
         \multirow{2}{*}{Classical}  & $[50, 30]$ & 2960 &$28.32 \pm 0.88$ & $8.52 \pm 0.09$ & $3.06 \pm 0.45$ & $2.73 \pm 0.29$ \\
         & $[100, 50]$ & 7660 & $27.36 \pm 1.51$  & $8.57 \pm 0.04$ & $2.49 \pm 0.63$ & $2.81 \pm 0.68$ \\ \hline
    \end{tabular} 
    \caption{\textbf{Training results of \model and the classical GAN for FashionMNIST dataset.} Number of parameters used in the generator and different metrics (averaged over 10 runs) to compare the performance of \model and the corresponding classical GAN using 10,000 generated images (best results highlighted in bold). For FID and JSD, the lower, the better and for IS, the higher, the better. We observe that with a similar model size ($\approx 3k $ parameters), \model outperforms the classical GAN for all metrics, except for IS.  Note that our results are among the best results obtained with the classical SOTA generative models~\cite{FashionMNIST_FID}, close to the FID of 21.73~\cite{Wei2021duelGAN} and 28.0~\cite{Bohm2022PAE}.}
    \label{tab:metrics_Fashion}
\end{table*}
\begin{table*}[t]
    \centering
    \begin{tabular}{c|c|c|c|c|c|c}
         & $G_{\bm{\theta}}$ config. & $\Np$ & FID $\downarrow$  & IS $\uparrow$   & JSD (features/$10^{-2}$) $\downarrow$ & JSD (images/$10^{-2}$) $\downarrow$\\
         \hline
         LaSt-QGAN & Circ.~3 ($d = 2$) &  3300 & $168.28 \pm 2.06$ &  $3.57 \pm 0.01$ &  $1.26 \pm 0.21$ & $2.07 \pm 0.27$  \\
         \hline
         \hline
         Classical & $[100, 50]$ & 7660 & $172.6 \pm 5.02$ & $3.5 \pm 0.03$ &  $6.99 \pm 1.13$  & $4.25 \pm 0.65$ \\ \hline
    \end{tabular} 
    \caption{\textbf{Training results of \model and the classical GAN for SAT4 dataset.} Number of parameters used in the generator and different metrics (averaged over 10 runs) to compare the performance of \model and the corresponding classical GAN using 10,000 generated images. For FID and JSD, the lower, the better and for IS, the higher, the better. We observe that \model outperforms the classical benchmark for all metrics by using only half the number of parameters. Note that the highest IS value for the SAT4 dataset is 4, as it consists of 4 classes.}
    \label{tab:metrics_SAT4}
\end{table*}

\figref{fig:tsne} visualizes the distribution of features generated by the classical and style-based quantum generators, downsampled using t-distributed Stochastic Neighbor Embedding (t-SNE)~\cite{Matten2008tsne, scikit-learn} for MNIST and Fashion MNIST dataset. Each feature is labeled after classifying the generated images using the ResNet50 pre-trained on the real image dataset. Although the separation of the generated features is not as clear as that of the real training set, the clustering of samples within a class reveals that the underlying structure of the data distribution is preserved during the data generation. 

\begin{figure}[!h]
    \centering
    \subfloat[\label{fig:Progress_MNIST}MNIST]{\includegraphics[width = 0.496\textwidth]{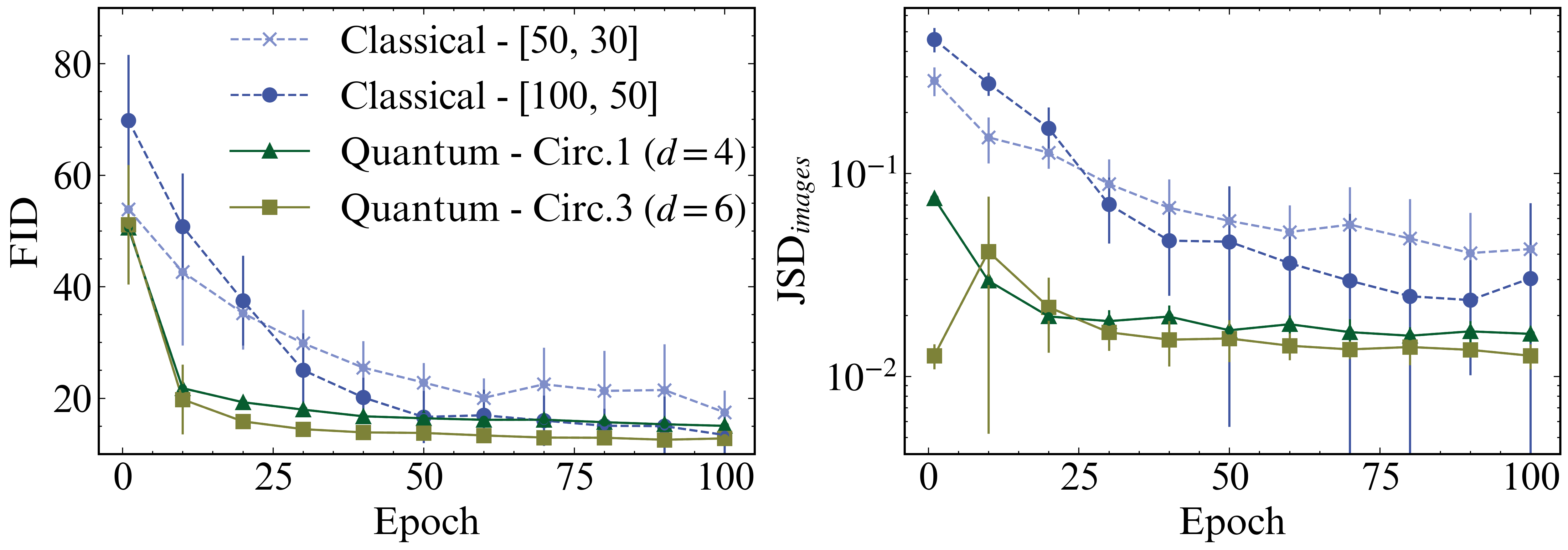}}     
    \newline
    \subfloat[\label{fig:Progress_FashionMNIST}FashionMNIST]{\includegraphics[width = 0.496\textwidth]{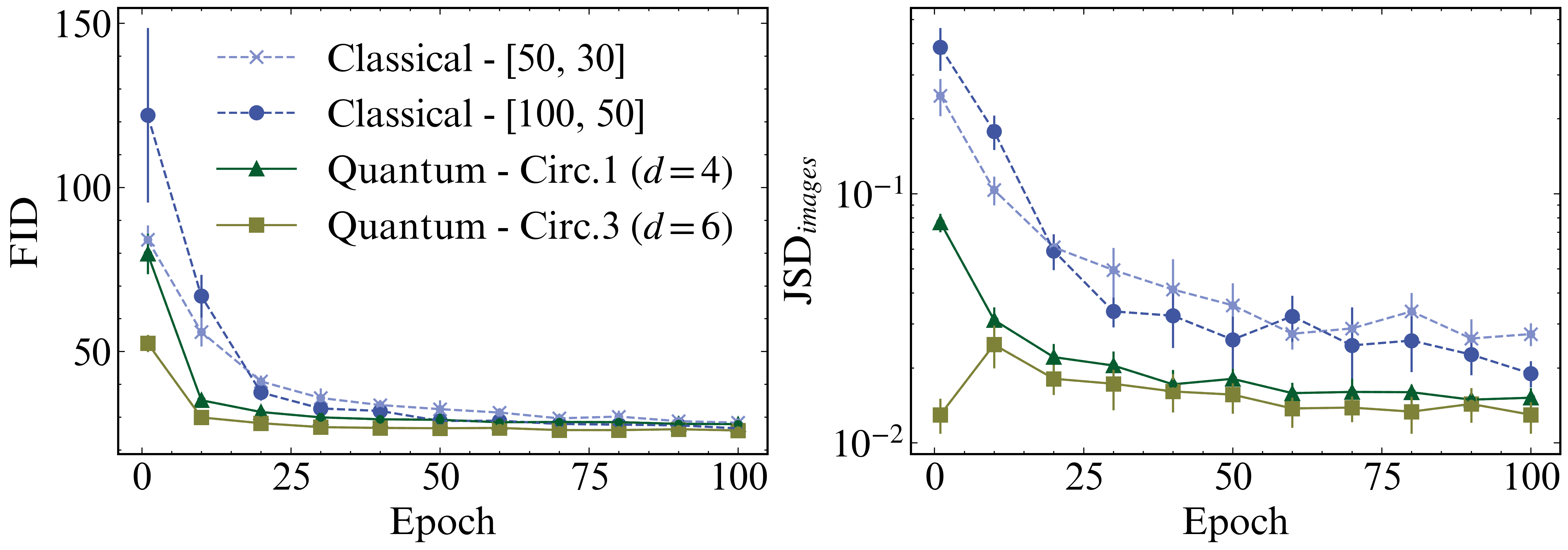}} 
    \caption{\textbf{Evaluation of training dynamics and stability (averaged over 10 runs) for \model and classical GANs.} The metrics computed over 10,000 samples generated during the training of \model and the classical GAN for MNIST and FashionMNIST dataset. We observe faster convergence and higher stability with \model than the classical model for both datasets using a similar number of parameters. Furthermore, for all tested models, \model reaches lower JSD compared to the classical model, highlighting its power to learn the hidden data distribution.}
    \label{fig:progress}
\end{figure}

The performance of the GAN training is also quantified in terms of the metrics introduced in \secref{sec:metrics}. 
In \tabref{tab:metrics_MNIST}, \ref{tab:metrics_Fashion} and  \ref{tab:metrics_SAT4}, we present the comparison of the top-performing results of \model with different quantum generator architectures and the classical GAN, using FID, IS and JSD computed on 10,000 samples. In particular, for JSD, we assess the performance by analyzing both the generated features and reconstructed images to gauge its ability to mimic the original data distribution before and after image reconstruction. We see that with a similar number of parameters, \model outperforms the classical benchmark for all types of datasets not only in terms of quality (FID, IS) but also in terms of diversity (JSD) in both features and images, showing that the model can successfully learn the hidden distribution of the real data. Notably, our \model achieves the FID value close to the state-of-the-art GAN techniques for the MNIST dataset, which are 7.87~\cite{Wei2021duelGAN} and 12.884~\cite{Lazcano2021hypGAN}. 

The rate of convergence serves as another crucial aspect in GAN training. \figref{fig:progress} illustrates the progression of various evaluation metrics for \model and its classical counterpart with different model architectures. 
As empirically observed in the plot, faster convergence is exhibited for \model compared to the classical GAN for both MNIST and FashionMNIST datasets. Notably, for the MNIST dataset, we reach the FID value below 20 in fewer than 20 training epochs for all depth $d$, which is at least twice as fast as the classical one. One might argue that the faster convergence is due to the fact that the quantum generator has a low number of parameters. However, the faster convergence is also observed using Circuit3 with depth 6 in the \model which is composed of more parameters compared to the classical GAN, empirically showing that this is independent of the number of parameters.  
We further stress that small standard deviations reveal the stability of training with \model during the whole training process, solving the training instability, one of the major issues in GANs~\cite{Arjovsky2017stability}. 
This aligns with the previous studies on the beneficial capacity properties and faster training convergence, which were experimentally proven in the previous papers in the context of classification task~\cite{Abbas2021} and discrete QGAN~\cite{letcher2023qgan}.

\subsection{\label{sec:datasetsize}Dependence on the dataset size}
In this section, we train \model with smaller training sets for MNIST and FashionMNIST datasets, comparing the outcomes against the classical GAN to study the generalization power of the quantum generator. That is, we study how close the underlying distribution of a generator trained on a small set of training data is to a true target distribution of the original images as a function of a training data size. In particular, the models are trained on varying sample sizes, $N = 2^k\times1000,~k=0,...,5$ samples, as well as on the complete training set, $N = 60,000$. 
To maintain consistency in the number of updates per epoch, we employ batch sizes of $N_{bs} = 4^k$ for each $N = 2^k\times1000$, where $k=0,...,5$, and a batch size of $N_{bs} = 4^6 = 256$ for the entire training set. 

\begin{figure}[h]
\centering
    \subfloat[\label{fig:Ns_MNIST}MNIST]{\includegraphics[width = 0.496\textwidth]{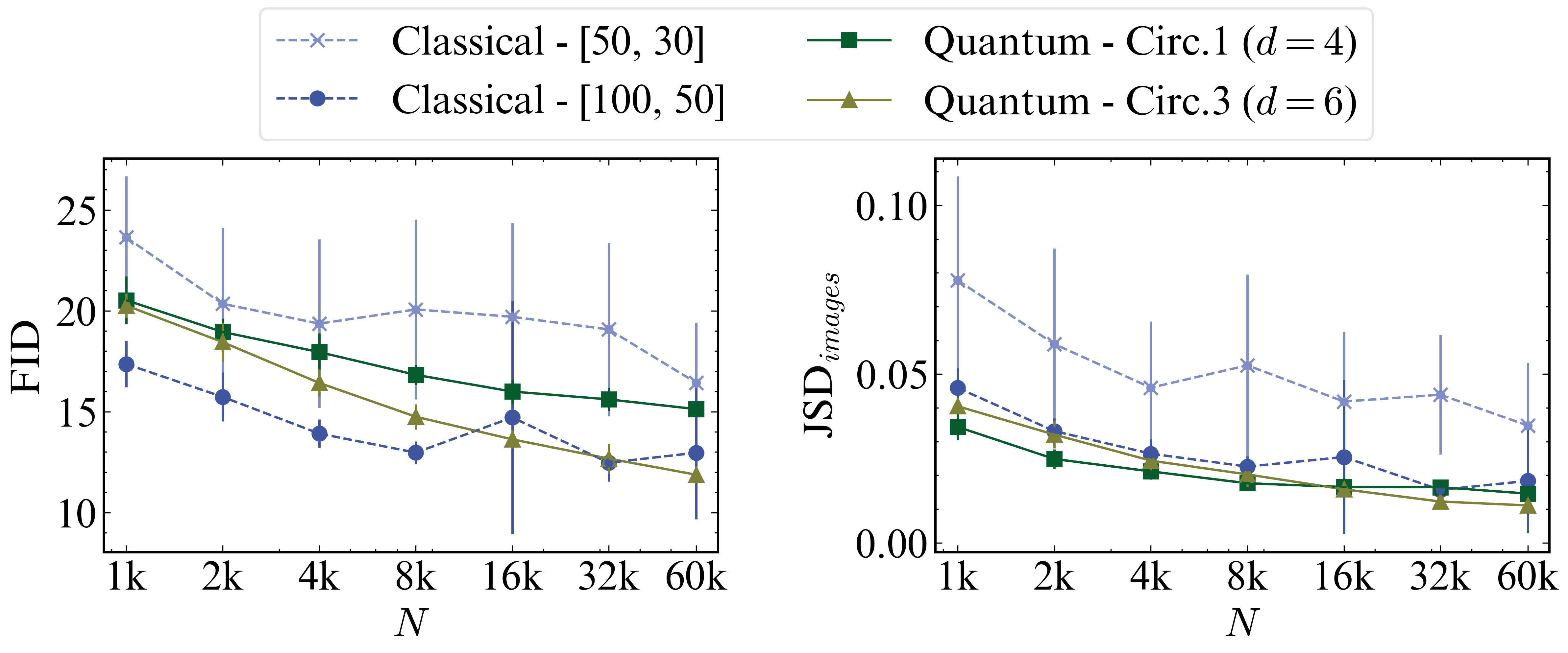}}     
    \newline
    \subfloat[\label{fig:Ns_FashionMNIST}FashionMNIST]{\includegraphics[width = 0.496\textwidth]{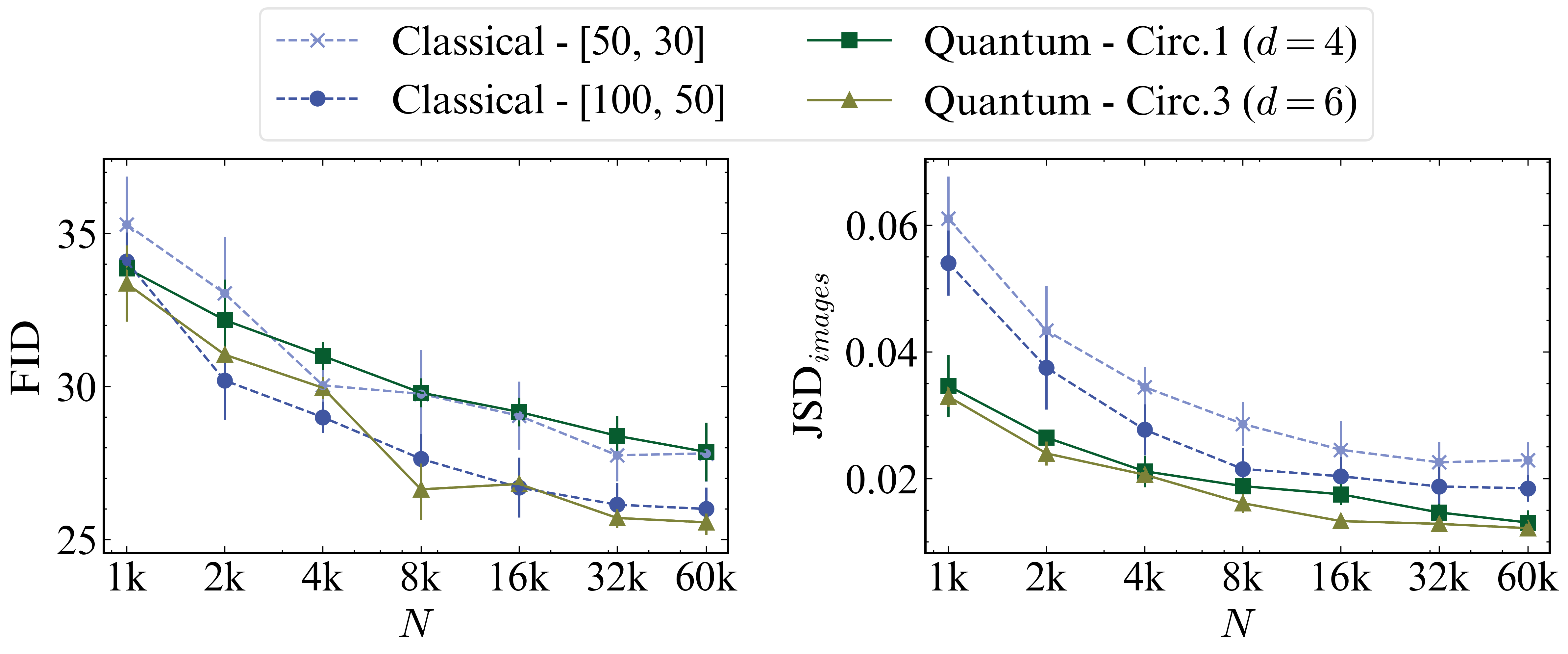}} 
    \caption{\textbf{Comparison of the generalization power in \model and its classical counterpart with varying dataset sizes, $N$.} Metrics are computed over 10,000 samples generated at the end of the training with different training set sizes and averaged over 10 runs. For small dataset sizes, we observe that \model using Circuit1 with $d=4$ consistently performs better than the classical GAN with $[50, 30]$ hidden nodes, which has a similar number of parameters (see \tabref{tab:metrics_MNIST}), indicating its ability to generalize from limited data. Additionally, the larger quantum model (Circ3 - $d = 6$) shows stronger performance with the FashionMNIST dataset, making it suitable for datasets with a more complex feature correlation.}
    \label{fig:datasetsize}
\end{figure}

\figref{fig:datasetsize} depicts the values of FID and JSD obtained at the end of the training with varying dataset sizes $N$ using 10,000 generated samples every time. For both datasets, the \model results in slightly better performance in terms of different evaluation metrics compared to the classical GAN with a similar number of parameters ($[h_1, h_2] = [50, 30]$) for smaller $N$, proving a higher generalization power with a small training set. In particular, the improvement is more pronounced with the MNIST dataset for the small generator size: we reach FID less than 20 only with $N = 4k$ samples with LaSt-QGAN. Conversely, with a larger generator size, the quantum generator demonstrates improved distribution learning capabilities for the FashionMNIST dataset. This dataset is distinguished by a strong correlation among latent features compared to the MNIST dataset, indicating the potential applicability of this architecture for datasets characterized by significant correlation. This observation can be elucidated through the measurements of Z and X observables, inherently correlated in their construction of outputs. It is also notable that in terms of JSD, we observe that it always outperforms the classical GAN for all model sizes, even with twice the number of parameters. Furthermore, lower standard deviations obtained in all cases with \model prove the stability of the quantum generator compared to the classical one. 

\section{\label{sec:statistics}Robustness against statistical noise}
Up to this section, \model has been trained analytically, under the \textit{infinite number of shots assumption}. Nonetheless, in practical application, the quantum states are sampled with a finite number of shots and one might argue that the resulting statistical noise might potentially degrade the quality of images in the real-case scenario. In this section, we demonstrate that the model is robust against the statistical fluctuation coming from the finite number of shots. 

We denote $\tilde{\mathbf{x}}_{\infty}$ and $\tilde{\img}_{\infty}$ the feature and the image generated analytically, and $\tilde{x}^i$ the $i^{th}$ component in the sample $\tilde{\mathbf{x}}$. For simplicity, we use the parameters of \model pre-trained analytically and generate the features $\tilde{\mathbf{x}}_{shots}$ to reconstruct images  $\tilde{\img}_{shots}$ using varying numbers of shots, $N_{shots} = 2^k$ for $k = 4,...,13$.  
\begin{figure}[!h]
    \centering
    \subfloat[\label{fig:shot_diff}Euclidean distance between $\tbfx_{shots}$ and $\tbfx_{\infty}$. ]{\includegraphics[width = 0.45\textwidth]{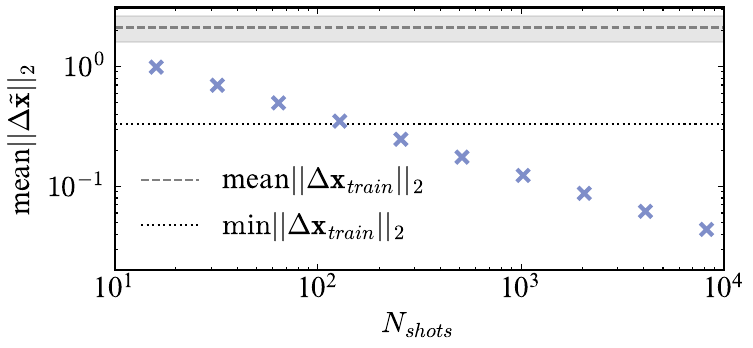}} 
    \hspace*{\fill}
    \newline 
    \subfloat[\label{fig:shot_hist}KL-divergence between $\tilde{H}_{shots}$ and $H_{train}$.]{\includegraphics[width = 0.45\textwidth]{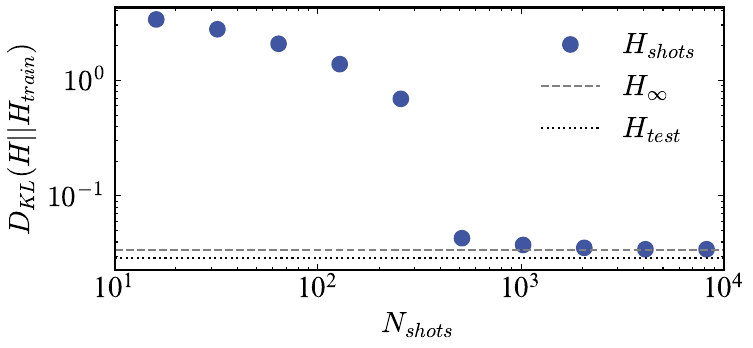}}  
    \hspace*{\fill}
    \caption{\label{fig:shot_features}\textbf{Quality of the features generated with different numbers of measurements.} (a) The $L_2$ distance between $\tilde{\mathbf{x}}_{shots}$ and $\tilde{\mathbf{x}}_{\infty}$ for MNIST dataset. The dashed line and the dotted line represent the mean and the minimum separation between samples inside the training set, i.e., $\| \mathbf{x}_{i} - \mathbf{x}_{j} \|_{\mathbf{x}_i, \mathbf{x}_j \in \mathcal{X}_{train}}$.  (b) The KL-divergence $D_{KL}$ calculated between the histograms $\tilde{H}^i_{shots}$ and $H^i_{train}$ over $x^i$ and $\tilde{x}^i_{train}$ using 500 bins. The final values are averaged over all the components $i = 1,..., 20$. Unlike $\|\Delta \tilde{\mathbf{x}}\|_2$ which decays exponentially with respect to the number of shots, $D_{KL}$ converges from $N_{shots} =  256$. }
\end{figure}

On \figref{fig:shot_diff}, we plot the Euclidean distance $\| \Delta \tbfx \| =  \|\tbfx_{shots} - \tbfx_{\infty}\|$, where $\tbfx_{shots}$ and $\tbfx_{\infty}$ are generated with the same input noise, averaged over 10,000 samples for MNIST dataset. Furthermore, as a reference, we indicate the mean and minimum separation between two samples in the training set, i.e. $\| \Delta \bfx_{train}\| =  \| \bfx_i - \bfx_j \|_{\bfx_i, \bfx_j \in \mathcal{X}_{train}}$.  It is noteworthy that $\| \Delta \tbfx \|$ drops below $\min \| \Delta \bfx_{train}\| $ after $N_{shots} = 256$. This implies that the features generated with more than 256 shots are close enough to the analytical features, positioning them within the vicinity of the corresponding $\tilde{\bfx}_{\infty}$ to differentiate them from other samples. 

To understand the general statistics over the generated features, we construct the histograms $H_{shot}^i$ for $\tilde{x}_{shot}^i$ and $H^i_{train}$ for $\tilde{x}_{train}^i$ to compute the KL-divergence between them. \figref{fig:shot_hist} displays the KL-divergence averaged over $i = 1,..., 20$. This underlines that, despite an exponential number of shots required for the exact outcomes, the overall statistics of each feature converge towards those of the training set with a finite number of shots larger than $N_{shots} = 512$. 

\begin{figure}[!h]
    \centering
    \subfloat[MNIST]{\includegraphics[width = 0.49\textwidth]{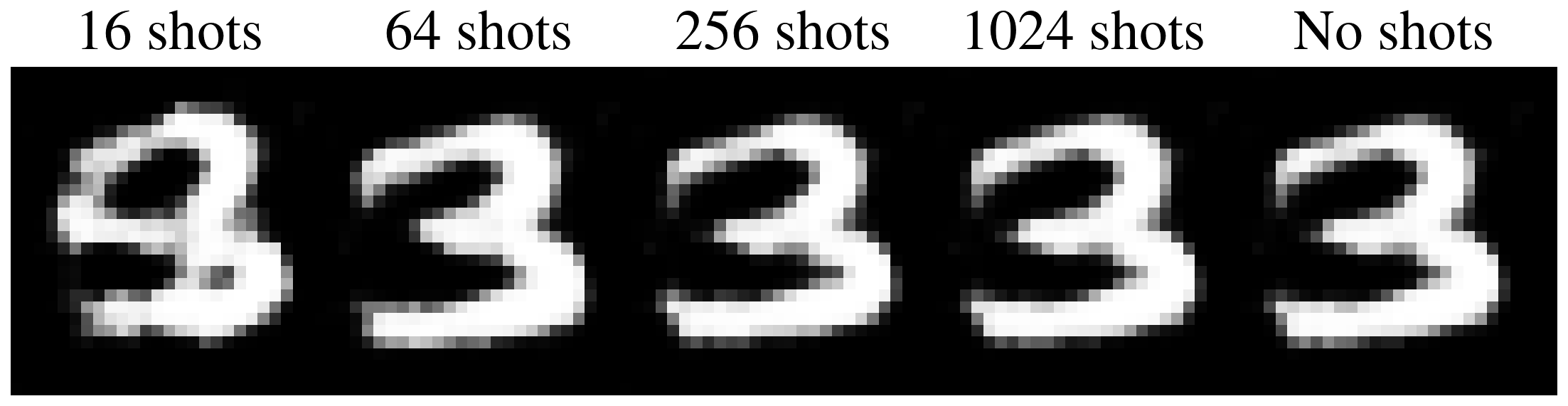}}
    \newline
    \subfloat[FashionMNIST]{\includegraphics[width = 0.49\textwidth]{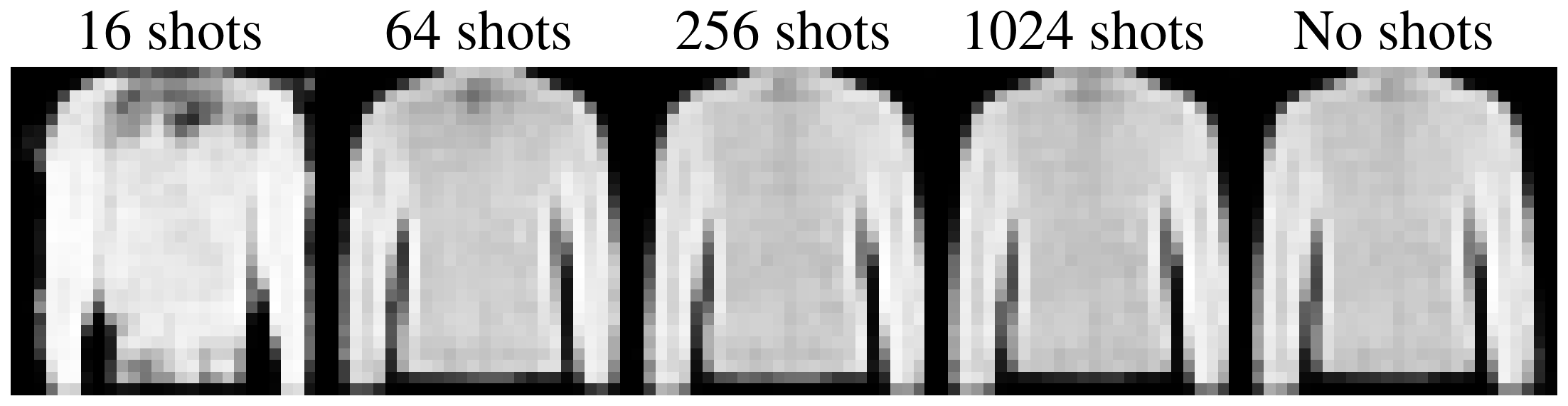}}
    \caption{\textbf{(a) MNIST and (b) FashionMNIST images generated with various number of measurements.} We can observe that the images get closer to the $\img_{\infty}$ from 256 shots. }
    \label{fig:shot_images}
\end{figure}

\begin{figure}[!h]
    \centering
    \subfloat[\label{fig:shot_img_diff}Pixel-by-pixel $L_2$ distance between $\tilde{\img}_{shots}$ and $\tilde{\img}_{\infty}$. ]{\includegraphics[width = 0.45\textwidth]{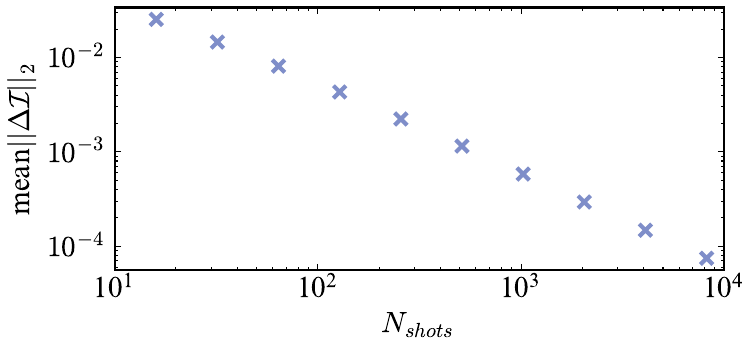}} 
    \newline 
    \subfloat[\label{fig:shot_img_fid}FID calculated for $\tilde{\img}_{shots}$.]{\includegraphics[width = 0.45\textwidth]{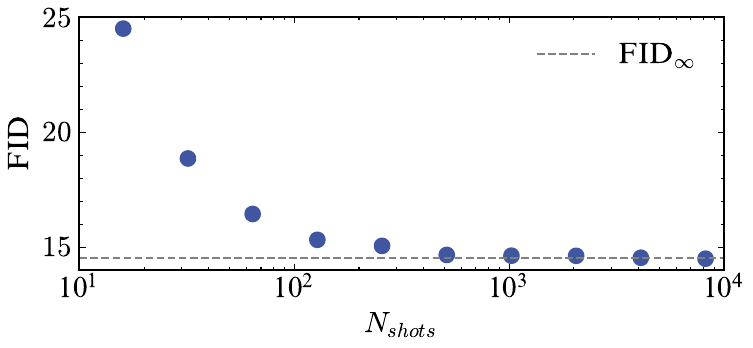}}  
    \caption{\label{fig:shot_image_diff_fid}\textbf{Quality of the images generated with different numbers of measurements.} (a) The pixel-by-pixel $L_2$ distance between $\tilde{\img}_{shots}$ and $\tilde{\img}_{\infty}$ for MNIST dataset. The exponential decay in $\| \Delta \tbfx\|$ shown on \figref{fig:shot_diff} is also leveraged for $\| \Delta \tilde{\img}\|$. (b) FID value computed for $\img_{shots}$ with different number of shots. Despite an exponential decay in the absolute pixel-by-pixel difference between the $\img_{shots}$ and $\img_{\infty}$, the FID converges to FID$_\infty$ after $N_{shots} = 512$.}
\end{figure}

To make this line of argument more concrete, we analyze the impact of a finite number of measurements on the generated images. \figref{fig:shot_images} shows the images generated with different numbers of shots for the MNIST and the FashionMNIST datasets. With bare eyes, we observe that the quality of images becomes already faithful with $N_{shots} = 256$, which aligns with the threshold observed for $\| \Delta \tbfx \|$. This can be confirmed with a quantitative analysis of the images using FID metrics. As shown on \figref{fig:shot_img_diff}, the absolute pixel-by-pixel difference between $\tilde{\img}_{shot}$ and $\tilde{\img}_{\infty}$ decreases exponentially with the number of shots, which might lead the readers to confirm the necessity of the infinite number of shots. However, on contrary, the FID value shown on \figref{fig:shot_img_fid} converges to the FID of $\tilde{\img}_\infty$ from $N_{shots} = 512$. Indeed, in practical implementation, the standard methods used to evaluate the quality of the images do not detect the difference occurring by statistical fluctuation. This empirically demonstrates that with a good construction of the classical autoencoder responsible for post-processing, the impact of shot noise can be alleviated, alluding to the feasibility of using a finite number of shots for image generation. 

It is important to acknowledge that when training on actual quantum hardware, the gradient computation will also be affected by finite shots, impacting the quality of the resulting features at the end of the training. Nevertheless, this study still provides valuable insight into mitigating the fluctuations in the features thanks to postprocessing, as long as the features converge towards real values within a certain range.

\section{\label{sec:bp}Mitigating barren plateaus}
One of the main challenges in PQC training is the problem of the exponentially vanishing loss gradients, also known as \textit{barren plateaus}~\cite{McClean2018BP, larocca2024review}. In particular, consider a loss function $\mcl(\bm{\theta})$ of the form
\begin{align}\label{eq:generic-loss}
    \mcl(\bm{\theta}) = \langle \psi_0 | U^\dagger(\bm{\theta}) O U(\bm{\theta}) | \psi_0 \rangle \;,
\end{align}
where $U(\bm{\theta})$ is some parametrized circuit, $O$ is some observable and $|\psi_0\rangle$ is some initial state. We say that the loss function $\mcl(\bm{\theta})$ 
exhibits a barren plateau if, for all the parameters $\theta_\nu$, there exists $b > 1$ such that : 
\begin{equation}
    \Var_{\bm{\theta}}[\partial_\nu \mcl (\bm{\theta})] \in \mathcal{O}\left(\frac{1}{b^n}\right)\;,
\end{equation} 
where we introduce a shorthand notation $\partial_\nu \mcl (\bm{\theta}) := \partial \mcl (\bm{\theta})/\partial \theta_\nu$.
Note that the definition of the barren plateau is also equivalent to showing that $\Var_{\bm{\theta}}[\mcl(\bm{\theta})] \in \mathcal{O} (\frac{1}{b'^n})$ for some $b'>1$, which implies the exponentially flat loss landscape~\cite{arrasmith2021equivalence}. Consequently, the number of measurement shots required to navigate through the flat region scales exponentially with the number of qubits, posing a serious scaling problem for trainability of PQCs.

Recently, it has been argued that various sources of barren plateaus previously discovered~\cite{McClean2018BP,Holmes2022ExprBP,Cerezo2021CostBP, Wang2021NoiseBP,marrero2020entanglement,patti2020entanglement,larocca2021diagnosing,holmes2020barren,Thanasilp2023QMLBP} can be unified under one key concept of \textit{the curse of dimensionality} whereby quantum states in the exponentially large Hilbert space are inappropriately handled~\cite{Cerezo2023BP, larocca2024review, Ragone2023UnifiedBP, fontana2023theadjoint, diaz2023showcasing}. While initially discussed in the setting of the loss function in Eq.~\eqref{eq:generic-loss}, the studies of BPs have largely been extended to various QML frameworks which take into account training data and non-linear loss functions~\cite{Thanasilp2023QMLBP, tangpanitanon2020expressibility, letcher2023qgan, Rudolph2022, leone2022practical, barthe2023gradients} -- even quantum models that are trained solely on classical computers~\cite{thanasilp2022exponential,xiong2023fundamental, suzuki2022quantumfisher, suzuki2023effect, huang2021power}. Of our particular interest, Ref.~\cite{letcher2023qgan, Rudolph2022} investigates barren plateau in quantum generative models~\cite{letcher2023qgan, Rudolph2022}, but only in the case of the discrete models. 

In this section, we study the barren plateau phenomena in the \model by analyzing the generator loss given by \eqnref{eq:loss_G}. Although the generator loss does not take the form of an expectation value as shown in \eqnref{eq:generic-loss}, it can be seen as a post-processing of expectation values. 
We begin by empirically investigating the variance of the partial derivative $\partial_\nu \mcl_G$ of the generator loss. 
Here, the derivative is only with respect to the generator parameters (since those are parameters in the quantum circuits) and the variance is taken over both the generator and the discriminator parameters, $\bm{\Theta}$ and $\bm{\phi}$. For the quantum generator, the weights $W$ and the biases $\mathbf{b}$ are initialized randomly from a uniform distribution $[-\delta, \delta]$ and the input noises $\mathbf{z}$ are sampled from a normal distribution, $\mathcal{N}(0, 1)$. In addition, the rotation angles are rescaled with respect to the latent space dimension $\Dz$ i.e.,
\begin{equation}
    \bm{\theta} = \frac{1}{\sqrt{\Dz}} W\mathbf{z} + \mathbf{b} \;.
\end{equation}
Due to the Central Limit Theorem~\cite{rychlik1976central}, each element of $\bm{\theta}$ independently follows a normal distribution, centered around 0 with a standard deviation $\sigma\approx \delta$. 

\begin{figure}[h]
    \centering
    \includegraphics[width = 0.45\textwidth]{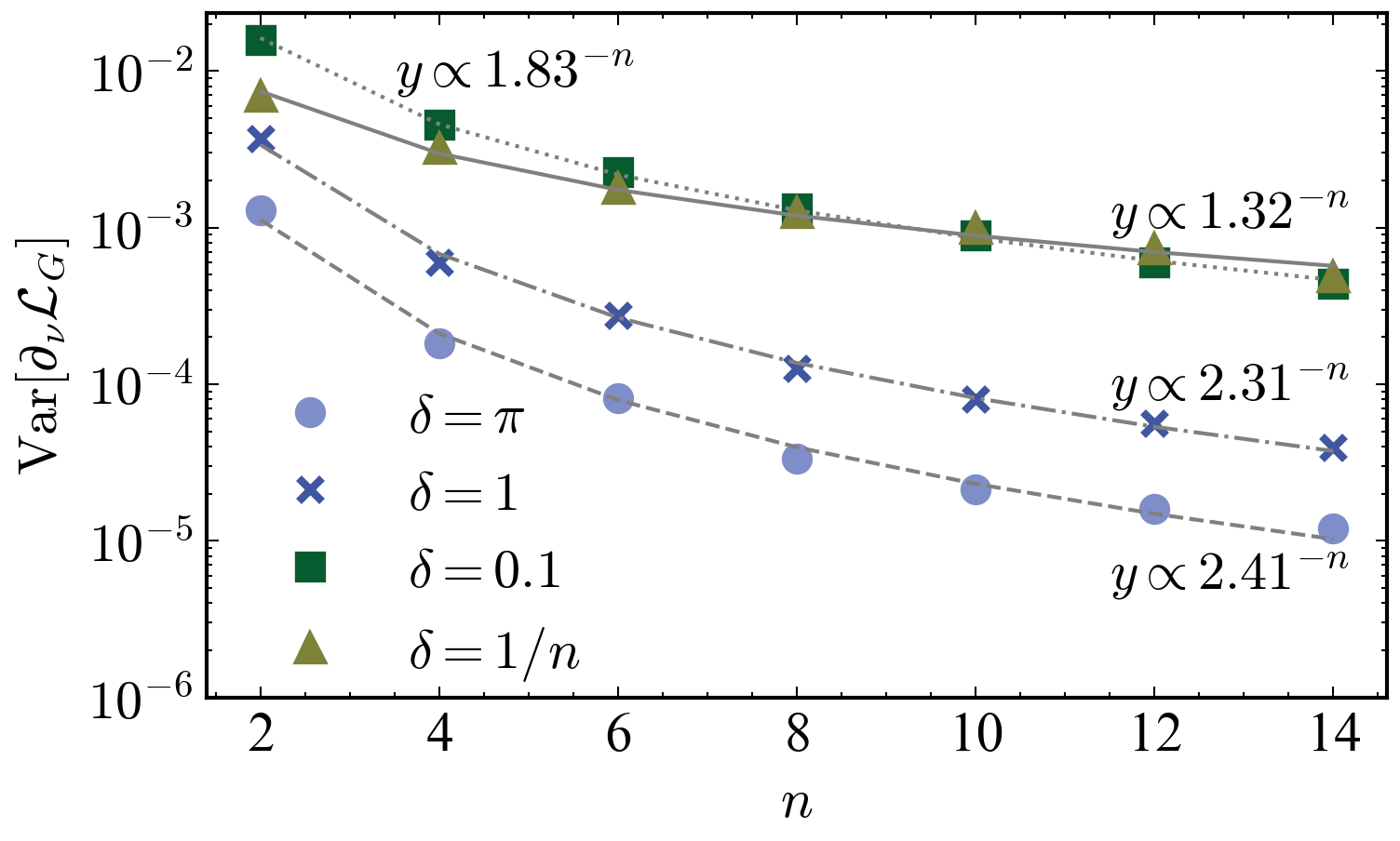}
    \caption{\textbf{Variance of the partial derivative of $\mcl_{G}$ versus the number of qubits $n$ using logarithmic depth quantum circuit.} The variance is computed with $\Dz = n$ for different initialization ranges, $\delta$, and averaged over the parameters of the first layer. The quantum generator consists of Circuit1, with logarithmic depth, $d = \lfloor \log(n) \rfloor$. Regardless of $\delta$, $\mcl_G$ does not exhibit BP with polynomially decaying variance, as the loss function only contains local observables, with zero initial state.}
    \label{fig:grads_logdepth}
\end{figure}

To understand the behavior of the gradients, we numerically compute the variance of partial derivatives $\partial_\nu \mcl_G$ with respect to the number of $n$ for different initialization bounds $\delta$ as shown on \figref{fig:grads_logdepth}. Here, we note that a quantum circuit is said to be free of the barren plateau if $\partial_\nu \mcl_G$ decays polynomially \textit{at least} with respect to one of the parameters. Therefore, in our analysis, we focus on calculating the derivatives with respect to the parameters in the first layer of the generator circuit and take an average over them. 
On \figref{fig:grads_logdepth}, the fitting curves clearly prove that $\Var_{\bm{\Theta}, \bm{\phi}}[\partial_\nu \mcl_G ]$ decays polynomially with $n$, i.e., $\Var_{\bm{\Theta}, \bm{\phi}}[\partial_\nu \mcl_G ] \in \mathcal{O}(1/n^b)$ with $b > 1$, although $b$ increase with $n$. 
This polynomial decay indicates an absence of barren plateaus and is indeed expected from the fact that the quantum generator only consists of single-qubit local observables together with limited expressivity of log-depth circuits~\cite{Cerezo2021CostBP}. 
\begin{figure}[h]
    \centering
    \includegraphics[width = 0.45\textwidth]{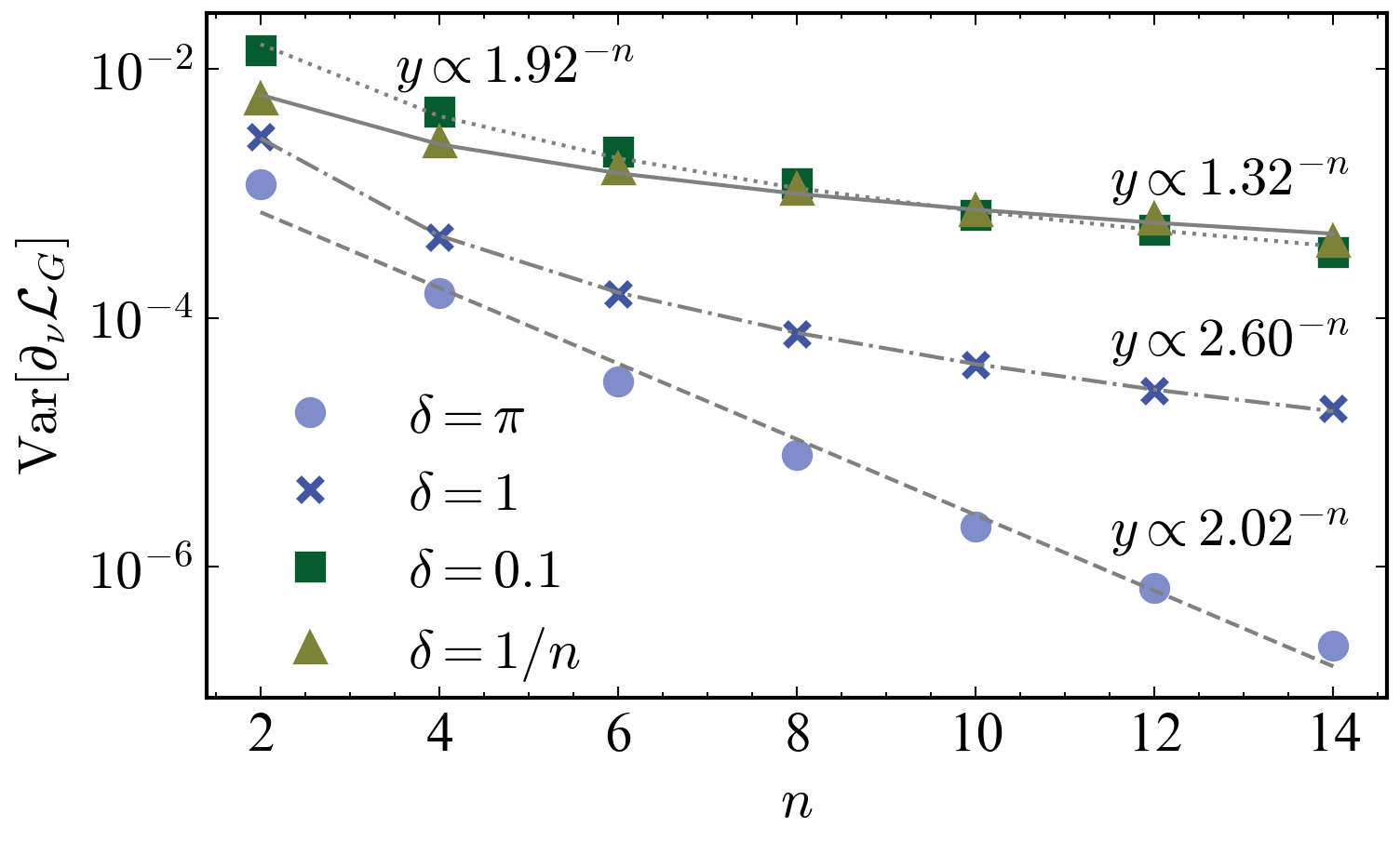}
    \caption{\textbf{Variance of the partial derivative of $\mcl_{G}$ versus the number of qubits $n$ using polynomial depth quantum circuit.} The variance is computed with $\Dz = n$ for different initialization ranges, $\delta$, and averaged over the parameters of the first layer. The quantum generator consists of Circuit1, with logarithmic depth, $d = \lfloor \text{poly}(n) \rfloor$. We observe a clear existence of the BP for $\delta = \pi$. However, using warm start with $\delta=0.1$ and $\delta = 1$, $\mcl_G$ decays polynomially, showing BP free regime.}
    \label{fig:grads_polydepth}
\end{figure}

We extend the study to the polynomial depth scenario with different initialization range $\delta$. In this regime, the circuit is sufficiently expressive to give rise to barren plateaus. 
As shown in \figref{fig:grads_polydepth}, when randomly initializing the parameters with $\delta = \pi$, the variance of the loss gradients vanishes exponentially in the number of qubits. 
On the other hand, in the case where we use a small angle initialization with the initial parameters sampled from a certain range with $\delta = 1$, $\delta = 0.1$ and $\delta = 1/n$ around the circuit identity, $\Var_{\bm{\Theta}, \bm{\phi}}[\partial_\mu \mcl]$ is empirically observed to decay polynomially, which mitigates the effect of barren plateaus. For larger system size, while there is no analytical guarantee on the scaling with fixed small $\delta$ and one could in principle expect the scaling to turn into exponential, we can analytically guarantee that the variance scaling remains polynomial with $\delta$ scaling with the system size.    

To further probe this with some analytics, we first note that, since the loss is a post-processing of expectation values which essentially are of the form \eqnref{eq:generic-loss}, the loss function does not suffer from barren plateaus if these expectation values do not exponentially concentrate over the parameters. For more details, Appendix.~\ref{adx:proof_bp} gives an insight regarding the lower bound on the loss concentration in the polynomial depth circuit with the small angle initialization initialization~\cite{Grant2019initialization, zhang2022escaping} using EfficientSU2 ansatz~\cite{IBMansatz} for both local and general observables. It supplements the prior research on normal initialization~\cite{zhang2022escaping} by providing a tight lower bound and a comprehensive insight into the behavior of the loss function based on the initial quantum state and final measurement. In particular, for the EfficientSU2 ansatz, if the initial range $\delta$ scales as $1/n$, the loss function $\mcl$ decays as : 
\begin{equation}
  \Var_{\bm{\theta}}[\mcl({\bm{\theta}})] \gtrsim \frac{1}{n^b} \;, \;\; b > 2\;. 
  \label{eq:delta_scale}
\end{equation}
On \figref{fig:grads_logdepth} and \ref{fig:grads_polydepth}, we plot as well $\Var_{\bm{\Theta}, \bm{\phi}}[\partial_\nu \mcl_G ]$ with varying $\delta = 1/n$. In this plot, we use Circuit1 as the quantum generator instead of EfficientSU2 ansatz, but we observe that the variance also decays polynomially as expected by \eqnref{eq:delta_scale}, indicating the mitigation of barren plateau within a certain range.

Lastly, we remark that while the circuits with log-depth or the small angle initialization are shown to evade barren plateaus in our model, the loss landscape can be classically simulable as discussed in the recent study~\cite{Cerezo2023BP}. The subtle difference between the two cases is that the circuit with log-depth leads to a classical simulability of the loss at any point of an entire landscape while only a small region around identity initialization can be classically simulated on average with the deep circuit. Although there is no guarantee of achieving the optimal solution, the small angle initialization will allow at least reaching the local or suboptimal minimum in the loss landscape.  

\section{\label{sec:conclusion}Discussion}
Quantum generative modeling has attracted much recent attention as one of the promising applications of quantum computers in data analysis. Nonetheless, there remain fundamental and practical challenges before a practical quantum advantage could be achieved. One of which is how to generate images with a dimension comparable to those generated by classical generative models.

In this paper, we introduced \model, which combines a classical latent embedding and a quantum GAN under a unified framework. By using the latent technique, images are mapped into a latent space with smaller dimensions where a style-based quantum GAN is trained to learn the latent representation of the images. This combined approach enables us to larger image generation with a hybrid quantum-classical generative model. 

Our empirical results demonstrate that the model can effectively synthesize images with a better quality level than the classical counterpart using approximately the same resources. 
In particular, from various quantitative evaluation metrics, we empirically observe that for these specific learning tasks, the quantum GAN is capable of achieving, and in some cases even surpassing, the performance of classical GAN in terms of both quality and diversity of the generated samples across all tested datasets while maintaining a similar number of trainable parameters. Furthermore, we investigated the performance of the models under varying dataset sizes and observed that \model reaches a comparable level of performance to the classical GAN, even when using a smaller dataset size, as well as the effect of shot noise on the resolution of the generated images. 
These empirical findings constitute a first step to demonstrate our model's potential for practical applications. Nevertheless, since our model relies on the classical autoencoders to amplify the outputs from a quantum GAN, it is a fundamental open question to see whether the interplay between the classical and quantum parts can be quantified.

We also study the barren plateau phenomena in the continuous generative models. Crucially, by using a mix of analytical and numerical tools, we show that \model with a polynomially deep generator circuit can be trained with a small angle initialization around the identity. Despite the loss landscape being exponentially flat on average, the strategy allows us to initialize on a region with substantial gradients and train towards some local minimum. Nevertheless, while providing a temporary remedy to a barren plateau problem, the strategy has certain drawbacks. We cannot guarantee the quality of the local minimum if the circuit itself contains no inductive bias that aligns with the target distribution. In addition, the small region around the initialization can be classically simulable on average. Crucially, we note that our barren plateau results here are also directly applicable to other continuous quantum generative models based on sampling expectation values such as the original style-based quantum GANs.

To go beyond the identity initialization, one could consider a warm-start strategy, i.e., a smart initialization that incorporates the problem structure into consideration~\cite{Cerezo2023BP, valls2024warmstart, rudolph2022synergy, mele2022avoiding}. Recently, a warm start in the context of variational quantum simulation has been analytically studied in Ref.~\cite{valls2024warmstart}, showing the potential of a warm-start strategy to circumvent barren plateaus but at the same time highlight additional challenges for achieving global minimum. Further investigation of warm-starts in the generative modeling setting is of particular importance for both fundamental and practical aspects. 

Lastly, it is crucial to remark that the role of a quantum circuit in the continuous generative model as a feature map shares a great similarity in the supervised quantum machine learning with classical data. This implies some of the pieces of knowledge in the literature can be applied to the generative setting. For example, one fundamental concern is the risk of the continuous generative models being classical surrogatable by similar techniques such as random Fourier feature~\cite{landman2022classically}. On the other hand, a provable quantum advantage based on cryptographic hardness~\cite{Liu2021Speedup, nietner2023average} strongly suggests the existence of classically hard continuous quantum generative models. Since the fundamental natures between discriminative and generative models do not perfectly align, to what extent one can apply the results from one field to another remains unanswered, leaving a great opportunity for future research.

\begin{acknowledgments}
We would like to acknowledge Zo\"e Holmes for insightful discussions. 
SC is supported by the quantum computing for earth observation (QC4EO) initiative of ESA $\Phi$-lab, partially funded under contract 4000135723/21/I-DT-lr, in the FutureEO program. ST acknowledges support from the Sandoz Family Foundation-Monique de Meuron program for Academic Promotion. MG is supported by CERN through the CERN Quantum Technology Initiative.
\end{acknowledgments}

%

\onecolumngrid
\appendix 
\section{\label{sec:related} Related Work}

In this section, we provide a brief summary of the recent research on classical and quantum generative models. Especially, we underline the difference between the discrete and the continuous quantum GAN. \tabref{tab:discrete_continuous} summarize the characteristics of the two different generative models for comparison.

\begin{table*}[h]{
    \fontfamily{\sfdefault}\selectfont
    \footnotesize
    \centering
    \begin{tabular}{c||c|c }
    \rowcolor{LightCyan} 
          & \textbf{Discrete Quantum Generative Models} & \textbf{Continuous Quantum Generative Models}  \\
          \hline
          \hline
          Task  & Encode a probability distribution over discrete values\hspace{1pt}&  Generate continuous outputs   \\ 
         
            \hline 
        Outputs & Discrete bit strings &  Continuous values \\ 
        \hline
        \hspace{0.3cm}Sample Complexity~~~& One measurement per sample & Set of measurements per samples \\ 
        \hline
        \multirow{2}{*}{Randomness } &  Use the probabilistic nature of quantum physics  & Sampled from a classical random distribution   \\ 
          & (quantum randomness)  & (classical randomness) \\ 
         \hline
        Projector & $\ket{x}\bra{x}$& Estimate a vector of expectation values \\ 
        \hline 
         Output size & $\mathcal{O}(2^n)$ &  $\mathcal{O}(n) $\\   
         \hline
         \multirow{2}{*}{Examples }&\hspace{1pt}Quantum Circuit Born Machine (QCBM)~\cite{Benedetti2019QCBM}\hspace{1pt} &\hspace{1pt}Variational Quantum Generator (VQG)~\cite{Romero2019}  \\
            &\hspace{1pt}Quantum Generative Adversarial Networks (qGAN)~\cite{Zoufal2019}\hspace{1pt} &\hspace{1pt}Style-based Quantum GAN~\cite{BravoPrieto2022}
    \end{tabular}
    \caption{\label{tab:discrete_continuous} \textbf{Comparison between discrete and continuous quantum generative models.} The discrete generative model treats each quantum measurement as a single output, focusing on learning the probability distribution across computational basis states. In contrast, the continuous model uses expectation values obtained from multiple shots, embedding external classical noise into the quantum circuit to generate samples.  }
    
}
\end{table*}

\textbf{Classical generative models for image generation.} GANs were proposed by I. Goodfellow in 2014 as an effective way of learning to generate data which follow a given distribution~\cite{Goodfellow2014GAN}. It consists of two neural networks competing with each other: a generator and a discriminator of fake data. Being a successful generative model for the creation of realistic data and images, variations of GANs were also explored, such as conditional GAN~\cite{Mirza2014cGAN} to generate data of a given class, the more stable Wasserstein GAN~\cite{Arjovsky2017wGAN}, and Style-GAN~\cite{Karrs2017Progan} for detailed image generation. 

Recently, Diffusion Models (DMs) proved to be powerful alternative generative models, trained by injecting noise into the images and then learning the reverse process to remove it~\cite{Jonathan2020DMs, Dickstein2015DM}. 
In the recent work, Rombach \etal introduced Latent DM, which operate on the latent space instead of the image space by mapping the image to a lower-dimensional latent space and learning the latent representation to reduce the complexity of the model and improve the visual fidelity~\cite{Rombach_2022_LDM_CVPR}.

\textbf{Quantum GANs for discrete data} The introduction of quantum GANs by Ref.~\cite{Lloyd2018} has suggested the possibility of learning the hidden statistics of a quantum or classical data set based on the intrinsically probabilistic nature of quantum systems. For example, C. Zoufal \etal introduced a hybrid GAN model with a classical discriminator and a $n$-qubit quantum generator which can efficiently learn a classical probability distribution over $2^n$ discrete variables with QNNs~\cite{Zoufal2019}.  They have demonstrated that the model can be used for an efficient initialization of a quantum state with an arbitrary probability distribution, one of the most crucial challenges in quantum computing, or even used for a realistic use case such as finance. This discrete quantum GAN handles the computational basis of the quantum circuit Hilbert space as discrete data and explicitly constructs 
 the probability distribution over them by performing a set of measurements at the end of the quantum generator. A similar strategy was applied to mimic the prototypical Bars-and-Stripes dataset images in Ref.~\cite{Zeng2019GAN}.  Assouel \etal  also proposed a quantum GAN model, which is called QuGAN, for discrete data generation in the context of finance but using a quantum discriminator directly connected to the quantum generator~\cite{Assouel2022}.  

\textbf{Quantum generative models for continuous data generation}. As a complement to the previously presented studies, which focus on reproducing the probability distribution over discrete data, several papers studied constructing quantum GANs to learn the hidden data distribution over continuous data. Romero and Aspuru-Guzik introduce the idea of a quantum generator to learn a continuous distribution with latent noise embedded via rotation gates~\cite{Romero2019}. Unlike the discrete quantum GAN which explicitly generates the probability distribution over the discrete data, the continuous GANs work in a similar way as the classical GAN by generating samples at the end of the generator. The classical latent noises are embedded into the quantum circuit by an encoding process, the so-called \textit{quantum feature map} or \textit{quantum encoding}~\cite{Schuld2019featuremap}. We measure the expectation value of quantum observables such as Pauli operators ($\sigma_x, \sigma_y, \sigma_z$), which are continuous by definition, to construct the output sample for each latent noise. This leads to learning the hidden data distribution from the continuous latent noise distribution in an implicit way.  
In Ref.~\cite{BravoPrieto2022}, C. Bravo Prieto \etal  employed a style-based quantum generator for Monte Carlo event generation, proving that the quantum GAN is able to reproduce a highly correlated multi-dimensional probability distribution. The particularity of this architecture is that the latent noises are embedded in the rotation angles of the learning layers via an affine transformation, with the weights and biases updated during the training. 
Instead of using a purely quantum generator, the possibility of using a hybrid generator (HG) has also been proposed by J. Li \etal  for small molecule drug discovery~\cite{Li2021drug}. The proposed QGAN-HG architecture consists of a quantum circuit attached to a classical layer with the latent noise embedded with single qubit gates. It showed a learning accuracy comparable to the classical MolGAN with 98\% reduced parameters. 

\textbf{Quantum generative models for image generation}. In the realm of image generation, quantum patch GAN has been proposed to generate the patches in images using a set of subgenerator sequence~\cite{Huang2021qGAN, Tsang2022patch}. However, their applications were only tested to a limited number of classes in MNIST~\cite{Huang2021qGAN} or in FashionMNIST  dataset~\cite{Tsang2022patch}. 

Instead of using a quantum circuit as a data generator, certain studies propose the possibility of using it as an additional component in GAN to improve performance. For example, Rudolph \etal suggested a hybrid GANs schema where the prior distribution of the classical GAN is generated by a Quantum Circuit Born Machine (QCBM)~\cite{Rudolph2022}. The architecture showed an ability to generate high-quality MNIST images on discretized latent space with up to $2^{16} =  65536$ samples using 8 qubits. 

\section{\label{adx:autoencoder}Architecture of Classical Autoencoder}
\begin{figure}[!h]
    \centering
    \includegraphics[width=\textwidth]{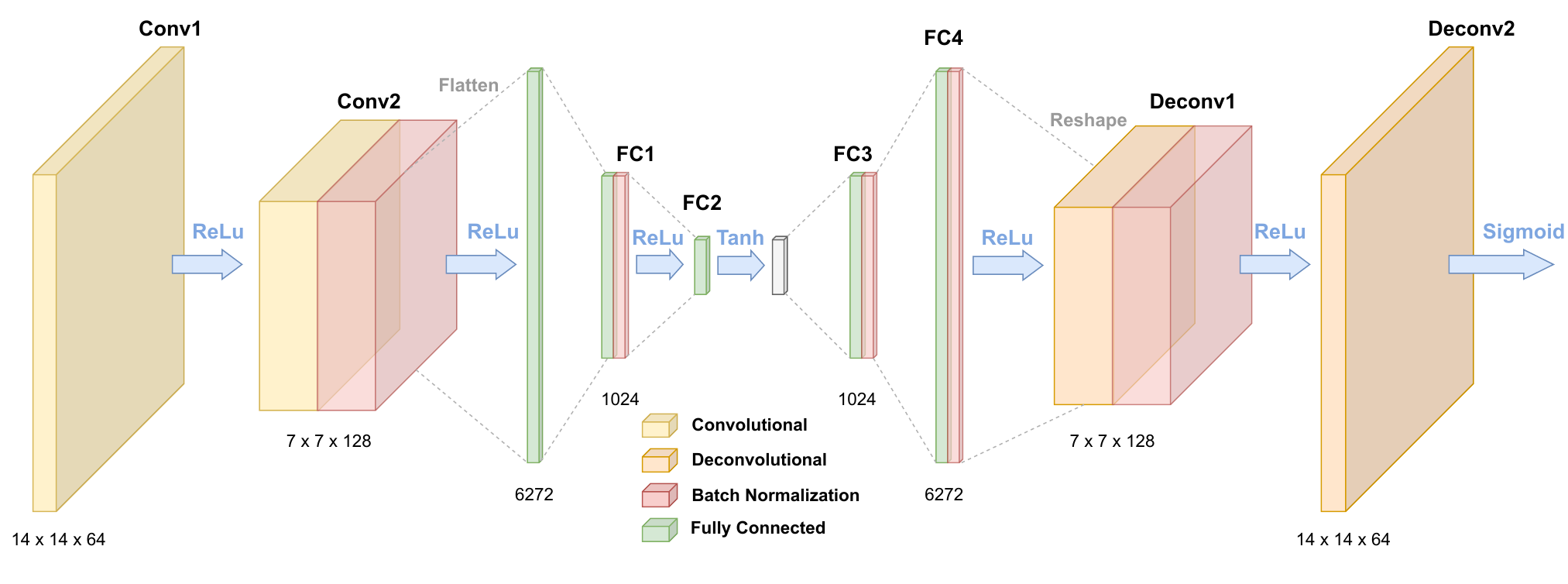}
    \caption{\label{fig:autoencoder} \textbf{Architecture of the convolutional autoencoder used in the paper.} We apply the Tanh activation function at the end of the encoder to ensure that the latent features are confined within the range of $[-1, 1]$. The autoencoder is pre-trained on the original image datasets with $28 \times 28 \times 1$ pixels, following the format of $width \times height \times channel$.}

\end{figure}

On \figref{fig:autoencoder}, we display the classical convolutional autoencoder used to reduce the image dimension. 
The model is trained based on the Mean Squared Error (MSE) loss using the Adam optimizer with a learning rate of 0.001 for 100 epochs. 

\begin{figure}[!h]
\centering
    \subfloat[MNIST]{\includegraphics[width = 0.24\textwidth]{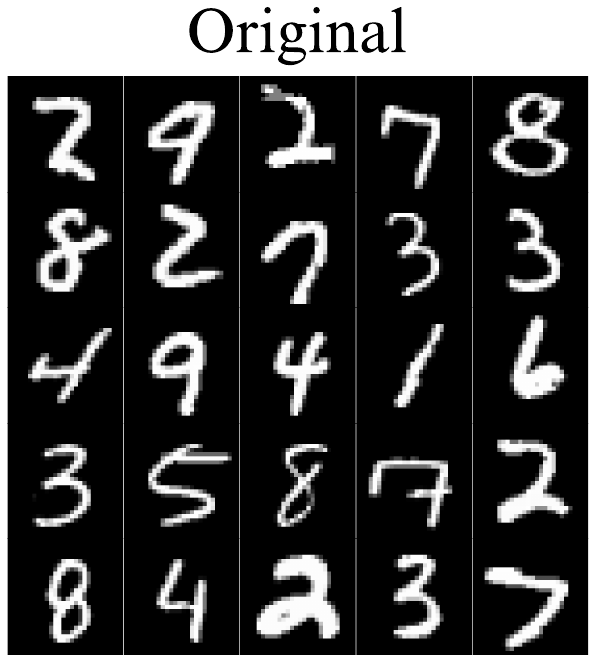} 
        \includegraphics[width = 0.24\textwidth]{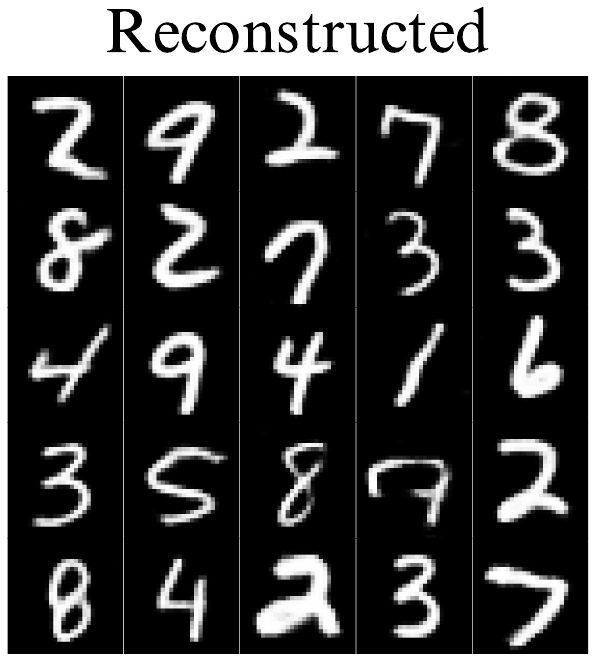} } 
    \subfloat[FashionMNIST]{\includegraphics[width = 0.24\textwidth]{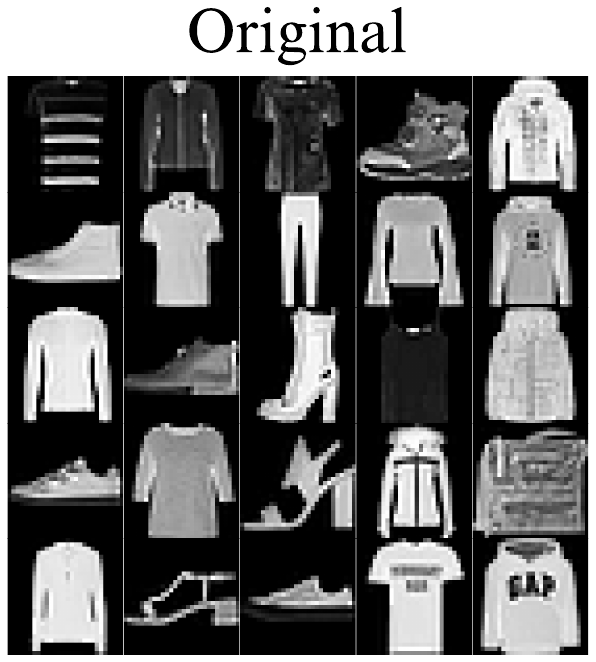}
    \includegraphics[width = 0.24\textwidth]{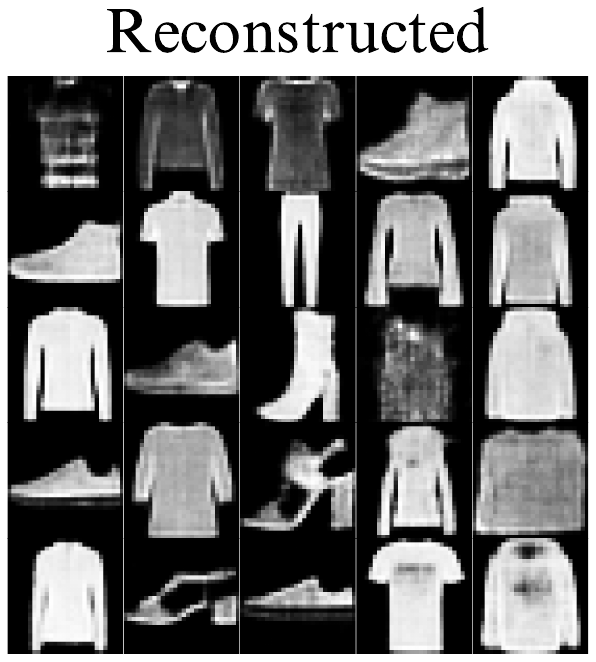}} 
    \caption{\label{fig:ae_images} \textbf{Comparison of original and reconstructed images utilizing the autoencoder in \figref{fig:autoencoder}.} Employing a 20-dimensional latent space, the reconstructed images maintain their overall form, with a loss of finer details.
  }
\end{figure}

\figref{fig:ae_images} display the original and the reconstructed images obtained using the suggested autoencoder architecture with latent space of dimension 20 for MNIST and FashionMNIST datasets. The reconstructed images are slightly blurry compared to the original ones, with a loss of details, but their general shape is recovered.

\section{\label{adx:proof_bp}Proof on Absence of Barren Plateau}
In \secref{sec:bp} of the main text, we have numerically demonstrated that the identity initialization with normally distributed quantum circuit parameters mitigates the barren plateau of the generator loss function of the form : 
\begin{equation}
    \mcl_G = -\mathop{\mathbb{E}}_{\mathbf{z}\sim \mathbb{P}_\mathbf{z}}[D_{\bm{\phi}}\left(G_{\bm{\theta}}(\mathbf{z})\right)]  
\end{equation}
where $G_{\bm{\theta}}(\mathbf{z})$ is a vector of the expectation value with a classical input $\mathbf{z}_i$. 
In this section, we provide analytical insight into the absence of a barren plateau using the small angle initialization mentioned in \secref{sec:bp}. To do so, we leverage the most general form of the quantum circuit, without any style-based architecture. The generator loss function can be seen as a function depending on the generator outputs, post-processed with the classical discriminator. Hence, under the realistic assumption that the classical discriminator does not exhibit any vanishing gradient, it is enough to prove that the generator outputs do not decay exponentially to imply the absence of barren plateau for the generator loss function $\mcl_G(\bm{\theta})$. In other words, we need to find a polynomial decaying lower bound for an expectation value of the quantum circuit with a certain observable. Furthermore, in \secref{sec:bp}, while we sampled the trainable parameters $W_\ell$ and $\mathbf{b}_\ell$ from a uniform distribution, the final rotation angles $\bm{\theta}_\ell$ in the style-based quantum circuits are normally distributed due to the Central Limit Theorem~\cite{rychlik1976central}. Consequently, the analysis of the barren plateau phenomenon in the style-based architecture can be considered equivalent when the parameters are initialized using a normal distribution.

From now on, we take the loss function of the form $\mcl(\bm{\theta}) =  \langle \psi_0 | U^\dagger(\bm{\theta}) O U(\bm{\theta}) | \psi_0 \rangle  =\Tr(O U(\bm{\theta})\rho U^\dagger(\bm{\theta} ))$ with $U(\bm{\theta})$ an arbitrary quantum circuit, $O$ the observable, $\ket{\psi_0}$ the initial state, and $\rho$ its corresponding density matrix. 
We evaluate the lower bound on the variance of the loss function for a normal initialization $\mathcal{N}(0, \sigma)$ with zero mean and standard deviation $\sigma$, following an argument similar to that given in the appendix of Ref.~\cite{letcher2023qgan} for a uniform initialization $[-\pi, \pi]$. Normal initialization has been suggested as a strategy to escape barren plateau in the prior study~\cite{zhang2022escaping}. As an extension of this research, we will introduce a lower bound on the variance, linked to the types of the initial states and the measurement operator in general circuits, building a more rigorous understanding of the loss decay. In particular, we compute a tight lower bound for a specific quantum circuit ansatz, EfficientSU2~\cite{IBMansatz}, and prove it numerically. 

Let us denote $P_{\bm{\alpha}} \in \mathbf{P}_n$ the Pauli string which consists of $n$ single-qubit Pauli matrices written as $\sigma_i \in \{\mathbb{I}, X, Y,Z\} = \{\sigma_0, \sigma_1, \sigma_2, \sigma_3\}$ :
\begin{equation}
    P_{\bm{\alpha}} = \bigotimes_{i=1}^n \sigma_{\alpha_i},~~~\bm{\alpha} \in \{0, 1,2,3\}^n\;, 
\end{equation}
where $\alpha_i$ is the $i$-th component of $\bm{\alpha}$. We use a bold symbol $\bm{\alpha}$ in order to clarify that it is a vector of $n$ indices. Similarly, the Pauli string rotation gates $P_{\bm{\alpha}}(\theta)$ with shorthand notation: 
\begin{equation}
    P_{\bm{\alpha}}(\theta) := R_{P_{\bm{\alpha}}}(\theta) = \exp(-iP_{\bm{\alpha}} \frac{\theta}{2}) = \cos(\theta) \mathbb{I} - i \sin(\theta) P_{\boldsymbol{\alpha}}\;,
\end{equation}
for some $\theta \in \mathbb{R}$. We consider a general quantum ansatz $U_L(\bm{\Theta})$~\cite{letcher2023qgan} with trainable parameters $\bm{\Theta}$, which consists of two orthogonal layers of single-qubit rotations for state initialization, $V_1$ and $V_2$, and an entangling layer $W_L$ of depth $L$  : 
\begin{equation}
    U_L(\bm{\Theta}) = W_L(\bm{\theta} ) V_2(\bm{\phi}) V_1(\bm{\omega})\;,~~\bm{\Theta} = (\bm{\theta}, \bm{\phi}, \bm{\omega})\;.  
    \label{eq:def_U}
\end{equation}
with 
\begin{equation}
    V_1(\bm{\omega}) = \bigotimes_{i=1}^n \sigma_{\mu_i}(\omega_i)\;, ~~~ V_2(\bm{\phi}) = \bigotimes_{i=1}^n \sigma_{\nu_i}(\phi_i)\;,~~~W(\bm{\theta}) = \prod_{\ell =1}^L \tilde{W}_{\ell(K_\ell)}(\bm{\theta}_{\ell})=  \prod_{\ell =1}^L \left(\prod_{i = 1}^{K_\ell} P_{\ell, i}(\theta_{\ell, i}) \right)C_\ell
    \label{eq:def_VW}
\end{equation}
where $C_\ell$ are $n$-qubit Clifford gates, $P_{\ell, i}$ the $i^{th}$ rotation gate and  $K_\ell$ the number of Pauli rotation gate at layer $\ell$. It is important to note that $\bm{\Theta}$ being the collection of rotation angles in $U_L$ does not correspond to the generator parameters defined in \secref{sec:theory} for the style-based architecture.  By definition of orthogonality, we ensure that $\mu_i, \nu_i \in \{1,2,3\}$ with $\mu_i \ne \nu_i$ for all $i$. 
Additionally, we take into account the most general form of the loss function with the observable $O = \sum_{\bm{\alpha}} a_{\bm{\alpha}} P_{\bm{\alpha}}$: 
\begin{equation}
    \mcl = \Tr(O U(\bm{\Theta})\rho U^\dagger(\bm{\Theta} )) = \sum_{\bm{\alpha}} a_{\bm{\alpha}} \Tr(P_{\bm{\alpha}}  U(\bm{\Theta})\rho   U^\dagger(\bm{\Theta} )) 
    = \sum_{\bm{\alpha}} a_{\bm{\alpha}} \mcl_{\bm{\alpha}}\;, 
    \label{eq:mcl1}
\end{equation} 
where $\rho$ is the initial density matrix, decomposed as a sum of Pauli strings : 
\begin{equation}
    \rho = \frac{1}{2^n} \sum_{\bm{\lambda}}  c_{\bm{\lambda}} P_{\bm{\lambda}}\;,~~c_{\bm{\lambda}} \in \mathbb{R}\;.
\end{equation}
In particular, if $\rho$ is a product state, $c_{\bm{\lambda}}$ should be less or equal to 1 for all $\bm{\lambda}$. 

\subsection{\label{adx:expectation}Expectation value with normal initialization}
\subsubsection{$L=0$ case}
To start with, let us calculate the expectation value of $\mcl$ over the initial parameters $\bm{\Theta}\sim \mathcal{N}(0, \sigma)$\footnote{To avoid potential confusion, we clarify the notation used in this paper: $\sigma_{\alpha_i}$ denotes the Pauli matrices, while $\sigma$ (without subscript) represents the standard deviation. The presence of the subscript $\alpha_i$ distinguishes the Pauli matrix notation from the standard deviation symbol $\sigma$. }. By linearity of the expectation, we have $\E_{\bm{\Theta}}[\mcl] = \sum_{\bm{\alpha}}a_{\bm{\alpha}}\E_{\bm{\Theta}}[\mcl_{\bm{\alpha}}]$ from \eqnref{eq:mcl1}, and thus, it is enough to compute $\E_{\bm{\Theta}}[\mcl_{\bm{\alpha}}]$ to find the final formula. We first consider the base case with $L = 0$ without any entangling layer $W_L$. The loss function $\mcl_{\bm{\alpha}}$ can be explicitly written as a product of the loss functions on each qubit : 
\begin{align}
    \mcl_{\bm{\alpha}} & = \Tr(P_{\bm{\alpha}}  U(\bm{\Theta})\rho   U^\dagger(\bm{\Theta} )) = \frac{1}{2^n} \sum_{\bm{\lambda}} c_{\bm{\lambda}} \Tr( U^\dagger(\bm{\Theta} ) P_{\bm{\alpha}} U(\bm{\Theta}) P_{\bm{\lambda}}  ) \nonumber \\   & = \sum_{\bm{\lambda}}  c_{\bm{\lambda}} \prod_i \frac{1}{2}\Tr(\sigma_{\mu_i}(-\omega_i) \sigma_{\nu_i}(-\phi_i) \sigma_{\alpha_i} \sigma_{\nu_i}(\phi_i) \sigma_{\mu_i}(\omega_i) \sigma_{\lambda_i} )  = \sum_{\bm{\lambda}}  c_{\bm{\lambda}} \prod_i\mcl^i_{{\bm{\alpha}}{\bm{\lambda}}},
    \label{eq:L_alpha_lambda}
\end{align}
where $\mcl^i_{{\bm{\alpha}}{\bm{\lambda}}}$ is the loss defined on each qubit $i$. On one hand, if $\sigma_{\alpha_i} = \mathbb{I}$ \textit{or} $\sigma_{\lambda_i} = \mathbb{I}$ with $\alpha_i \ne \lambda_i$, it is evident that $\mcl^i_{{\bm{\alpha}}{\bm{\lambda}}} = 0$. On the other hand, if $\sigma_{\alpha_i} = \sigma_{\lambda_i} = \mathbb{I}$, we have  $\mcl^i_{{\bm{\alpha}}{\bm{\lambda}}} = 1$.  Thus, we only need to consider the qubits where $\alpha_j$ and $\lambda_j$ are both non-trivial.

In case of the uniform initialization over $[-\pi, \pi]$, it is straightforward to show that $\E_{\bm{\Theta}}[{\mcl_{\bm{\alpha} \bm{\lambda}}^i}]$ vanishes due to the periodicity of the trigonometric functions~\cite{letcher2023qgan}. However, the calculation requires a bit more effort in the case of the normal initialization as the results vary depending on $\mu_j, \nu_j, \alpha_j$ and $\lambda$. 
First of all, if $\alpha_j = \nu_j$, then $\sigma_{\alpha_j}$ commutes with $\sigma_{\nu_j}(\phi_j)$ and \eqnref{eq:L_alpha_lambda} simplifies  into 
\begin{align}
     \mcl^j_{\bm{\alpha}\bm{\lambda}} & = \frac{1}{2}\Tr\big(\sigma_{\mu_j}(-\omega_j) \sigma_{\alpha_j} \sigma_{\mu_j}(\omega_j) \sigma_{\lambda_j}\big)  = \frac{1}{2}\Tr\big(\sigma_{\mu_j}(-2\omega_j) \sigma_{\alpha_j} \sigma_{\lambda_j}\big) \nonumber \\& = \frac{1}{2}\cos(\omega_j)\Tr\big(\sigma_{\alpha_j} \sigma_{\lambda_j}\big) 
     + \frac{i}{2}\sin(\omega_j) \Tr\left(\sigma_{\mu_j} \sigma_{\alpha_j} \sigma_{\lambda_j}\right)
\label{eq:mcl2}
\end{align}
recalling that $\sigma_{\beta}\sigma_{\gamma}(\omega)  = \sigma_{\gamma}(-\omega) \sigma_{\beta}$ for $\beta \ne \gamma$ . 
We have three different cases : 
\begin{equation}
\mcl^j_{\bm{\alpha} \bm{\lambda}}  = 
\begin{cases}
\frac{1}{2}\cos(\omega_j)\text{Tr}\big( \mathbb{I} \big) +   \frac{i}{2}\sin(\omega_{j}) \Tr\big(\sigma_{\mu_j} \big)  
     = \cos(\omega_j)  & \alpha_j = \lambda_j \\
     -\frac{i}{2}\sin(\omega_j) \text{Tr}\big(\sigma_{\mu_j}\sigma_{\lambda_j}  \sigma_{\alpha_j}  \big)  = -\frac{i}{2}\sin(\omega_j) \Tr\big(\sigma_{\alpha_j} \big) = 0 &   \alpha_j \ne \lambda_j = \mu_j  \\ 
     -\frac{i}{2}\sin(\omega_j) \Tr\big(\sigma_{\mu_j}\sigma_{\lambda_j}  \sigma_{\alpha_j}  \big)  
     = -\frac{i}{2}\frac{1}{2}\sin(\omega_j) \text{Tr}\big(\pm 2i \sigma_{\mu_j} \sigma_{\mu_j}  \big)  = \pm \sin(\omega_j)  & 
      \alpha_j   \ne \lambda_j  \ne \mu_j
\end{cases}
\label{eq:mcl3}
\end{equation}
The third line is derived from the fact that $[\sigma_{\lambda_j}, \sigma_{\alpha_j}] = \pm 2i\sigma_{\mu_j}$ as $\alpha_j \ne \mu_j$ by orthogonality.
From the equation above, the loss function can be summarized as : 
\begin{equation}
    \mcl_{\bm{\alpha} \bm{\lambda}}^j = \cos(\omega_j) \delta_{\alpha_j \lambda_j } \pm \sin(\omega_j) \bar{\delta}_{\alpha_j \lambda_j }\bar{\delta}_{\lambda_j \mu_j}\;, 
    \label{eq:L_sum_1}
\end{equation}
where we denote $\bar{\delta}_{ij} = 1 - \delta_{ij} $, i.e. $\bar{\delta}_{ij} = 1$ if $i \ne j$ and $0$ if $i = j$. 

Let's consider the normal initialization of the parameters with the mean $\mu = 0$ and the standard deviation $\sigma$. If we take the expectation value $\E_{\omega_j}[\mcl_{\bm{\alpha} \bm{\lambda}}]$ over the parameter space $\omega_j$, the contribution of the $\sin(\omega_j)$ will cancel out, because $\sin(\omega_j)$ is an odd function at $\omega_j = 0$ over $\mathbb{R}$, while the Gaussian distribution is an even function, leaving only the contribution of the even function $\cos(\omega_j)$. Thus, $\E_{\omega_j}[ \mcl^j_{\bm{\alpha} \bm{\lambda}}]$ can be written as : 
\begin{equation}
    \E_{\omega_j}\left[ \mcl^j_{\bm{\alpha} \bm{\lambda}}\right] = \delta_{\alpha_j \lambda_j}\frac{1}{\sigma \sqrt{2\pi}} \int_{-\infty}^{\infty} \cos(\omega_j) e^{-\frac{\omega_j^2}{2\sigma^2}} d\omega_j = e^{-\frac{\sigma^2}{2}}\delta_{\alpha_j \lambda_j}\;.
\label{eq:Emcl1}
\end{equation}

Similarly, in the case of $\alpha_j \ne \nu_j$, each term in \eqnref{eq:L_alpha_lambda} is written as : 
\begin{equation}
    \mcl^j_{\bm{\alpha} \bm{\lambda}} = \frac{1}{2}\cos(\phi_j)\text{Tr}\big(\sigma_{\mu_j}(-\omega_j) \sigma_{\alpha_j} \sigma_{\mu_j}(\omega_j)  \sigma_{\lambda_j}\big) + \frac{i}{2}\sin(\phi_j)\text{Tr}\big(\sigma_{\mu_j}(-\omega_j) \sigma_{\nu_j} \sigma_{\alpha_j} \sigma_{\mu_j}(\omega_j)  \sigma_{\lambda_j}\big)\;.
\end{equation}
We should distinguish two different cases: 1) $\alpha_j = \mu_j$ and 2) $\alpha_j  \ne \mu_j$. In 1) $\sigma_{\alpha_j}$ and $\sigma_{\mu_j}(\omega_j)$ commute, hence : 

\begin{align}
   \mcl^j_{\bm{\alpha} \bm{\lambda}}&  = 
\frac{1}{2}\cos(\phi_j)\text{Tr}\big(\sigma_{\alpha_j}  \sigma_{\lambda_j}\big) + \frac{i}{2}\sin(\phi_j)\Tr(\sigma_{\mu_j}(-2\omega_j)\sigma_{\nu_j}\sigma_{\alpha_j}\sigma_{\lambda_j}) \nonumber \\  
&  = \frac{1}{2}\cos(\phi_j)\Tr\big(\sigma_{\alpha_j} \sigma_{\lambda_j}\big) + \frac{i}{2}\sin(\phi_j)\left(\cos(\omega_j)\Tr(\sigma_{\nu_j}\sigma_{\alpha_j}\sigma_{\lambda_j}) + i\sin(\omega_j)\Tr(\sigma_{\mu_j}\sigma_{\nu_j}\sigma_{\alpha_j}\sigma_{\lambda_j})\right) 
\nonumber \\ 
& = 
    \begin{cases}
\cos(\phi_j) & \alpha_j = \lambda_j \\
\sin(\phi_j)\sin(\omega_j) &   \alpha_j \ne \lambda_j = \nu_j \\ 
\pm \sin(\phi_j)\cos(\omega_j) & \alpha_j \ne \lambda_j \ne \nu_j  
\label{eq:L_al_2}
\end{cases} 
\end{align}
When we compute the expectation value, the last two cases vanish due to the parity over $\phi_j$. 
On the other hand, in 2) we have : 
\begin{align}
\mcl^j_{\bm{\alpha} \bm{\lambda}}&  =  \frac{1}{2}\cos(\phi_j)\Tr\big(\sigma_{\mu_j}(-2\omega_j) \sigma_{\alpha_j}\sigma_{\lambda_j}\big) + \frac{i}{2}\sin(\phi_j)\Tr\big(\sigma_{\nu_j} \sigma_{\alpha_j}\sigma_{\lambda_j}\big), \nonumber \\ 
    &  =  \frac{1}{2}\cos(\phi_j)\cos(\omega_j)\Tr(\sigma_{\alpha_j }\sigma_{\lambda_j}) +\frac{i}{2}\cos(\phi_j)\sin(\omega_j)\Tr(\sigma_{\mu_j }\sigma_{\alpha_j }\sigma_{\lambda_j})  
     + \frac{i}{2}\sin(\phi_j)\Tr(\sigma_{\nu_j }\sigma_{\alpha_j }\sigma_{\lambda_j})  \nonumber \\ 
      & = 
      \begin{cases}
          \cos(\phi_j) \cos(\omega_j) & \alpha_j = \lambda_j \\ 
          \pm \sin(\phi_j)  & \alpha_j \ne \lambda_j = \mu_j \\
          \pm \cos(\phi_j)\sin(\omega_j) & \alpha_j \ne \lambda_j = \nu_j 
      \end{cases}
     \label{eq:L_al_3}
\end{align}
Similar to before, the expectation value of the last two terms vanishes over $\phi_j$ and $\omega_j$ due to parity, leading to : 
\begin{equation}
    \mathbb{E}_{\omega_j, \phi_j}[\mcl^j_{\bm{\alpha} \bm{\lambda}}] =  
    \begin{cases}
        e^{-\frac{\sigma^2}{2}}\delta_{\alpha_j \lambda_j}  & \mu_j = \alpha_j \\
        \left(e^{-\frac{\sigma^2}{2}}\right)^2\delta_{\alpha_j \lambda_j}  &\mu_j  \ne \alpha_j 
    \end{cases}
    \label{eq:Emcl2}
\end{equation} 
Combining \eqnref{eq:Emcl1} and \eqnref{eq:Emcl2}, the expectation value $\E_{\bm{\Theta}[\mcl^i_{\bm{\alpha}}]}$ can be summarized as : 
\begin{equation}
    \E_{\bm{\Theta}}\left[\mcl^j_{\bm{\alpha} \bm{\lambda}}\right] = 
    \left(e^{-\frac{\sigma^2}{2}}\right)^{1 + \bar{\delta}_{\alpha_j \mu_j} \bar{\delta}_{\alpha_j \nu_j} }\delta_{\alpha_j \lambda_j}  \;. 
\end{equation}

From now on, we define $\mathcal{J}_{\bm{\alpha}} = \{j \, | \,\alpha_j \ne 0\}$ as a set of the qubit indices where the Pauli string $P_{\bm{\alpha}}$ has non-trivial Pauli matrices and the weight of $P_{\bm{\alpha}}$, denoted as $w(P_{\bm{\alpha}}) := |\mathcal{J}_{\bm{\alpha}}|$.  Then, the expectation of the final loss across the whole input state $\rho$ (c.f. \eqnref{eq:L_alpha_lambda}) will be : 
\begin{equation}
    \E_{\bm{\Theta}}\left[\mcl_{\bm{\alpha}}\right] = \sum_{\bm{\lambda}} c_{\bm{\lambda}} \prod_j \E_{\bm{\Theta}}\left[\mcl_{\bm{\alpha} \bm{\lambda}}^j\right] = \sum_{\bm{\lambda}} c_{\bm{\lambda}} \left(e^{-\frac{\sigma^2}{2}}\right)^{w(P_{\bm{\alpha}})}  
 \prod_{j \in \mathcal{J}_{\bm{\alpha}}} \delta_{\alpha_j \lambda_j} \left(e^{-\frac{\sigma^2}{2}}\right)^{\bar{\delta}_{\alpha_j \mu_j} \bar{\delta}_{\alpha_j \nu_j}} = K_{\bm{\alpha}}\left(c_{\bm{\lambda}}\right) \left(e^{-\frac{\sigma^2}{2}}\right)^{w(P_{\bm{\alpha}})}\;,  
 \label{eq:Emcl3}
\end{equation}
where $K_{\bm{\alpha}}\left(c_{\bm{\lambda}}\right) = c_{\bm{\lambda}^*}$ if there exists $\bm{\lambda}^*$ such that $\bm{\lambda}^* = \bm{\alpha}$, and $K_{\bm{\alpha}}\left(c_{\bm{\lambda}}\right) = 0$ otherwise. We underline that if there is at least one qubit $j$ such that $\alpha_j \ne \lambda_j$, the expectation value $\E_{\bm{\Theta}}\left[\mcl_{\bm{\alpha}\bm{\lambda}}\right]$ will be zero. In particular, if $\rho$ is a product state, we have $c_{\bm{\lambda}} \le 1$ for all $c_{\bm{\lambda}}$ and the inequality can be simplified as: 
\begin{equation}
    \E_{\bm{\Theta}}\left[\mcl_{\bm{\alpha}}\right] \le \left(e^{-\frac{\sigma^2}{2}}\right)^{w(P_{\bm{\alpha}})}\;.
\end{equation} 

\subsubsection{$L\geq 1$ case}
We use recursive steps to derive the expectation value for $L \ge 1$. For clarity, we will use superscript $\mcl^{(L)}$ to denote the loss function with $U_L(\bm{\Theta})$. Taking into account the definition of $U_L$ and $\tilde{W}_{L(K_L)}$ in \eqnref{eq:def_U} and \eqnref{eq:def_VW}, we write explicitly the loss function in terms of $U_{L - 1}$ and take out $P_{L, K_L}(\theta_{L, K_L})$ from $\tilde{W}_{L(K_L)}$ in order to make the recursion steps clear: 
\begin{align}
    \mcl^{(L)}_{\bm{\alpha}} & = \Tr(U_L^\dagger P_{\bm{\alpha}} U_L \rho)  = \Tr(U^\dagger_{L-1}\tilde{W}_{L(K_L)}^\dagger P_{\bm{\alpha}} \tilde{W}_{L(K_L)} U_L \rho) \nonumber \\  &
    = \Tr(U_{L-1}^\dagger \tilde{W}^\dagger_{L(K_L - 1)} P_{L, K_L}(-\theta_{L, K_L})  P_{\bm{\alpha}} P_{L, K_L}(\theta_{L, K_L})\tilde{W}_{L(K_L - 1)} U_{L-1}\rho)\;,
    \label{eq:induction1}
\end{align} 
where $\tilde{W}_{L(K_L -1)} = \left(\prod^{K_L-1}_{i=1} P_{L, i}(\theta_{L, i}) \right) C_L$. 
As $P_{\bm{\alpha}}$ is a Pauli string, it will either commute or anti-commute with $P_{L, i}$ which is also a Pauli string. 
In the former, $P_{L, K_L}(-\theta_{L, K_L})$ commutes with $P_{\bm{\alpha}}$ and \eqnref{eq:induction1} will be written as : 
\begin{equation}
    \mcl^{(L)}_{\bm{\alpha}} = \tilde{\mcl}^{(L)}_{\bm{\alpha}} := \Tr(U_{L-1}^\dagger  
 \tilde{W}^\dagger_{L(K_L - 1)}  P_{\bm{\alpha}} \tilde{W}_{L(K_L - 1)} U_{L-1}\rho)\; ,
    \label{eq:induction2_1}
\end{equation}
which has the same form as \eqnref{eq:induction1} but with $\tilde{W}_{L(K_L - 1)}$. 

On the other hand, if $P_{L, K_L}$ anti-commutes with $P_{\bm{\alpha}}$, we have $P_{\bm{\alpha}} P_{L, K_L}(\theta_{L, K_L}) = P_{L, K_L}(-\theta_{L, K_L}) P_{\bm{\alpha}}$. This leads to the following expression : 
 \begin{align}
     \mcl^{(L)}_{\bm{\alpha}} & = \Tr(U_{L-1}^\dagger \tilde{W}^\dagger_{L(K_L - 1)} P_{L, K_L}(-2\theta_{L, K_L})  P_{\bm{\alpha}} \tilde{W}_{L(K_L - 1)} U_{L-1}\rho)   \nonumber \\ 
    & = \cos(\theta_{L, K_L}) \Tr(U_{L-1}^\dagger \tilde{W}^\dagger_{L(K_L - 1)} P_{\bm{\alpha}} \tilde{W}_{L(K_L - 1)} U_{L-1} \rho) + i\sin(\theta_{L, K_L})\Tr(U_{L-1}^\dagger \tilde{W}^\dagger_{L(K_L - 1)}  P_{L, K_L}P_{\bm{\alpha}} \tilde{W}_{L(K_L - 1)} U_{L-1} \rho)
     \label{eq:induction2_2}
\end{align}
As both $P_{\bm{\alpha}}$ and $P_{L, K_L}$ are Pauli strings, there exists a Pauli string $P_{\bm{\tau}}$ such that $[P_{L, K_L}, P_{\bm{\alpha}}] = 2iP_{\bm{\tau}}$. Furthermore, due to the additivity of the trace, we have : 
\begin{equation}
    \Tr(A P_{L, K_L} P_{\bm{\alpha}} B) = -\Tr(AP_{\bm{\alpha}} P_{L, K_L} B) =  = \frac{1}{2} \Tr(A [P_{L, K_L}, P_{\bm{\alpha}}] B)  = i \Tr(A P_{\bm{\tau}} B)\;, 
\end{equation}
with $A$ and $B$ two matrices. While taking into account the definition of $\tilde{\mcl}^{(L)}_{\bm{\alpha}}$ in \eqnref{eq:induction2_1}, we can simplify \eqnref{eq:induction2_2} as following: 
\begin{equation}
   \mcl^{(L)}_{\bm{\alpha}} = \cos(\theta_{L, K_L}) \tilde{\mcl}^{(L)}_{\bm{\alpha}} + \sin(\theta_{L, K_L}) \tilde{\mcl}^{(L)}_{\bm{\tau}}\;.
\end{equation}
From its expression, we notice that $\mcl^{(L)}_{\bm{\alpha}}$ can be expressed as a weighted sum of two loss functions $\tilde{\mcl}^{(L)}_{\bm{\alpha}}$ and $\tilde{\mcl}^{(L)}_{\bm{\tau}}$ for the observable $P_{\bm{\alpha}}$ and $P_{\bm{\tau}}$ while excluding $P_{L, K_L}$ from $U_{L}$. Hence, the loss function can be extended as a product of cosine and sine functions of $\bm{\theta}_L$ by repeating  \eqnref{eq:induction2_1} and \eqnref{eq:induction2_2} for all $P_{L, i}$ with $i = K_L-1,...,1$ depending on whether $P_{L, i}$ and $P_{\bm{\alpha}}$ commute. 

For the following calculations, we define $\mathcal{N}^\ell_A(P_{\bm{\alpha}})$ the set of indices of the Pauli string in $\ell$-th layer which anticommutes with $P_{\bm{\alpha}}$, i.e,. 
\begin{equation}
    \mathcal{N}^\ell_A\left(P_{\bm{\alpha}}\right) = \{\,k \mid \, [P_{\ell, k}, P_{\bm{\alpha}}] \ne 0\;, k \in \{1,..., K_\ell \} \, \}\;, 
\end{equation}
and $n^\ell_A\left(P_{\bm{\alpha}}\right)$ its cardinality. 
At the end of the recursive steps, we obtain: 
\begin{align}
    \mcl^{(L)}_{\bm{\alpha}} = & \prod_{i_j \in \mathcal{N}^L_A\left(P_{\bm{\alpha}}\right) } \cos(\theta_{L, i_j}) \Tr(U^\dagger_{L-1}C^\dagger_L P_{\bm{\alpha}} C_L U_{L-1} \rho) \nonumber \\ &   + \sum_{i_j   \in \mathcal{N}^L_A\left(P_{\bm{\alpha}}\right) }\sin(\theta_{L, i_j}) \prod_{\substack{i_{k} \in \mathcal{N}^L_A\left(P_{\bm{\alpha}}\right) \\ i_{k} > i_j} }  \cos(\theta_{L, i_{k}}) \prod_{\substack{i_{m}\in \mathcal{N}^L_A\left([P_{L, i_j}, P_{\bm{\alpha}}]\right) \\ i_{m} < i_j} }  \cos(\theta_{L, i_{m}}) \Tr(U^\dagger_{L-1}C^\dagger_L \frac{i}{2}  [P_{L, i_j}, P_{\bm{\alpha}}] C_L U_{L-1} \rho) 
    \nonumber \\ & + \mathcal{O}(\sin(\theta_{L, i_j})\sin(\theta_{L, i_{k}}))\;.
    \label{eq:induction3}
\end{align}
\begin{figure}[h]
    \centering
    \includegraphics[width = 0.5\textwidth]{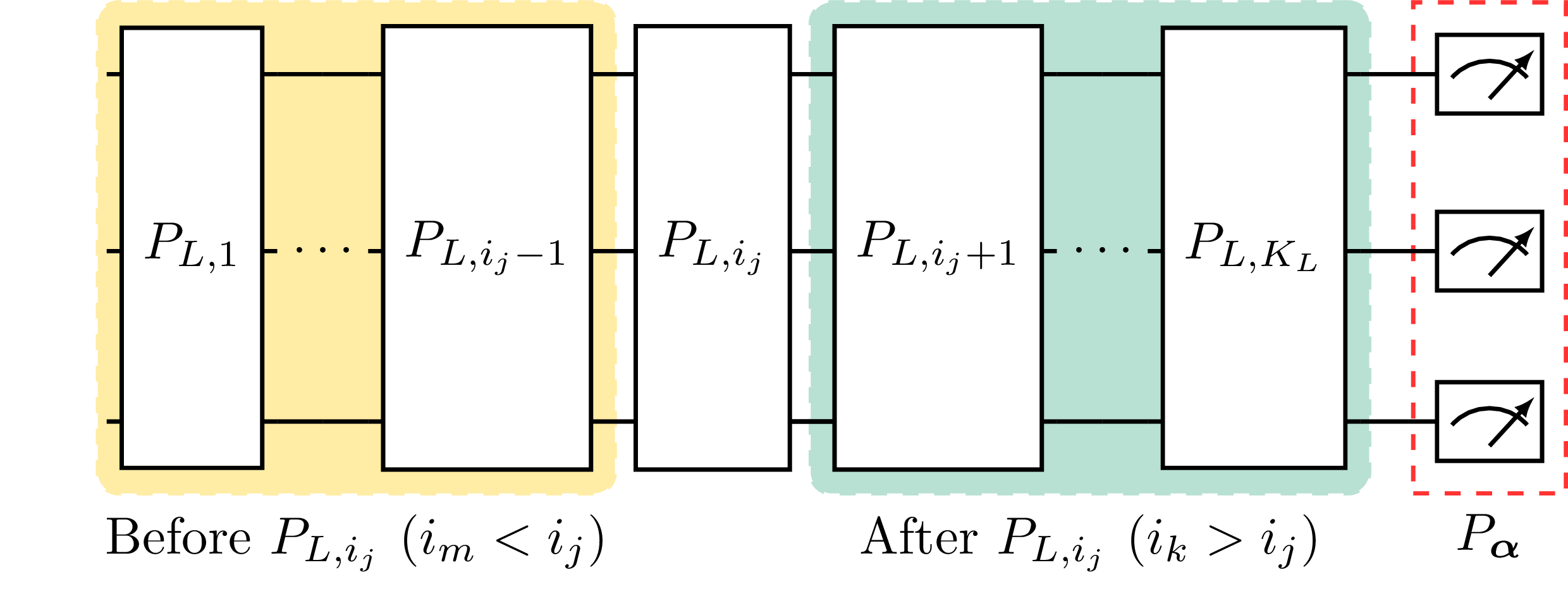}
    \caption{\textbf{Schematic diagram of $\tilde{W}_L$ to illustrate the second term in \eqnref{eq:induction3}.} The figure visualizes the Pauli gates that are located \textit{before} $P_{i_j}$ (in yellow) and \textit{after} $P_{i_j}$ (in green). The second term of \eqnref{eq:induction3} consists of $\cos(\theta_{i_k})$ for $i_k\in \mathcal{N}_A^L(P_{\bm{\alpha}})$ located \textit{after} $P_{i_j}$ and $\cos(\theta_{i_m})$ for $i_m\in \mathcal{N}_A^L(P_{\bm{\alpha}})$ located \textit{before} $P_{i_j}$. }
    \label{fig:P_ij_circuit}
\end{figure}
More explicitly, the first term in \eqnref{eq:induction3} captures the contributions involving only the cosine terms for the Pauli strings $P_{i_j}$ that anti-commute with $P_{\bm{\alpha}}$. The second term comprises all the sine contributions of $P_{i_j}$. Specifically, each term in the sum captures the cosine contribution from $P_{i_k}(\theta_{i_k})$ with $i_k > i_j$ that anti-commute with $P_{\bm{\alpha}}$ and from $P_{i_m}(\theta_{i_m})$ with $i_m < i_j$ that anti-commute with $[P_{i_j}, P_{\bm{alpha}}]$. \figref{fig:P_ij_circuit} is provided to aid in understanding this term. Finally, the last term in \eqnref{eq:induction3} is dependent on the higher order of $\sin(\theta)$ and contains all the possible combinations of $\cos(\theta_{i_j})$ and $\sin(\theta_{i_j})$, multiplied the nested commutation relations. Since we are interested in small angle initialization, $\sin(\theta_{i_j})$ is much smaller compared to $\cos(\theta_{i_j})$. Hence, the last term is negligible compared to the first two terms. This natural emergence of nested commutators in the loss function is fundamentally related to the \textit{Dynamical Lie Algebra} (DLA) that determines the expressivity of the circuit.

although the equation above looks overwhelming, taking an expectation over it significantly simplifies the final equation. As $C_L$ is a Clifford gate, there exists a Pauli string $P_{\bm{\gamma}}\in \mathbf{P}_n$ such that $C_L^\dagger P_{\bm{\alpha}} C_L = P_{\bm{\gamma}}$. Thus, the trace in the first term can be rewritten as 
\begin{equation}
    \Tr(U^\dagger_{L-1}C^\dagger_L P_{\bm{\alpha}} C_L U_{L-1} \rho) =   \Tr(U^\dagger_{L-1}P_{\bm{\gamma}} U_{L-1} \rho) := \mcl^{(L-1)}_{\bm{\gamma}}\; , 
\end{equation} 
where $\mcl^{(L-1)}_{\bm{\gamma}}$ corresponds to the loss function of the observable $P_{\bm{\gamma}}$ with $L-1$ layers. 
Similar to the previous justifications, all the expectation values except the first term in \eqnref{eq:induction3} vanish due to the parity of the sine function. Thus, we have: 
\begin{align}
    \E_{\Theta}\left[\mcl^{(L)}_{\bm{\alpha}} \right]  & = \E_{\bm{\theta}_L}\left[\prod_{i_j \in \mathcal{N}^L_A\left(P_{\bm{\alpha}} \right)} \cos(\theta_{i_j})\right]  \E_{\bm{\Theta} / \bm{\theta}_L}\left[\mcl^{(L-1)}_{\bm{\gamma}}\right] \nonumber \\ & = \prod_{i_j \in \mathcal{N}^L_A\left(P_{\bm{\alpha}} \right)} \E_{\bm{\theta}_L}\left[\cos(\theta_{i_j})\right]  \E_{\bm{\Theta} / \bm{\theta}_L}\left[\mcl^{(L-1)}_{\bm{\gamma}}\right] 
    \nonumber \\ 
    & =  e^{-n^L_A\left(P_{\bm{\alpha}}\right)\frac{\sigma^2}{2}} \E_{\bm{\Theta} / \bm{\theta}_L}\left[\mcl^{(L-1)}_{\bm{\gamma}}\right]\;,  
\end{align}
where we use the commutativity between the product and expectation for independent variables in the second line. 

Now, let us call $C^{i_n}_{i_1} :=  C_{i_n} \cdots C_{i_1} $,  $C^{ i_1\dagger}_{i_n} :=  C^\dagger_{i_1} ... C^\dagger_{i_n}$ and $C^{i_1 \dagger }_{i_n}P_{\bm{\alpha}} C^{i_n}_{i_1}$, the Pauli observable $P_{\bm{\alpha}}$, conjugated between $C^\dagger_i$ and $C_i$ for all $i = i_1,...,i_n$. By definition of the Clifford gates, $C^{\ell\dagger}_{L}P_{\bm{\alpha}} C^L_\ell$ is also a Pauli string for all $\ell = 1,..., L$ with $C^{L+1\dagger}_{L}P_{\bm{\alpha}} C^L_{L+1} = P_{\bm{\alpha}}$, and hence, there exists a Pauli string $P_{\bm{\eta}}$ such that $P_{\bm{\eta}} = C^{1\dagger}_LP_{\bm{\alpha}} C^L_1$. Then, we have the final equation for the expectation value of the loss : 
\begin{equation}
    \E_{\bm{\Theta}}\left[\mcl^{(L)}_{\bm{\alpha}}\right]  = \left(e^{-\frac{\sigma^2}{2}}\right)^{N^L_A} \E_{\bm{\Theta}}\left[\mcl^{(0)}_{\bm{\eta}}\right] =  K_{\bm{\eta}}\left(c_{\bm{\lambda}}\right)\left(e^{-\frac{\sigma^2}{2}}\right)^{N^L_A + w(P_{\bm{\eta}})}\;,
    \label{eq:final_E1}
\end{equation}
with $N^L_A = \sum_{\ell = 1}^L n_A^\ell\left(C^{\ell + 1\dagger}_LP_{\bm{\alpha}} C^L_{\ell + 1} \right)$. 

In particular, if $\rho$ is a product state and there exists at least one $P_{\ell, i}$ which anti-commutes with $C^{\ell+1\dagger}_{ L} P_{\bm{\alpha}} C^L_{\ell + 1}$ for all $\ell$\;, $N_A^L$ will scale linearly with respect to $L$ and the expectation value will decay exponentially with respect to the number of qubits for a polynomial depth circuit : 
\begin{equation}
    \E_{\bm{\Theta}}\left[\mcl^{(L)}_{\bm{\alpha}}\right]  \le \left(e^{-\frac{\sigma^2}{2}}\right)^{\text{poly}(n) + w(P_{\bm{\eta}})}\;. 
    \label{eq:final_E2}
\end{equation}
This will be the case of EfficientSU2 ansatz which will be presented in \adxref{adx:case_study}. 

\subsection{\label{adx:variance_normal}Variance with normal initialization}
\subsubsection{$L=0$ case}
In this section, we calculate the lower bound for the variance of the loss function, $\Var_{\bm{\Theta}}[\mcl]$.  For simplicity, we are interested in the scenario where the covariance between the loss functions for two different Pauli strings, $\mcl_{\bm{\alpha}}$ and $\mcl_{\bm{\beta}}$ vanishes, i.e. $\Var_{\bm{\Theta}}[\mcl_{\bm{\alpha}}\mcl_{\bm{\beta}}]=0$ for ${\bm{\alpha}} \ne {\bm{\beta}}$. Then, $\Var_{\bm{\Theta}}[\mcl]$ can be written as a sum : 
\begin{equation}
    \Var_{\bm{\Theta}}[\mcl] = \sum_{\bm{\alpha}} a_{\bm{\alpha}} \Var_{\bm{\Theta}}[\mcl_{\bm{\alpha}}] = \sum_{\bm{\alpha}}a_{\bm{\alpha}} \left(\E_{\bm{\Theta}}\left[\mcl_{\bm{\alpha}}^2\right] - \E_{\bm{\Theta}}\left[\mcl_{\bm{\alpha}}\right]^2\right)\;. 
    \label{eq:var1}
\end{equation}
with $\mcl_{\bm{\alpha}}^2$ that can be decomposed explicitly as :   
\begin{equation}
    \mcl_{\bm{\alpha}}^2 = \sum_{\bm{\lambda}} \sum_{\bm{\lambda}'} c_{\bm{\lambda}} c_{\bm{\lambda}'} \prod_i \mcl_{\bm{\alpha} \bm{\lambda}}^i \prod_j  \mcl_{\bm{\alpha} \bm{\lambda}'}^j = \sum_{\bm{\lambda}} \sum_{\bm{\lambda}'} c_{\bm{\lambda}} c_{\bm{\lambda}' } \prod_i \mcl_{\bm{\alpha} \bm{\lambda}}^{i} \mcl_{\bm{\alpha} \bm{\lambda}'}^{i}\;.
\end{equation}

Now, we will explicitly compute $\E_{\bm{\Theta}}\left[{\mcl_{\bm{\alpha} \bm{\lambda}}^{i}} \mcl_{\bm{\alpha} \bm{\lambda}'}^{i} \right]$ by distinguishing different cases of $\alpha_j \ne 0$. To begin with, let us consider the case where $\alpha_j = \nu_j$. With \eqnref{eq:L_sum_1}, we have : 
\begin{equation}
    \mcl^j_{\bm{\alpha} \bm{\lambda}} \mcl^j_{\bm{\alpha} \bm{\lambda}'}  = \cos^2(\omega_j) \delta_{\alpha_j \lambda_j} \delta_{\lambda_j \lambda'_j} 
    +  \sin^2 (\omega_j) \bar{\delta}_{\alpha_j \lambda_j}
 \bar{\delta}_{\lambda_j\mu_j }\delta_{\lambda_j\lambda_j'}  + \text{odd}\;, 
 \label{eq:mcl2_1}
\end{equation} 
where the last term indicates the term which is proportional to $\cos(\omega_j)\sin(\omega_j)$, thus odd at $\omega_j = 0$ over $\mathbb{R}$.  This term cancels out as it is an odd function integrated with an even probability distribution. 
By integrating the first two terms over normal probability distribution for $\omega_j$, we obtain : 
\begin{equation}
    \E_{\omega_j}\left[{\cos^2(\omega_j)}\right] = \frac{1}{\sigma \sqrt{2\pi}} \int_{-\infty}^{\infty} \cos^2(\omega_j) e^{-\frac{\omega_j^2}{2\sigma^2}} d\omega_j = e^{-\sigma^2} \cosh{\sigma^2} = \frac{1 + e^{-2\sigma^2}}{2}\;,
    \label{eq:cos2_exp}
\end{equation}

\begin{equation}
    \E_{\omega_j}\left[{\sin^2(\omega_j)}\right] = \frac{1}{\sigma \sqrt{2\pi}} \int_{-\infty}^{\infty} \sin^2(\omega_j) e^{-\frac{\omega_j^2}{2\sigma^2}} d\omega_j = e^{-\sigma^2} \sinh{\sigma^2} = \frac{1 - e^{-2\sigma^2}}{2}\;.
    \label{eq:sin2_exp}
\end{equation}
Therefore, we can write the expectation value of \eqnref{eq:mcl2_1} as : 
\begin{equation}
    \E_{\omega_j}\left[\mcl^j_{\bm{\alpha} \bm{\lambda}} \mcl^j_{\bm{\alpha} \bm{\lambda}'}\right] =  \frac{1 + e^{-2\sigma^2}}{2}  \delta_{\alpha_j \lambda_j} \delta_{\lambda_j \lambda'_j} + \frac{1 - e^{-2\sigma^2}}{2}  \bar{\delta}_{\alpha_j \lambda_j}
 \bar{\delta}_{\lambda_j\mu_j }\delta_{\lambda_j\lambda_j'}\;.
 \label{eq:mcl2_2}
\end{equation}

On the other hand, if  $\alpha_j \ne \nu_j$, there are two different cases : 1) $\alpha_j = \mu_j$ and  2) $\alpha_j \ne \mu_j$. 
We start with the first case of $\alpha_j = \mu_j$.  Using \eqnref{eq:L_al_2}, we can expand the expression of $\mcl_{\bm{\alpha} \bm{\lambda}}^j\mcl_{\bm{\alpha} \bm{\lambda}'}^j$ as: 
\begin{equation}
        \mcl_{\bm{\alpha} \bm{\lambda}}^j \mcl_{\bm{\alpha} \bm{\lambda}'}^j = \cos^2(\phi_j)\delta_{\alpha_j \lambda_j}\delta_{\lambda_j \lambda_j'} + \sin^2(\omega_j)\sin^2(\phi_j) \bar{\delta}_{\alpha_j \lambda_j}\delta_{\lambda_j \nu_j}\delta_{\lambda_j \lambda_j'} + \cos^2(\omega_j)\sin^2(\phi_j) \bar{\delta}_{\alpha_j \lambda_j}\bar{\delta}_{\lambda_j \nu_j}\delta_{\lambda_j \lambda_j'} + \text{odd}\;,
        \label{eq:mcl2_3}
    \end{equation}
which gives the following expectation value $\E_{\omega_j, \phi_j}\left[\mcl^j_{\bm{\alpha} \bm{\lambda}} \mcl^j_{\bm{\alpha} \bm{\lambda}'}\right] $ using \eqnref{eq:cos2_exp} and \eqnref{eq:sin2_exp}:
\begin{equation}
    \E_{\omega_j, \phi_j}\left[\mcl^j_{\bm{\alpha} \bm{\lambda}} \mcl^j_{\bm{\alpha} \bm{\lambda}'}\right] =  \frac{1 + e^{-2\sigma^2}}{2}  \delta_{\alpha_j \lambda_j}\delta_{\lambda_j\lambda_j'} 
    + \left(\frac{1 - e^{-2\sigma^2}}{2}\right)^2\bar{\delta}_{\alpha_j \lambda_j}\delta_{\lambda_j \nu_j}\delta_{\lambda_j \lambda_j'}  + \frac{1 - e^{-4\sigma^2}}{4}\bar{\delta}_{\alpha_j \lambda_j}\bar{\delta}_{\lambda_j \nu_j}\delta_{\lambda_j \lambda_j'}\;.   
\end{equation}
Now, let's consider the case with $\alpha_j \ne \mu_j$. 
With the similar justification as before, $\mcl^j_{\bm{\alpha} \bm{\lambda}} \mcl^j_{\bm{\alpha} \bm{\lambda}'}$ can be written as : 
\begin{align}
\mcl_{\bm{\alpha} \bm{\lambda}}^j \mcl_{\bm{\alpha} \bm{\lambda}'}^j = \cos^2(\omega_j)\cos^2(\phi_j)\delta_{\alpha_j \lambda_j}\delta_{\lambda_j \lambda_j'} + \sin^2(\phi_j) \bar{\delta}_{\alpha_j \lambda_j}\delta_{\lambda_j \mu_j}\delta_{\lambda_j \lambda_j'} + \cos^2(\phi_j)\sin^2(\omega_j) \bar{\delta}_{\alpha_j \lambda_j}\delta_{\lambda_j \nu_j}\delta_{\lambda_j \lambda_j'} + \text{odd}\;.  
\end{align}
where the last odd term is proportional to $\cos(\omega_j)\sin(\omega_j)$ and $\cos(\phi_j)\sin(\phi_j)$. This leads to :
\begin{equation}
     \E_{\omega_j, \phi_j}\left[\mcl^j_{\bm{\alpha}\bm{\lambda}} \mcl^j_{\bm{\alpha}\bm{\lambda}'}\right] =  \left(\frac{1 + e^{-2\sigma^2}}{2}\right)^2  \delta_{\alpha_j \lambda_j}\delta_{\lambda_j \lambda_j'} 
    + \frac{1 - e^{-2\sigma^2}}{2}\bar{\delta}_{\alpha_j \lambda_j}\delta_{\lambda_j \mu_j}\delta_{\lambda_j \lambda_j'}  + \frac{1 - e^{-4\sigma^2}}{4}\bar{\delta}_{\alpha_j \lambda_j}\delta_{\lambda_j \nu_j}\delta_{\lambda_j \lambda_j'}\;.   
\end{equation}

We can notice that in all cases, $\E_{\omega_j, \phi_j}\left[\mcl^j_{\bm{\alpha}\bm{\lambda}} \mcl^j_{\bm{\alpha}\bm{\lambda}'}\right]$ cancels out if $\lambda_j \ne \lambda_j'$. Thus, the final expression will be summarized as a sum over $\bm{\lambda}$. Combining the expressions above, we have the expression for the expectation value $\E[\mathcal{L}^2_{\bm{\alpha}}]$: 
\begin{equation}
     \E_{\bm{\Theta}}\left[\mcl^2_{\bm{\alpha}}\right] = \sum_{\bm{\lambda}} c^2_{\bm{\lambda}} \prod_j\E_{\bm{\Theta}}\left[\left(\mcl^j_{\bm{\alpha}\bm{\lambda}} \right)^2\right] = \sum_{\bm{\lambda}} c^2_{\bm{\lambda}}\prod_{j \in \mathcal{J}_\alpha}  \tilde{c}_{\alpha_j \mu_j \nu_j \lambda_j}  \left(  \frac{1 + e^{-2\sigma^2}}{2}\delta_{\alpha_j \lambda_j } +  \frac{1 - e^{-2\sigma^2}}{2}\bar{\delta}_{\alpha_j \lambda_j } \right)\;, 
     \label{eq:mcl2_4}
\end{equation}
where $\tilde{c}_{\alpha_j \mu_j \nu_j \lambda_j}$ is the coefficient indicating different possibilities depending on the value of $\alpha_j, \mu_j, \nu_j$ and $\lambda_j$ as follows: 
\begin{align}
    \tilde{c}_{\alpha_j \mu_j \nu_j \lambda_j}  = & \left(\delta_{\alpha_j \nu_j} + \bar{\delta}_{\alpha_j \nu_j} \delta_{\alpha_j \mu_j} + \bar{\delta}_{\alpha_j \nu_j} \bar{\delta}_{\alpha_j \mu_j} \right) \delta_{\alpha_j \lambda_j} +\frac{1 + e^{-2\sigma^2}}{2} \bar{\delta}_{\alpha_j \lambda_j}\delta_{\alpha_j\nu_j}\bar{\delta}_{\lambda_j\mu_j}  \nonumber \\ 
     &  +  \bar{\delta}_{\alpha_j \lambda_j}\bar{\delta}_{\alpha_j\nu_j} \left(   \bar{\delta}_{\alpha_j\mu_j} \delta_{\lambda_j \mu_j} + \frac{1 - e^{-2\sigma^2}}{2}\delta_{\alpha_j\mu_j} \delta_{\lambda_j \nu_j} + \frac{1 + e^{-2\sigma^2}}{2} (\delta_{\alpha_j\mu_j}\bar{\delta}_{\lambda_j\nu_j} + \bar{\delta}_{\alpha_j\mu_j}\delta_{\lambda_j\nu_j})\right)\;. 
\end{align}
\eqnref{eq:mcl2_4} implies that $\E_{\bm{\Theta}}\left[\mcl_{\bm{\alpha}}^2\right]$ scales differently depending on whether the Pauli matrix in $P_{\bm{\alpha}}$ is the same as the one in the initial state $\rho$  at each qubit $j$ and this will highly influence the lower bound of the variance calculated in the following steps. 

\subsubsection{$L\geq1$ case}
To calculate the lower bound of $\E_{\bm{\Theta}}\left[(\mcl_{\bm{\alpha}}^{(L)})^2\right]$ for $L > 0$, we can follow a recursive approach similar to the previous steps. 
 We begin by squaring Eq.~\eqref{eq:induction3} to compute $\left(\mcl^{(L)}_{\alpha}\right)^2$:
\begin{align}
    \left(\mcl^{(L)}_{\bm{\alpha}}\right)^2 = & \prod_{i_j \in \mathcal{N}^L_A\left(P_{\bm{\alpha}}\right) } \cos(\theta_{L, i_j})^2 \left(\mcl^{(L-1)}_{\bm{\gamma}}\right)^2 \nonumber \\ & + \sum_{i_j \in \mathcal{N}^L_A\left(P_{\bm{\alpha}}\right)} \sin^2(\theta_{i_j})  \prod_{\substack{i_{k} \in \mathcal{N}^L_A\left(P_{\bm{\alpha}}\right) \\ i_{k} > i_j} }  \cos^2(\theta_{L, i_{k}}) \prod_{\substack{i_{m}\in \mathcal{N}^L_A\left([P_{L, i_j}, P_{\bm{\alpha}}]\right) \\ i_{m} < i_j} }  \cos^2(\theta_{L, i_{m}}) \left(\mcl^{(L-1)}_{\bm{\delta}_{i_j}}\right)^2  \nonumber \\  
    & + \mathcal{O}(\sin^4(\theta)) + \text{odd}
    \label{eq:mcl_square}
\end{align}
where $P_{\bm{\delta}_{i_j}}= \frac{i}{2} C^\dagger [P_{L, i_j}, P_{\bm{\alpha}}]C \in \mathcal{G}$ and the odd term containing all the terms depending on $\cos(\theta_{L, i})\sin(\theta_{L, i})$ and $\sin(\theta_{L, i})\sin(\theta_{L, k})$ for $i \ne k$.  Taking the expectation value over \eqnref{eq:mcl_square} and cancelling out all the odd terms leads to :  
\begin{align}
\E_{\bm{\Theta}}[(\mcl_{\bm{\alpha}}^{(L)})^2]  
=&   \left(\frac{1 + e^{-2\sigma^2}}{2}\right)^{n^L_A(P_{\bm{\alpha}})} \E_{\bm{\Theta}/\bm{\theta}_L}\left[\left(\mcl^{(L-1)}_{\gamma}\right)^2\right]  \nonumber \\  & + \sum_{i_j \in \mathcal{N}^L_A\left(P_{\bm{\alpha}}\right)} \left( \frac{1 - e^{-2\sigma^2}}{2}\right) \left(\frac{1 + e^{-2\sigma^2}}{2} \right)^{\tilde{n}^L_A(i_j)}\E_{\bm{\Theta}/\bm{\theta}_L}\left[\left(\mcl^{(L-1)}_{\bm{\delta}_{i_j}}\right)^2\right] \nonumber \\ & + \mathcal{O}\left( \left(\frac{1 - e^{-2\sigma^2}}{2}\right)^2 \right) ,\; 
\label{eq:mcl2_5}
\end{align}
where $\tilde{n}^L_A(i_j)$ is the number of cosine terms multiplied to $\sin(\theta_{i_j})$ in \eqnref{eq:mcl_square}. Rigorously, it can be mathematically written as:
\begin{equation}
    \tilde{n}^\ell_A(i_j) = \lvert\{i_k \in \mathcal{N}^L_A\left(P_{\bm{\alpha}}\right) \, \vert \, i_k > i_j\,  \}  \rvert  + \lvert\{i_m \in \mathcal{N}^L_A\left([P_{\ell, i_j}, P_{\bm{\alpha}}] \right) \, \vert \,  i_m < i_j\,  \}  \rvert  
\end{equation}
Indeed, it counts all the Pauli gates in $\tilde{W}_\ell$ that anti-commute with $P_{\bm{\alpha}}$ if they are located after $P_{\ell, i_j}$, and those that anti-commute with $[P_{\ell, i_j}, P_{\bm{\alpha}}]$ if they are located before $P_{\ell, i_j}$ (see \figref{fig:P_ij_circuit}).

As $\E_{\bm{\Theta}}[(\mcl_{\bm{\alpha}}^{(L)})^2]$ depends on $P_{\bm{\alpha}}$ and all possible nested commutators including $P_{L, i}$, it is extremely complicated to generalize the exact equation. Therefore, for the general case, we will compute an exact lower bound depending on $\sigma$. However, this bound does not provide much information about the scaling with respect to the system's size without further assumption on the circuit architecture. In \adxref{adx:case_study}, we consider an EfficientSU2 architecture as a specific example and provide some \textit{approximate} bounds which show the polynomial scaling of the variance for the leading order in $\sigma$. These bounds, despite being approximate, are rather tight as supported by our numerics.

First of all, let us consider the case with $\sigma \ll 1 $. As $e^{-2\sigma^2}$ is close to $1$, $\frac{1 + e^{-2\sigma^2}}{2}$ dominates over $\frac{1 - e^{-2\sigma^2}}{2}$, and therefore, the first term will mainly contribute in the lower bound : 
\begin{equation}
    \E_{\bm{\Theta}}[(\mcl_{\bm{\alpha}}^{(L)})^2] > \big(\frac{1 + e^{-2\sigma^2}}{2}\big)^{n_A^L\left(P_{\bm{\alpha}}\right)}\E_{\bm{\Theta}}[(\mcl^{(L-1)}_{\bm{\gamma}})^2] > \big(\frac{1 + e^{-2\sigma^2}}{2}\big)^{N_A^L} \E_{\bm{\Theta}}[(\mcl^{(0)}_{\bm{\eta}})^2],
    \label{eq:general_E2}
\end{equation}
where we obtain the last inequality by induction. Combining \eqnref{eq:final_E1} and \eqnref{eq:general_E2}, we have the lower bound for the variance $\Var_{\bm{\Theta}}[\mcl_{\bm{\alpha}}^{(L)}]$ : 
\begin{equation}
    \Var[\mcl_{\bm{\alpha}}] = \E[(\mcl_{\bm{\alpha}})^2] - \E[\mcl_{\bm{\alpha}}]^2 \ge  \Big(\frac{1 + e^{-2\sigma^2}}{2}\Big)^{N^L_A}\E_{\bm{\Theta}}[(\mcl^{(0)}_{\bm{\eta}})^2] - (e^{-\sigma^2})^{N^L_A}\E_{\bm{\Theta}}[(\mcl^{(0)}_{\bm{\eta}})]^2. 
\end{equation}
In particular, if there exist a $\bm{\lambda}^*$ such that $\eta_j = \lambda_j$ for all $\eta_j \ne 0$, we have : 
\begin{equation}
    \Var[\mcl_{\bm{\alpha}}] = \E[(\mcl_{\bm{\alpha}})^2] - \E[\mcl_{\bm{\alpha}}]^2 >  c_{ \bm{\lambda}^*}^2\Big(\frac{1 + e^{-2\sigma^2}}{2}\Big)^{N^L_A + w(P_{\bm{\eta}})} -  c_{ \bm{\lambda}^*}(e^{-\sigma^2})^{N^L_A + w(P_{\bm{\eta}})}, 
    \label{eq:var_alpha}
\end{equation}

otherwise, $\E_{\bm{\Theta}}[\mcl^{(L)}_{\bm{\eta}}] = 0$, and thus, 
\begin{equation}
    \Var[\mcl_{\bm{\alpha}}] = \E[(\mcl_{\bm{\alpha}})^2] > \sum_{\bm{\lambda}}c_{\bm{\lambda}}^2\Big(\frac{1 + e^{-2\sigma^2}}{2}\Big)^{N^L_A + \sum_{j\in \mathcal{J_{\bm{\eta}}}} \delta_{\eta_j \lambda_j}}\Big(\frac{1 - e^{-2\sigma^2}}{2}\Big)^{\sum_{j\in \mathcal{J_{\bm{\eta}}}} \bar{\delta}_{\eta_j \lambda_j}}.
    \label{eq:var_beta}
\end{equation}
On the other hand, if $\sigma \gg 0$, the identity initialization is not valid anymore, and a $\text{poly}(n)$ depth unstructured circuit is sufficiently expressive to induce exponential concentration of the loss function. 

\subsection{\label{adx:case_study}Case study: Absence of barren plateau in EfficientSU2 ansatz}
\begin{figure}[!h]
    \centering
    \subfloat[Pairwise entanglement]{\includegraphics[height = 0.15\textheight]{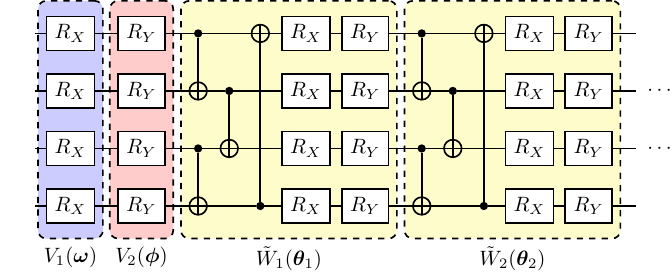}}
    \subfloat[Circular entanglement]{\includegraphics[height = 0.15\textheight]{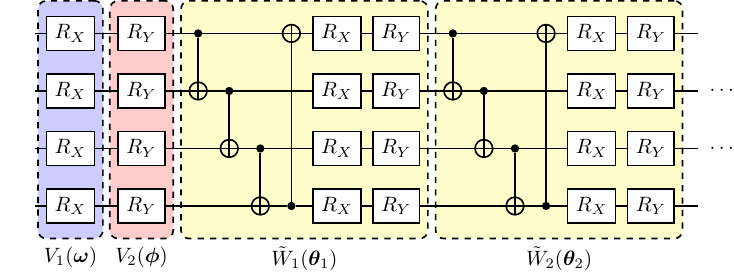}}
    \caption{\label{fig:EfficientSU2}\textbf{Circuit architecture of the EfficientSU2 ansatz~\cite{IBMansatz}.} The circuit is initialized with two orthogonal layers of single-qubit Pauli rotation gates and $L$ layers of entanglement layers with CNOT gates and single-qubit rotation gates. The entanglement gates can have either (a) a pairwise structure, or (b) a circular structure. }
    \end{figure}
In this section, we will analyze the decay of the gradient variance in the EfficientSU2 ansatz shown on \figref{fig:EfficientSU2}, which consists only of single-qubit Pauli rotations and CNOT gates. The single-qubit rotations can be chosen in different combinations, and the entanglement map can be either \textit{pairwise} or \textit{circular}. Although we only use the EfficientSU2 ansatz for the study, the result can also be generalized to other types of quantum circuits. Our results show that with small $\sigma$ the variance of the loss scales polynomially with the number of qubits. We note that while we rely on some approximations to obtain these \textit{approximate} lower bounds of the variance (i.e., the bounds are written in the leading order of $\sigma$), these bounds are rather tight, as supported by our numerics.

We will analyze how the loss function behaves depending on the type of observables and the entanglement map for both local and global observables.
Let us start with two different local observables $P_{\bm{\alpha}}= \sigma_{\alpha_0}\otimes \mathbb{I}^{\otimes n-1}$ and $P_{\bm{\beta}} = \sigma_{\beta_0}\otimes \mathbb{I}^{\otimes n-1}$ such that $\alpha_0 \ne \beta_0$. Furthermore, we assume that $\rho$ only consists of $P_{\bm{\lambda}}$ such that $\lambda_j \in \{0, \alpha_0\}$ for all $\bm{\lambda}$. This is the case if we have zero initial state $\rho = \ket{0}\bra{0}^{\otimes n} = (\mathbb{I} + Z)^{\otimes n}/2^n$ with $\sigma_{\alpha_0} = Z$ and $\sigma_{\beta_0} = X$.

 \tabref{tab:Loss_summary} summarizes $\mcl_{\bm{\alpha}}$ and $\mcl_{\bm{\beta}}$ for all possible combinations of $\mu_0$ and $\nu_0$ in $L = 0$ case. 
\begin{table}[!h]
    \centering
    \begin{tabular}{|c|c|c| c|c|c|c|c| }
    \hline
           &  $\mcl_{\alpha_0}$ & $\mcl_{\beta_0}$\  &  $\E[\mcl_{\alpha_0}]$ & $\E[\mcl_{\beta_0}]$\  &  $\E[\mcl_{\alpha_0}^2]$ & $\E[\mcl_{\beta_0}^2]$ &   $\E[\mcl_{\alpha_0}\mcl_{\beta_0}]$  \\
           \hline
         $\alpha_0 = \nu_0, \beta_0 =  \mu_0$ & $\cos(\omega_j)$ & $\sin(\omega_j)\sin(\phi_j)$  & $e^{-\frac{\sigma^2}{2}}$ & 0 & $\frac{1 + e^{-2\sigma^2}}{2}$ & $\left(\frac{1 - e^{-2\sigma^2}}{2}\right)^2$  & 0  \\ 
         $\alpha_0  = \nu_0, \beta_0 \ne \mu_0$ & $\cos(\omega_j)$ & $\sin(\omega_j) \cos(\phi_j)$ & $e^{-\frac{\sigma^2}{2}}$ & 0 & $\frac{1 + e^{-2\sigma^2}}{2}$  & $\frac{1 - e^{-4\sigma^2}}{4}$ & 0   \\  
         $\alpha_0 \ne \nu_0, \beta_0 =  \mu_0$ & $\cos(\omega_j)\cos(\phi_j)$ &   $\cos(\omega_j)\sin(\phi_j)$   & $e^{-\sigma^2}$ & 0 & $\left(\frac{1 + e^{-2\sigma^2}}{2}\right)^2$ &  $\frac{1 - e^{-4\sigma^2}}{4}$ & 0  \\  
         $\alpha_0 = \mu_0, \beta_0 =  \nu_0$ & $\cos(\phi_j)$ & $0$ & $e^{-\frac{\sigma^2}{2}}$ & 0 &$\frac{1 + e^{-2\sigma^2}}{2}$   & $\frac{1 - e^{-2\sigma^2}}{2}$ &  0 \\ 
         $\alpha_0 \ne \mu_0, \beta_0 =  \nu_0$ & $\cos(\phi_j)\cos(\omega_j) $ & $\sin(\omega_j) $  & $e^{-\sigma^2}$ &0  & $\left(\frac{1 + e^{-2\sigma^2}}{2}\right)^2$ & $\frac{1 - e^{-2\sigma^2}}{2}$  &  0 \\  
         $\alpha_0 = \mu_0, \beta_0 \ne  \nu_0$ & $\cos(\phi_j)$ & $\sin(\phi_j)$  & $e^{-\frac{\sigma^2}{2}}$ & 0& $\frac{1 + e^{-2\sigma^2}}{2}$  & $\frac{1 - e^{-4\sigma^2}}{4}$ & 0 \\  
    \hline
    \end{tabular}
    \caption{\label{tab:Loss_summary}\textbf{Summary of  the loss functions $\mcl_\alpha$ and $\mcl_{\beta}$ for different circuit architectures.} The covariance $\E_{\bm{\Theta}}\left[\mcl_{\alpha_0}\mcl_{\beta_0}\right]$ vanish for all cases as $\mcl_{\alpha_0}\mcl_{\beta_0}$ is always odd with respect to $\omega_j$ or $\phi_j$ over $\mathbb{R}$.  }
\end{table}
As the table shows, the covariance between $\mcl_{\bm{\alpha}}$ and $\mcl_{\bm{\beta}}$ vanishes, as assumed in \adxref{adx:variance_normal} and thus, \eqnref{eq:var1} holds. 

It is more straightforward to show the lower bound of $\Var_{\bm{\Theta}}[\mcl_{\bm{\alpha}}]$, as the $\E_{\bm{\Theta}}[\mcl^{(0)}_{\bm{\alpha}}]$ scales following the dominant term, $(1 + e^{-2\sigma^2})/{2}$. As EfficientSU2 ansatz only consists of CNOT gates among the Clifford gates, the Pauli strings $C^{\ell \dagger}_L P_{\bm{\alpha}} C^L_\ell$, in particular, $C^{1 \dagger}_L P_{\bm{\alpha}} C^L_1 = P_{\bm{\eta}}$\;, only consists of $\sigma_{\alpha_0}$. Furthermore, the ansatz only contains single-qubit Pauli rotation gates, hence, the number of anti-commuting Pauli operators, $n^\ell_A\left(C^{\ell + 1\dagger}_LP_{\bm{\alpha}} C^L_{\ell + 1} \right)$ will be proportional to the weight $w\left(C^{\ell+1\dagger}_{ L} P_{\bm{\alpha}} C^L_{\ell + 1}\right)$ of the Pauli string at each layer $\ell$ and, as a result, the total number of anti-commuting gates will be proportional to the sum of weights, $N^L_A \approx \sum_{\ell = 1}^L w\left(C^{\ell+1\dagger}_{ L} P_{\bm{\alpha}} C^L_{\ell + 1}\right)$. 

As $N_A^L$ depends on the position of the observable and the types of the entanglement map, it is complex to find a general formula. Therefore, we will assume that the non-trivial Pauli matrices can span over $n$-qubits for $C^{\ell\dagger}_{L}P_{\bm{\alpha}} C^L_\ell,  \ell = 1,...,n$\;, i.e. for all $d \in \{1, ..., n\}$, there exists an $\ell \in \{1,...,n\}$ such that $w(C^{\ell\dagger}_{L}P_{\bm{\alpha}} C^L_\ell) = d$. Under this assumption, we take the average over $w(C^{\ell\dagger}_{L}P_{\bm{\alpha}}C^L_\ell)$ to have $w(P_{\bm{\eta}}) \approx n/2$ and $N^L_A \approx (n^2 + n)/2$.  In addition, we also consider the extreme-case scenario for the polynomial depth circuit where the $N^L_A \approx Ln = n^2$\;, and $w(P_{\bm{\eta}}) = n$. For $\sigma \ll 1$\;, \eqnref{eq:var_alpha} and \tabref{tab:Loss_summary} lead to : 

\begin{align}
    \Var[\mcl_{\bm{\alpha}}] & \ge \min\left[c_{\bm{\lambda}^*}^2\left(\frac{1 + e^{-2\sigma^2}}{2}\right)^{\frac{n^2}{2}+ n} - c_{\bm{\lambda}^*}\left(e^{-\sigma^2}\right)^{\frac{n^2}{2} + n} , c_{\bm{\lambda}^*}^2\left(\frac{1 + e^{-2\sigma^2}}{2}\right)^{n^2 + n} - c_{\bm{\lambda}^*}\left(e^{-\sigma^2}\right)^{n^2 + n} \right]\;,
    \label{eq:var_localZ}
\end{align}
with $\bm{\lambda}^*$ such that $\bm{\lambda}^* = \bm{\eta}$. In particular, in the case of the zero state, the equation simplifies with $c_{\bm{\lambda}^*} = 1$. 
It is important to consider both cases, because the interplay between $\E_{\bm{\Theta}}\left[\mcl_{\bm{\alpha}}\right]^2$  and $\E_{\bm{\Theta}}\left[(\mcl_{\bm{\alpha}})^2\right]$ depends greatly on the value of $\sigma$ and $n$.
Taking into account only the extreme case will make the calculation deviate too much from the real lower bound.

\begin{figure}[!h]
    \centering
    \includegraphics[width = 0.8\textwidth]{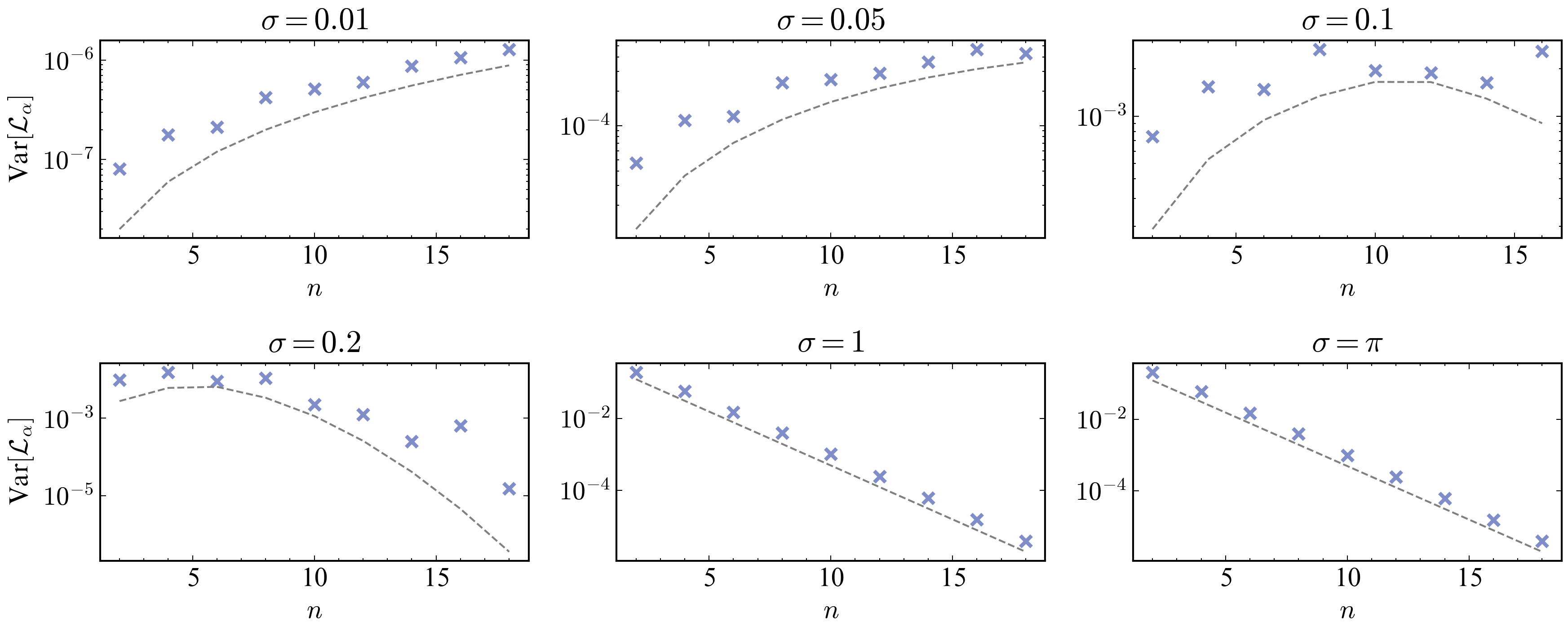}
    \caption{\textbf{Variance of $\mcl_{\bm{\alpha}}$ for the local Z observable and its lower bound in polynomial depth circuit with different initialization ranges.} The blue cross represents the simulated variance obtained in EfficientSU2 ansatz with $\text{poly}(n)$ depth and the gray dashed line its lower bound calculated with \eqnref{eq:var_localZ}.}
    \label{fig:local_Z_loss}
\end{figure}

\begin{figure}[!h]
    \centering
    \subfloat[Pairwise entanglement]{\includegraphics[width = 0.98\textwidth]{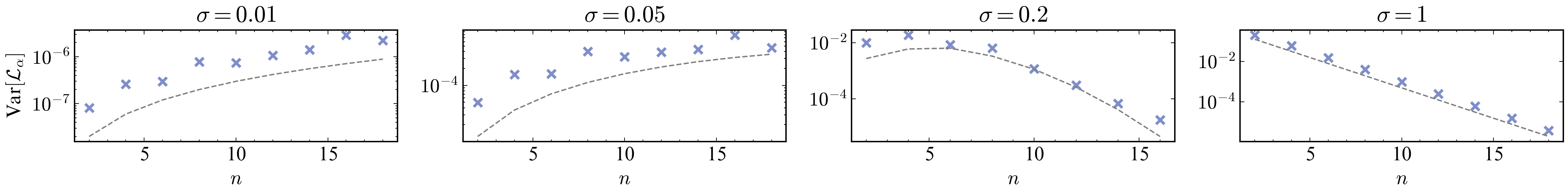}}
    \newline
    \subfloat[Circular entanglement]{\includegraphics[width = 0.98\textwidth]{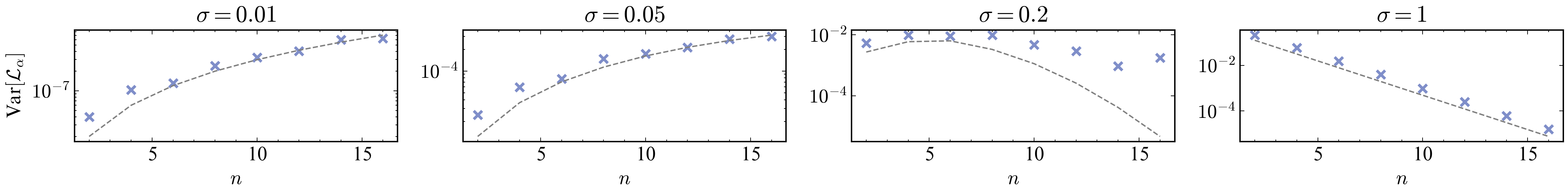}}
    \caption{\textbf{Variance of $\mcl_{\bm{\alpha}}$ for the global Z observable and its lower bound in the polynomial depth circuit with different initialization ranges.} The blue cross represents the simulated variance obtained in the EfficientSU2 ansatz with $\text{poly}(n)$ depth and the gray dashed line its lower bound calculated with \eqnref{eq:var_localZ}. We observe that $\Var_{\bm{\Theta}}[\mcl_{\bm{\alpha}}]$ for both (a) pairwise and (b) circular entanglement also follows the theoretical lower bound, and does not exhibit a barren plateau with an appropriate choice of initialization.}
    \label{fig:global_Z_loss}
\end{figure}

We also compute the scaling of the variance while assuming the initial zero state. Taking a Taylor expansion with respect to $\sigma$ around $0$ gives us: 
\begin{align}
    \Var[\mcl_{\bm{\alpha}}] 
    & > (1 -\sigma^2 + \sigma^4)^{n^2 + n}  - (1 - \sigma^2 + \frac{\sigma^4}{2})^{n^2 + n}\\ 
    & > \frac{1}{2}n(n+1) \sigma^4 - \frac{1}{2}(n^2(n+1))^2\sigma^6  \\ 
    & \approx \frac{1}{2}n^2\sigma^4(1 - n^2 \sigma^2) \\
    & > \frac{1}{n^b}\;. 
    \label{eq:var_localZ_taylor}
\end{align} 
with a $b > 1$ independent of $n$, which implies that $\Var_{\bm{\Theta}}[\mcl_{\bm{\alpha}}]$ decays polynomially. Rearranging \eqnref{eq:var_localZ_taylor}, we can conclude that $\Var_{\bm{\Theta}}\left[\mcl_{\bm{\alpha}}\right]$ will scale as $\mathcal{O}(1/n^b)$ with $b > 2$ if : 
\begin{equation}
    \sigma \in \Theta\left(\frac{1}{n}\right)\;.
\end{equation} 
Note that the $\max\left(w\left(C^{\ell\dagger}_{L}P_{\bm{\alpha}} C^L_\ell\right)\right) = n$ and $\max\left(N^L_A\right) = n^2$, and thus, the lower bound given by \eqnref{eq:var_localZ} also applies to the global observable case with $P_{\bm{\alpha}} = \sigma_{\alpha_0}^{\otimes n}$ as shown on \figref{fig:global_Z_loss}. On \figref{fig:local_Z_loss} and \ref{fig:global_Z_loss}, we display $\Var[\mcl^{(L)}_{\bm{\alpha}}]$ for $L = n$ and its lower bound calculated with \eqnref{eq:var_localZ} with respect to the number of qubits. Furthermore, for $\sigma \ge 1$, we take the lower bound as $\Var_{\bm{\Theta}}[\mcl_{\bm{\alpha}}] = 2^{-n}$ as justified in \adxref{adx:variance_normal}. The figures clearly show that the variance follows the computed lower bound for both local and global observables in the case of $\sigma \ll 1$ while it decays exponentially for large $\sigma$. 


On the other hand, understanding the variance of $\E_{\bm{\Theta}}[\mcl_{\bm{\beta}}]$ requires additional insight due to the presence of sine terms. As shown on \tabref{tab:Loss_summary}, $\E_{\bm{\Theta}}[\mcl_{\bm{\beta}}]$ vanishes regardless of the ansatz, thus it suffices to find the scaling of $\E_{\bm{\Theta}}[\mcl_{\bm{\beta}}^2]$. 
We start by rewriting \eqnref{eq:mcl2_5} for the EfficientSU2 ansatz with $K_\ell = 2$ (c.f. \figref{fig:EfficientSU2}) as follows : 
\begin{align}
    \E_{\bm{\Theta}}\left[\left(\mcl^{(L)}_{\bm{\beta}}\right)^2\right] =&   \left(\frac{1 + e^{-2\sigma^2}}{2}\right)^2 \E_{\bm{\Theta}}\left[\left(\mcl^{(L - 1)}_{\bm{\gamma}'}\right)^2\right] + \left(\frac{1 + e^{-2\sigma^2}}{2}\right)\left(\frac{1 - e^{-2\sigma^2}}{2}\right) \E_{\bm{\Theta}}\left[\left(\mcl^{(L - 1)}_{\bm{\delta}_1'}\right)^2\right] \nonumber \\  
    & + \left(\frac{1 + e^{-2\sigma^2}}{2}\right)\left(\frac{1 - e^{-2\sigma^2}}{2}\right) \E_{\bm{\Theta}}\left[\left(\mcl^{(L - 1)}_{\bm{\delta}_2'}\right)^2\right]  
    + \mathcal{O}\left(\frac{1 - e^{-2\sigma^2}}{2}\right)^2
    \label{eq:E_beta_2}
\end{align}
where we denote $P_{\bm{\gamma}'} = C^\dagger P_{\bm{\beta}} C$,  $P_{\bm{\delta}_1'} = -iC^\dagger \frac{1}{2}[P_{L, 1}, P_{\bm{\beta}}]C$ and $P_{\bm{\delta}_2'} = -iC^\dagger \frac{1}{2}[P_{L, 2}, P_{\bm{\beta}}]C$. 

\begin{figure}[!h]
    \centering
    \includegraphics[width = 0.8\textwidth]{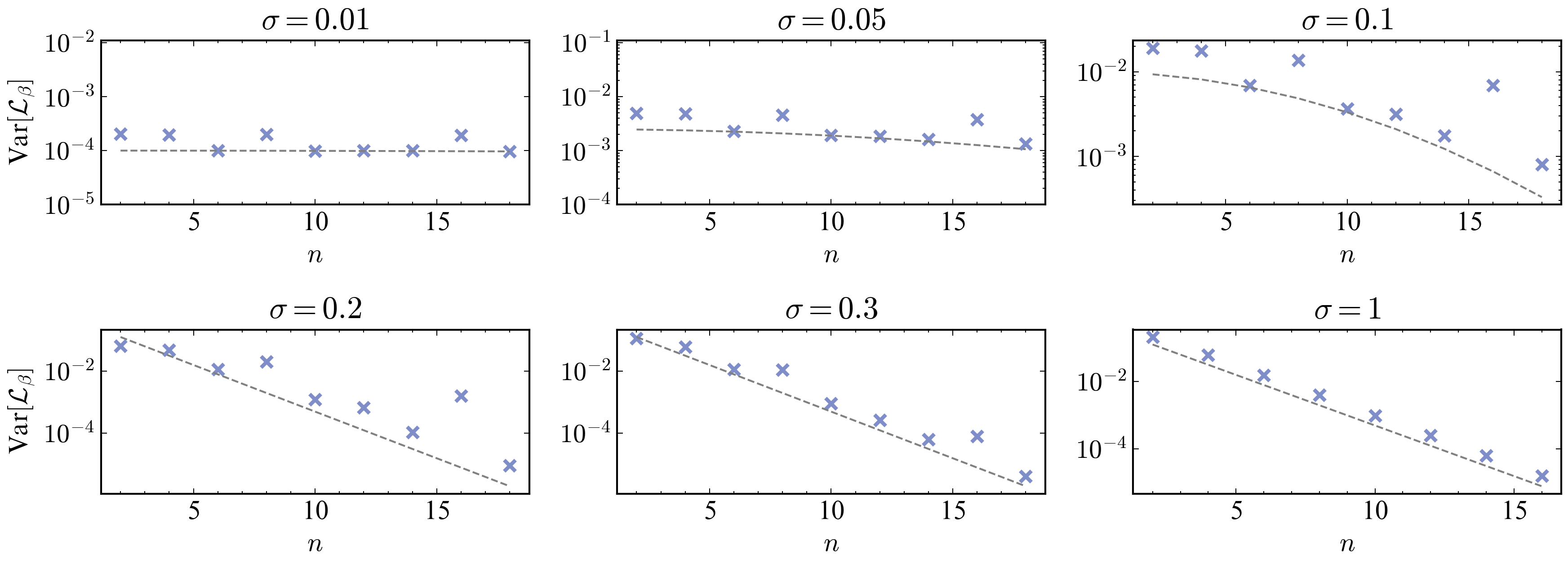}
    \caption{\textbf{Variance of $\mcl_{\bm{\beta}}$ for the local X observable and its lower bound in polynomial depth circuit with different initialization ranges.} The blue cross represents the simulated variance obtained in EfficientSU2 ansatz with $\text{poly}(n)$ depth and the pairwise entanglement map. The gray dashed line corresponds to the lower bound calculated with \eqnref{eq:var_localX}.}
    \label{fig:local_X_loss}
\end{figure}

Without losing generality, we assume that $[P_{L, i_j}, P_{\bm{\beta}}] = 2i\sigma_{\alpha_0}\otimes\mathbb{I}^{n-1}$ for $i_j \in \{1, 2\}$. This corresponds to the case of an $X$ observable with RZ or RY rotations in $\tilde{W}_{\ell}$. This leads $\E_{\bm{\Theta}}\left[\left(\mcl^{(L - 1)}_{\bm{\delta}_{i_j}'}\right)^2\right]$ back to the previous case for $\mcl_{\bm{\alpha}}$, scaling as $((1 + e^{-2\sigma^2})/2)^{n^2 + n}$. 
On the other hand, since $P_{{\bm{\gamma}}'}$ only contains $\sigma_{\beta_0}$, in the worst case scenario, we have $w(C^{\ell \dagger}_L P_{\bm{\beta}} C^L_\ell) = n$, leading to the the first term scaling as $((1 - e^{-2\sigma^2})/2)^{n}$. Therefore, the second term will be the dominant term in case of the small angle initialization and we can conclude that : 
\begin{align}
    \Var_{\bm{\Theta}}[\mcl_{\bm{\beta}}^2] & > c_{\bm{\lambda}^{**}}\left(\frac{1 + e^{-2\sigma^2}}{2}\right)^{n^2 + n} \left(\frac{1 - e^{-2\sigma^2}}{2}\right),
    \label{eq:var_localX}
\end{align}
with $c_{\bm{\lambda}^{**}}$ such that $P_{\bm{\lambda}^{**}} =  C^{1\dagger}_{L}[P_{L, i_j}, P_{\bm{\beta}}] C^L_1$.
The equation is confirmed with \figref{fig:local_X_loss} showing $\Var[\mcl^{(L)}_{\bm{\beta}}]$ for $L = n$ and its lower bound calculated with \eqnref{eq:var_localX}.

For the zero initial state, by taking the Taylor expansion with respect to $\sigma$ around $0$, \eqnref{eq:var_localX} can be approximated as : 
\begin{equation}
    \Var_{\bm{\Theta}}[\mcl_{\bm{\beta}}^2] > (1 - \sigma^2)^{n^2 + n}\sigma^2 
  > \sigma^2 - (n^2 + n)\sigma^4 > \sigma^2 - 2n^2\sigma^4 > \frac{1}{n^b}, 
\end{equation}
resulting in the same conclusion as before, that $\Var_{\bm{\Theta}}[\mcl_{\bm{\beta}}^2]$ decays as $\mathcal{O}(1/n^2)$ if $\sigma \in \Theta(1/n)$. 

\begin{figure}[!h]
    \centering
    \subfloat[\label{fig:global_X_loss_pairwise}Pairwise entanglement]{\includegraphics[width = 0.98\textwidth]{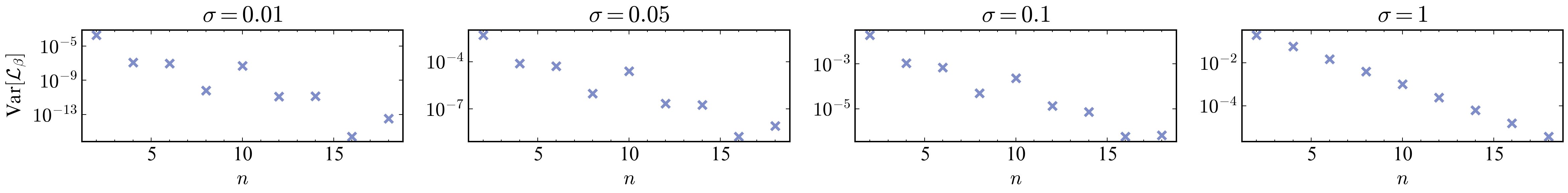}}
    \newline
    \subfloat[\label{fig:global_X_loss_circular}Circular entanglement]{\includegraphics[width = 0.98\textwidth]{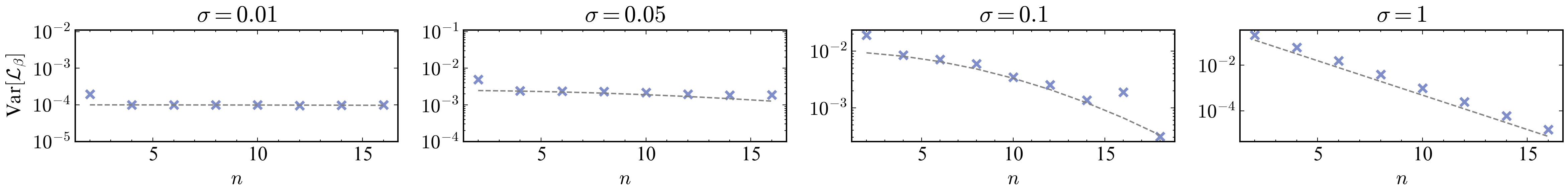}}
    \caption{\textbf{Variance of $\mcl_{\bm{\beta}}$ for the global X observable and its lower bound in the polynomial depth circuit with different initialization ranges.} The blue cross represents the simulated variance obtained in the EfficientSU2 ansatz with $\text{poly}(n)$ depth and the gray dashed line its lower bound calculated with \eqnref{eq:var_localX}. We observe that $\Var_{\bm{\Theta}}[\mcl_{\bm{\beta}}]$ decays exponentially for the (a) pairwise entanglement independent of $\sigma$, while it follows the theoretical lower bound given for (b) circular entanglement. This discrepancy comes from the fact that $C^{\ell \dagger}_L P_{\bm{\beta}} C^\ell_L$ behaves differently depending on the entanglement map.}
    \label{fig:global_X_loss}
\end{figure}

Unlike the Z observable, which exhibits similar behavior in both the global and local observables, the behavior of the global X observable varies depending on the type of entanglement. In \figref{fig:global_X_loss}, we observe that the variance decays exponentially regardless of $\sigma$ with the pairwise entanglement map. However, with the circular entanglement map, it follows the lower bound given by \eqnref{eq:var_localX}, which was computed for the local observable.

Although it is complicated to justify the result mathematically, we can explain this difference by analyzing $C^{\ell\dagger}_L P_{\bm{\beta}} C^L_\ell$. For the circular entanglement map, it is straightforward to see that $P_{\bm{\gamma}'} = C^\dagger X^{\otimes n} C = \mathbb{I}X\mathbb{I}^{\otimes(n -2)}$. This brings us back to the case of the local X observable with $\mathcal{L}^{(L-1)}_{\bm{\gamma}'}$ in \eqnref{eq:E_beta_2}, resulting in 
the same lower bound. 
Conversely, for the pairwise entanglement map, we have $P_{\bm{\gamma}'} = C^\dagger X^{\otimes n} C = (\mathbb{I}X)^{\otimes n/2}$, introducing an additional contribution of $((1-e^{-2\sigma^2})/2)^{n}$ in \eqnref{eq:var_localX}. Indeed, we can easily find that $w(C^{\ell\dagger}_L P_{\bm{\beta}} C^L_\ell) = n$ or $n/2$ for all $\ell = 1,...,L$. Therefore, this higher weight of the X observable leads to  the exponential decay of the variance.

\begin{figure}[!h]
    \centering 
    \subfloat[Local Z]{\includegraphics[width = 0.33\textwidth]{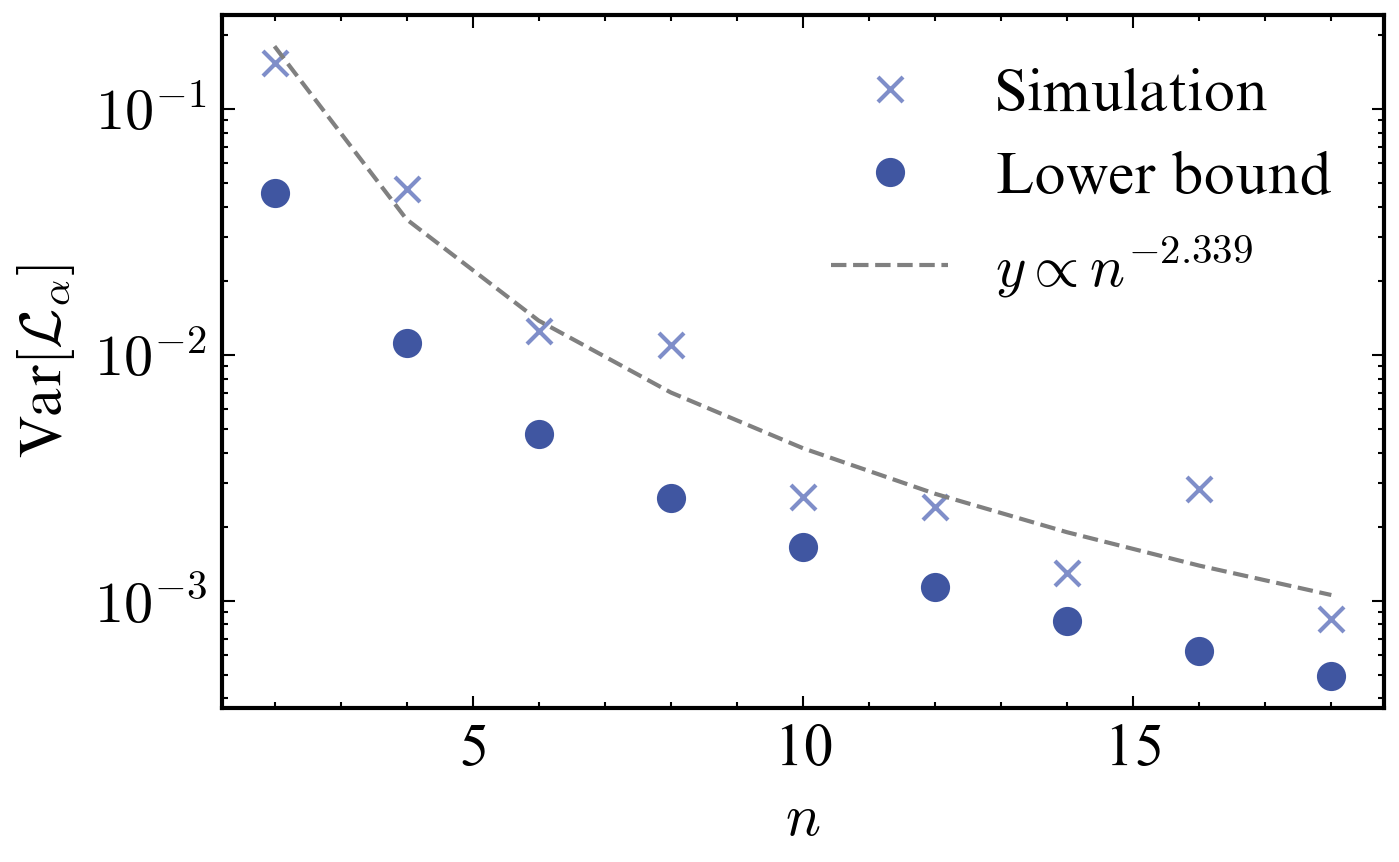}}
    \subfloat[Global Z]{\includegraphics[width = 0.33\textwidth]{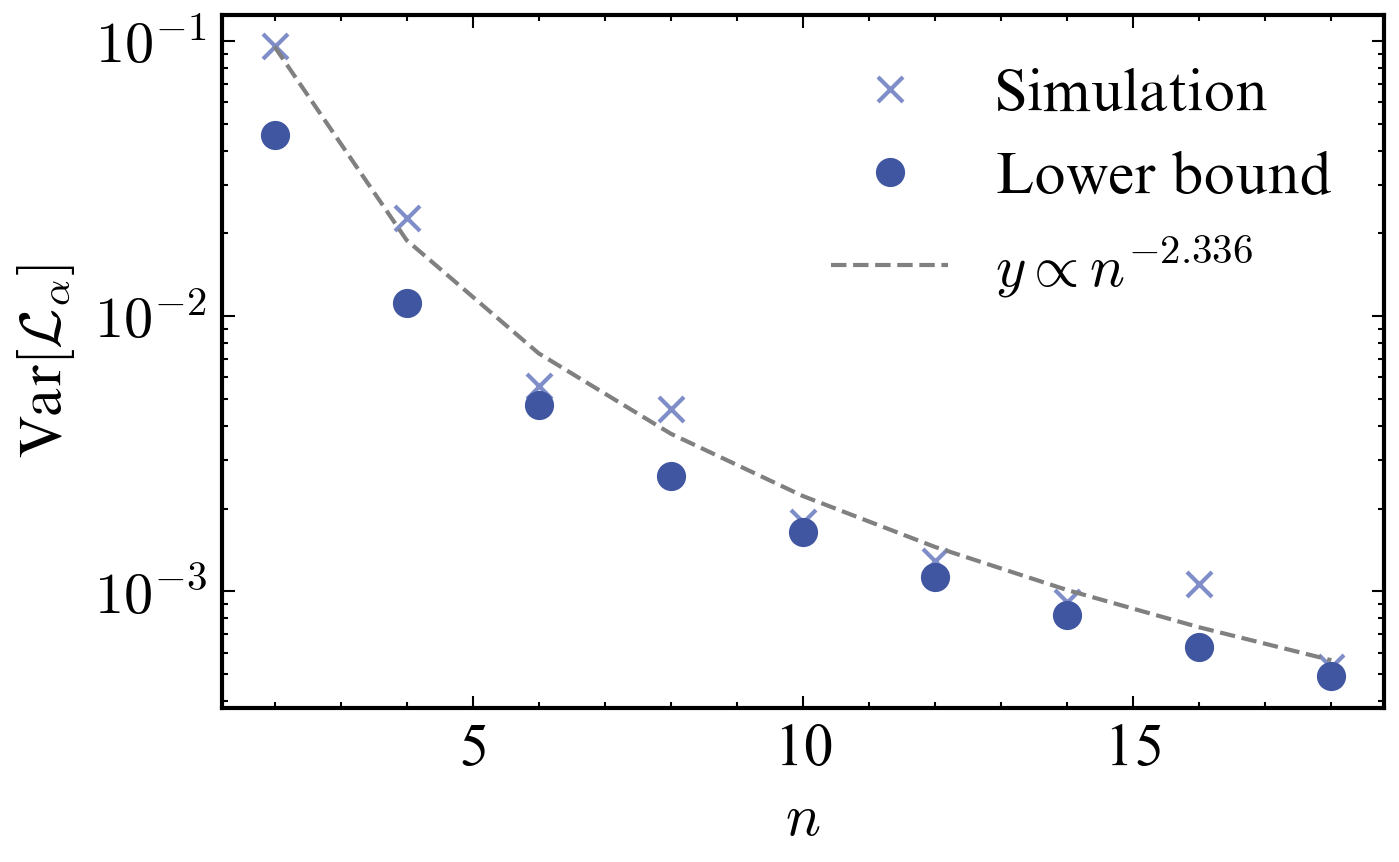}}
    \subfloat[Local X]{\includegraphics[width = 0.33\textwidth]{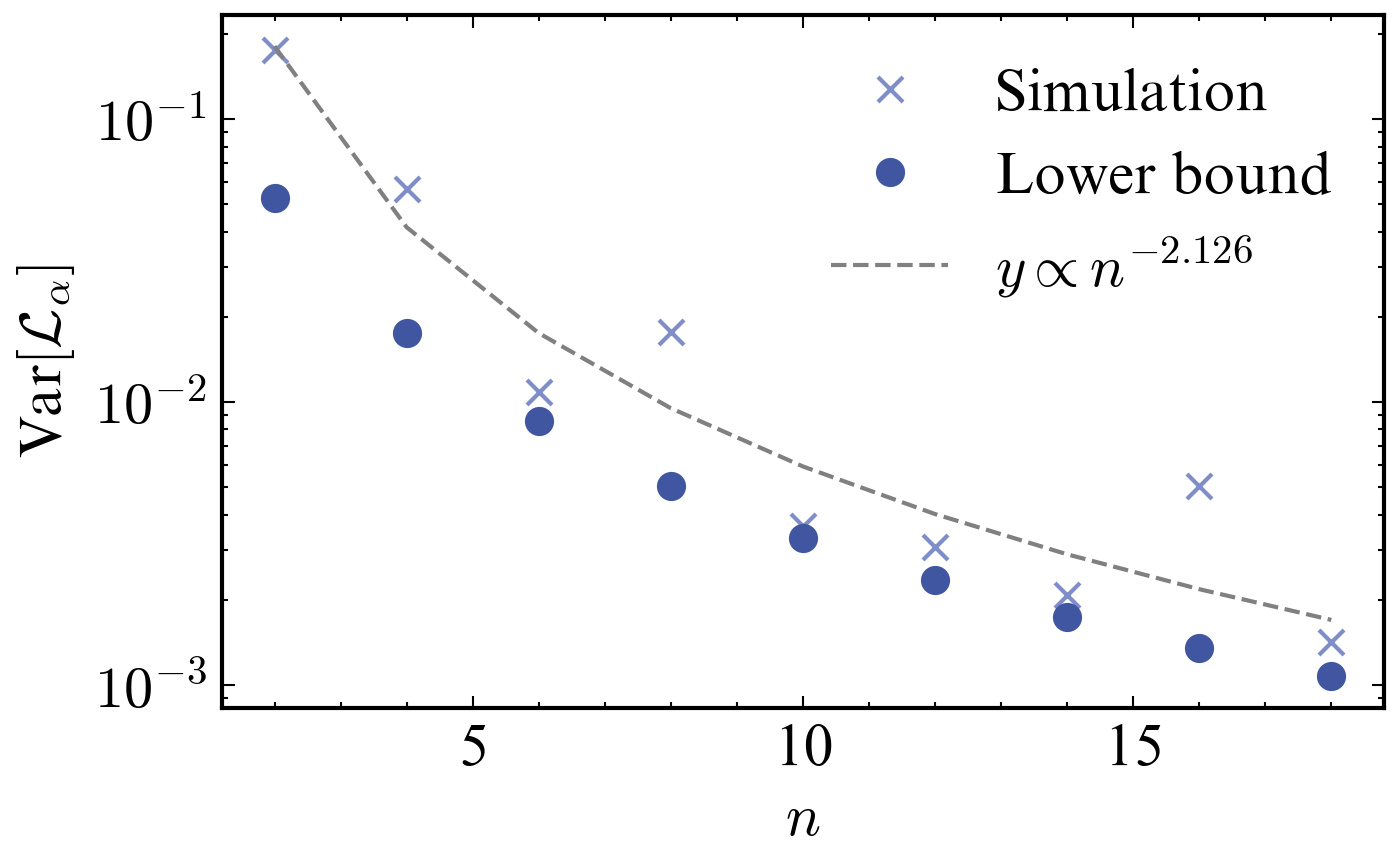}}
    \caption{\textbf{Variance of the loss function with respect to $n$ using $\sigma = 1/n$.} The blue cross corresponds to the simulated result, the dashed line its polynomial fit, and the green points the theoretical bounds computed with (a), (b) \eqnref{eq:var_localZ} and (c) \eqnref{eq:var_localX}. As predicted, the variance decays as $\mathcal{O}(1/n^2)$ with $\sigma$ scaling as $1/n$. }
    \label{fig:varying_delta}
\end{figure}

\figref{fig:varying_delta} shows the variance of the loss function $\Var_{\bm{\Theta}}[\mcl_{\bm{\alpha}}]$ of EfficientSU2 ansatz with $\text{poly}(n)$ depth versus the number of qubits $n$ using the initialization range varying as $\sigma = 1/n$ for Z and X observables. As expected, the loss function scales as $\mathcal{O}(1/n^b)$ with $b > 2$, clearly proving the mitigation of the barren plateau in the polynomial depth circuit.

\section{Study of BP with Other Circuits}
\begin{figure}[!h]
\centering
\subfloat[EfficientSU2]{\includegraphics[width = 0.33\textwidth]{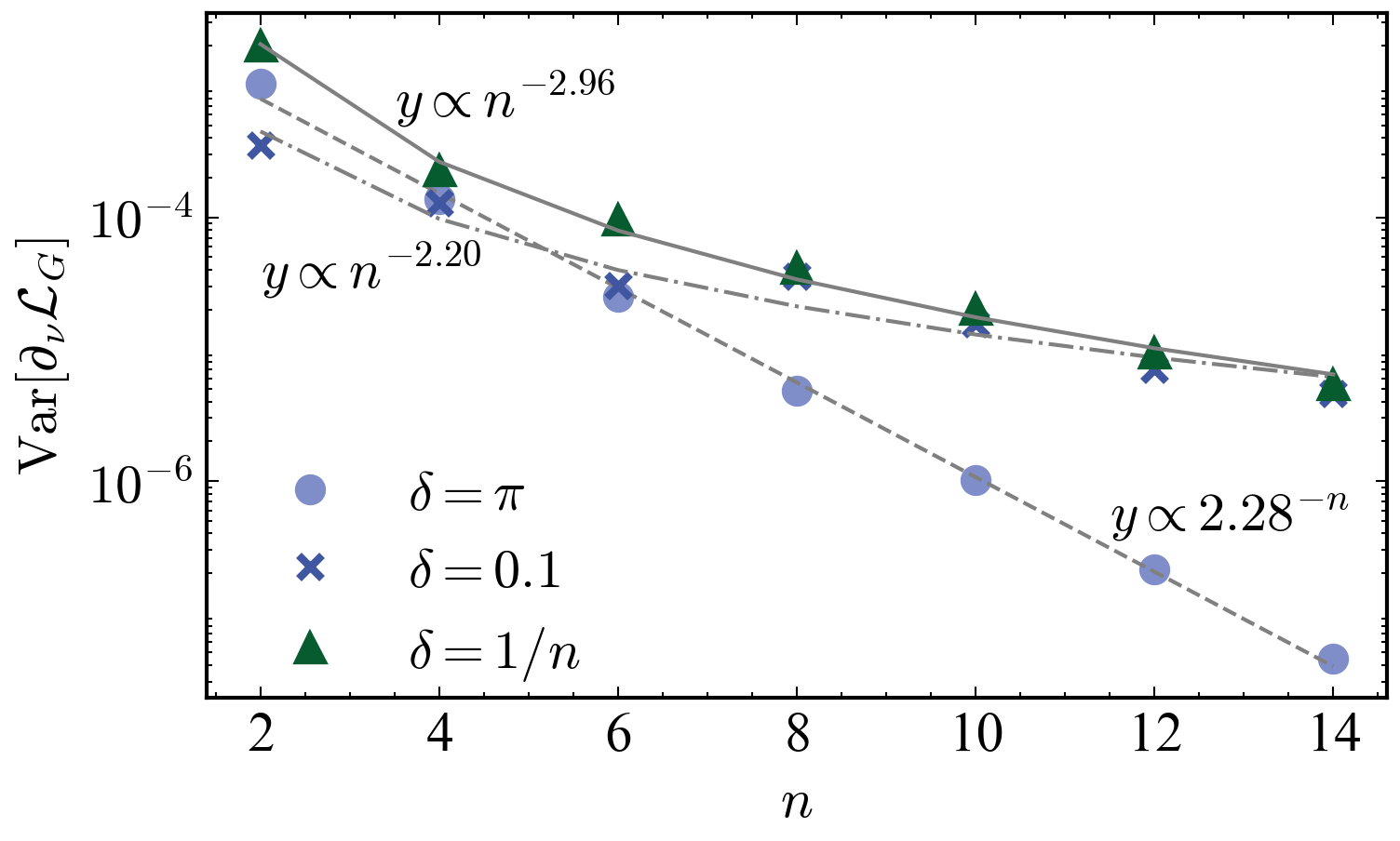}}
\subfloat[Circuit3]{\includegraphics[width = 0.33\textwidth]{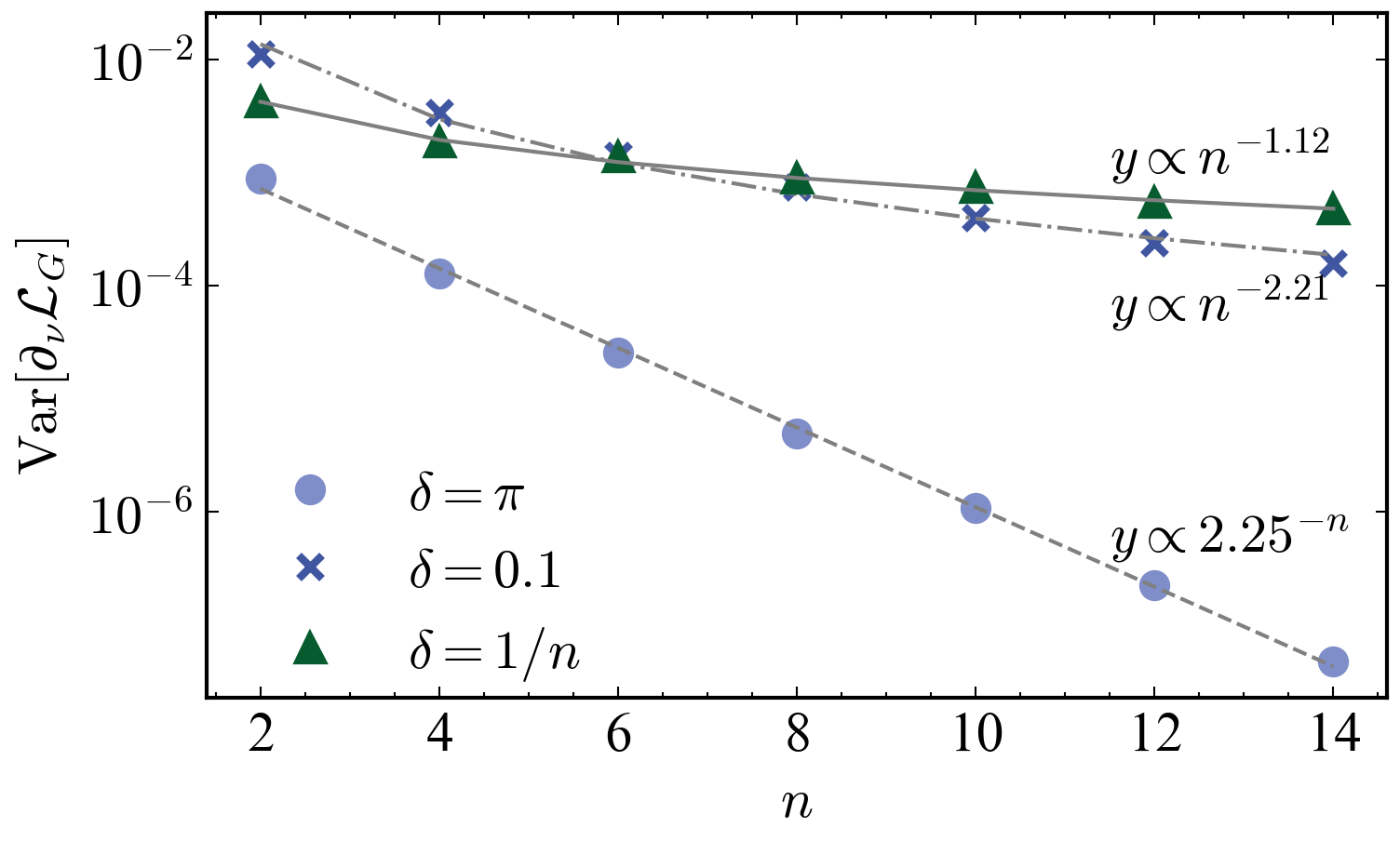}}
    \caption{\textbf{Variance of the partial derivative of $\mcl_{G}$ versus the number of qubits $n$ using polynomial depth quantum generator with different circuit architecture in \model.} The variance is computed with $\Dz = n$ for different initialization ranges, $\delta$, and averaged over the parameters of the first layer. The quantum generator consists of different circuits presented in \figref{fig:circuits} with polynomial depth, $d = \lfloor \log(n) \rfloor$.}
    \label{fig:variance_other}
\end{figure}

Previously, in \secref{sec:bp}, we only displayed the results for the variance of the gradients calculated with Circuit1 (see \figref{fig:circuits}) as a quantum generator in \model. In this section, we compute the variance for other types of circuits to confirm that this absence of barren plateau is not only limited to a specific type of circuit. As shown on \figref{fig:variance_other}, we observe that $\Var_{\bm{\Theta}, \bm{\phi}}[\partial_\nu \mcl_G ]$  decays polynomially with a small angle initialization for other circuits as well. Notably, the varying \(\sigma = 1/n\) found in \adxref{adx:case_study} ensures a polynomial decay, with the slope decreasing as the system size increases.

\begin{thebibliography}{112}%
\makeatletter
\providecommand \@ifxundefined [1]{%
 \@ifx{#1\undefined}
}%
\providecommand \@ifnum [1]{%
 \ifnum #1\expandafter \@firstoftwo
 \else \expandafter \@secondoftwo
 \fi
}%
\providecommand \@ifx [1]{%
 \ifx #1\expandafter \@firstoftwo
 \else \expandafter \@secondoftwo
 \fi
}%
\providecommand \natexlab [1]{#1}%
\providecommand \enquote  [1]{``#1''}%
\providecommand \bibnamefont  [1]{#1}%
\providecommand \bibfnamefont [1]{#1}%
\providecommand \citenamefont [1]{#1}%
\providecommand \href@noop [0]{\@secondoftwo}%
\providecommand \href [0]{\begingroup \@sanitize@url \@href}%
\providecommand \@href[1]{\@@startlink{#1}\@@href}%
\providecommand \@@href[1]{\endgroup#1\@@endlink}%
\providecommand \@sanitize@url [0]{\catcode `\\12\catcode `\$12\catcode `\&12\catcode `\#12\catcode `\^12\catcode `\_12\catcode `\%12\relax}%
\providecommand \@@startlink[1]{}%
\providecommand \@@endlink[0]{}%
\providecommand \url  [0]{\begingroup\@sanitize@url \@url }%
\providecommand \@url [1]{\endgroup\@href {#1}{\urlprefix }}%
\providecommand \urlprefix  [0]{URL }%
\providecommand \Eprint [0]{\href }%
\providecommand \doibase [0]{https://doi.org/}%
\providecommand \selectlanguage [0]{\@gobble}%
\providecommand \bibinfo  [0]{\@secondoftwo}%
\providecommand \bibfield  [0]{\@secondoftwo}%
\providecommand \translation [1]{[#1]}%
\providecommand \BibitemOpen [0]{}%
\providecommand \bibitemStop [0]{}%
\providecommand \bibitemNoStop [0]{.\EOS\space}%
\providecommand \EOS [0]{\spacefactor3000\relax}%
\providecommand \BibitemShut  [1]{\csname bibitem#1\endcsname}%
\let\auto@bib@innerbib\@empty
\bibitem [{\citenamefont {Ray}(2023)}]{RAY2023ChatGPT}%
  \BibitemOpen
  \bibfield  {author} {\bibinfo {author} {\bibfnamefont {P.~P.}\ \bibnamefont {Ray}},\ }\bibfield  {title} {\bibinfo {title} {Chatgpt: A comprehensive review on background, applications, key challenges, bias, ethics, limitations and future scope},\ }\href {https://doi.org/https://doi.org/10.1016/j.iotcps.2023.04.003} {\bibfield  {journal} {\bibinfo  {journal} {Internet of Things and Cyber-Physical Systems}\ }\textbf {\bibinfo {volume} {3}},\ \bibinfo {pages} {121} (\bibinfo {year} {2023})}\BibitemShut {NoStop}%
\bibitem [{\citenamefont {Daras}\ and\ \citenamefont {Dimakis}(2022)}]{Daras2022DALLE}%
  \BibitemOpen
  \bibfield  {author} {\bibinfo {author} {\bibfnamefont {G.}~\bibnamefont {Daras}}\ and\ \bibinfo {author} {\bibfnamefont {A.~G.}\ \bibnamefont {Dimakis}},\ }\href@noop {} {\bibinfo {title} {Discovering the hidden vocabulary of dalle-2}} (\bibinfo {year} {2022}),\ \Eprint {https://arxiv.org/abs/2206.00169} {arXiv:2206.00169 [cs.LG]} \BibitemShut {NoStop}%
\bibitem [{\citenamefont {Isola}\ \emph {et~al.}(2017)\citenamefont {Isola}, \citenamefont {Zhu}, \citenamefont {Zhou},\ and\ \citenamefont {Efros}}]{Isola2017pix}%
  \BibitemOpen
  \bibfield  {author} {\bibinfo {author} {\bibfnamefont {P.}~\bibnamefont {Isola}}, \bibinfo {author} {\bibfnamefont {J.}~\bibnamefont {Zhu}}, \bibinfo {author} {\bibfnamefont {T.}~\bibnamefont {Zhou}},\ and\ \bibinfo {author} {\bibfnamefont {A.~A.}\ \bibnamefont {Efros}},\ }\bibfield  {title} {\bibinfo {title} {Image-to-image translation with conditional adversarial networks},\ }in\ \href {https://doi.org/10.1109/CVPR.2017.632} {\emph {\bibinfo {booktitle} {2017 IEEE Conference on Computer Vision and Pattern Recognition (CVPR)}}}\ (\bibinfo  {publisher} {IEEE Computer Society},\ \bibinfo {address} {Los Alamitos, CA, USA},\ \bibinfo {year} {2017})\ pp.\ \bibinfo {pages} {5967--5976}\BibitemShut {NoStop}%
\bibitem [{\citenamefont {Paganini}\ \emph {et~al.}(2018)\citenamefont {Paganini}, \citenamefont {de~Oliveira},\ and\ \citenamefont {Nachman}}]{Paganini2018CaloGAN}%
  \BibitemOpen
  \bibfield  {author} {\bibinfo {author} {\bibfnamefont {M.}~\bibnamefont {Paganini}}, \bibinfo {author} {\bibfnamefont {L.}~\bibnamefont {de~Oliveira}},\ and\ \bibinfo {author} {\bibfnamefont {B.}~\bibnamefont {Nachman}},\ }\bibfield  {title} {\bibinfo {title} {Calogan: Simulating 3d high energy particle showers in multilayer electromagnetic calorimeters with generative adversarial networks},\ }\href {https://doi.org/10.1103/PhysRevD.97.014021} {\bibfield  {journal} {\bibinfo  {journal} {Phys. Rev. D}\ }\textbf {\bibinfo {volume} {97}},\ \bibinfo {pages} {014021} (\bibinfo {year} {2018})}\BibitemShut {NoStop}%
\bibitem [{\citenamefont {Kingma}\ and\ \citenamefont {Welling}(2022)}]{Kingma2022AE}%
  \BibitemOpen
  \bibfield  {author} {\bibinfo {author} {\bibfnamefont {D.~P.}\ \bibnamefont {Kingma}}\ and\ \bibinfo {author} {\bibfnamefont {M.}~\bibnamefont {Welling}},\ }\href@noop {} {\bibinfo {title} {Auto-encoding variational bayes}} (\bibinfo {year} {2022}),\ \Eprint {https://arxiv.org/abs/1312.6114} {arXiv:1312.6114 [stat.ML]} \BibitemShut {NoStop}%
\bibitem [{\citenamefont {Goodfellow}\ \emph {et~al.}(2014)\citenamefont {Goodfellow}, \citenamefont {Pouget-Abadie}, \citenamefont {Mirza}, \citenamefont {Xu}, \citenamefont {Warde-Farley}, \citenamefont {Ozair}, \citenamefont {Courville},\ and\ \citenamefont {Bengio}}]{Goodfellow2014GAN}%
  \BibitemOpen
  \bibfield  {author} {\bibinfo {author} {\bibfnamefont {I.~J.}\ \bibnamefont {Goodfellow}}, \bibinfo {author} {\bibfnamefont {J.}~\bibnamefont {Pouget-Abadie}}, \bibinfo {author} {\bibfnamefont {M.}~\bibnamefont {Mirza}}, \bibinfo {author} {\bibfnamefont {B.}~\bibnamefont {Xu}}, \bibinfo {author} {\bibfnamefont {D.}~\bibnamefont {Warde-Farley}}, \bibinfo {author} {\bibfnamefont {S.}~\bibnamefont {Ozair}}, \bibinfo {author} {\bibfnamefont {A.}~\bibnamefont {Courville}},\ and\ \bibinfo {author} {\bibfnamefont {Y.}~\bibnamefont {Bengio}},\ }\href@noop {} {\bibinfo {title} {Generative adversarial networks}} (\bibinfo {year} {2014}),\ \Eprint {https://arxiv.org/abs/1406.2661} {arXiv:1406.2661 [stat.ML]} \BibitemShut {NoStop}%
\bibitem [{\citenamefont {Sohl-Dickstein}\ \emph {et~al.}(2015)\citenamefont {Sohl-Dickstein}, \citenamefont {Weiss}, \citenamefont {Maheswaranathan},\ and\ \citenamefont {Ganguli}}]{Dickstein2015DM}%
  \BibitemOpen
  \bibfield  {author} {\bibinfo {author} {\bibfnamefont {J.}~\bibnamefont {Sohl-Dickstein}}, \bibinfo {author} {\bibfnamefont {E.}~\bibnamefont {Weiss}}, \bibinfo {author} {\bibfnamefont {N.}~\bibnamefont {Maheswaranathan}},\ and\ \bibinfo {author} {\bibfnamefont {S.}~\bibnamefont {Ganguli}},\ }\bibfield  {title} {\bibinfo {title} {Deep unsupervised learning using nonequilibrium thermodynamics},\ }in\ \href {https://proceedings.mlr.press/v37/sohl-dickstein15.html} {\emph {\bibinfo {booktitle} {Proceedings of the 32nd International Conference on Machine Learning}}},\ \bibinfo {series} {Proceedings of Machine Learning Research}, Vol.~\bibinfo {volume} {37}\ (\bibinfo  {publisher} {PMLR},\ \bibinfo {year} {2015})\ pp.\ \bibinfo {pages} {2256--2265}\BibitemShut {NoStop}%
\bibitem [{\citenamefont {Du}\ and\ \citenamefont {Mordatch}(2019)}]{Du2019EM}%
  \BibitemOpen
  \bibfield  {author} {\bibinfo {author} {\bibfnamefont {Y.}~\bibnamefont {Du}}\ and\ \bibinfo {author} {\bibfnamefont {I.}~\bibnamefont {Mordatch}},\ }\bibfield  {title} {\bibinfo {title} {Implicit generation and modeling with energy based models},\ }in\ \href {https://proceedings.neurips.cc/paper_files/paper/2019/file/378a063b8fdb1db941e34f4bde584c7d-Paper.pdf} {\emph {\bibinfo {booktitle} {Advances in Neural Information Processing Systems}}},\ Vol.~\bibinfo {volume} {32},\ \bibinfo {editor} {edited by\ \bibinfo {editor} {\bibfnamefont {H.}~\bibnamefont {Wallach}}, \bibinfo {editor} {\bibfnamefont {H.}~\bibnamefont {Larochelle}}, \bibinfo {editor} {\bibfnamefont {A.}~\bibnamefont {Beygelzimer}}, \bibinfo {editor} {\bibfnamefont {F.}~\bibnamefont {d\textquotesingle Alch\'{e}-Buc}}, \bibinfo {editor} {\bibfnamefont {E.}~\bibnamefont {Fox}},\ and\ \bibinfo {editor} {\bibfnamefont {R.}~\bibnamefont {Garnett}}}\ (\bibinfo  {publisher} {Curran Associates, Inc.},\ \bibinfo {year} {2019})\BibitemShut {NoStop}%
\bibitem [{\citenamefont {Papamakarios}\ \emph {et~al.}(2021)\citenamefont {Papamakarios}, \citenamefont {Nalisnick}, \citenamefont {Rezende}, \citenamefont {Mohamed},\ and\ \citenamefont {Lakshminarayanan}}]{Papamakarios2021Normalizing}%
  \BibitemOpen
  \bibfield  {author} {\bibinfo {author} {\bibfnamefont {G.}~\bibnamefont {Papamakarios}}, \bibinfo {author} {\bibfnamefont {E.}~\bibnamefont {Nalisnick}}, \bibinfo {author} {\bibfnamefont {D.~J.}\ \bibnamefont {Rezende}}, \bibinfo {author} {\bibfnamefont {S.}~\bibnamefont {Mohamed}},\ and\ \bibinfo {author} {\bibfnamefont {B.}~\bibnamefont {Lakshminarayanan}},\ }\bibfield  {title} {\bibinfo {title} {Normalizing flows for probabilistic modeling and inference},\ }\href@noop {} {\bibfield  {journal} {\bibinfo  {journal} {J. Mach. Learn. Res.}\ }\textbf {\bibinfo {volume} {22}} (\bibinfo {year} {2021})}\BibitemShut {NoStop}%
\bibitem [{\citenamefont {Gulrajani}\ \emph {et~al.}(2017)\citenamefont {Gulrajani}, \citenamefont {Ahmed}, \citenamefont {Arjovsky}, \citenamefont {Dumoulin},\ and\ \citenamefont {Courville}}]{Gulrajani2017wasserstein}%
  \BibitemOpen
  \bibfield  {author} {\bibinfo {author} {\bibfnamefont {I.}~\bibnamefont {Gulrajani}}, \bibinfo {author} {\bibfnamefont {F.}~\bibnamefont {Ahmed}}, \bibinfo {author} {\bibfnamefont {M.}~\bibnamefont {Arjovsky}}, \bibinfo {author} {\bibfnamefont {V.}~\bibnamefont {Dumoulin}},\ and\ \bibinfo {author} {\bibfnamefont {A.~C.}\ \bibnamefont {Courville}},\ }\bibfield  {title} {\bibinfo {title} {Improved training of wasserstein gans},\ }in\ \href {https://proceedings.neurips.cc/paper_files/paper/2017/file/892c3b1c6dccd52936e27cbd0ff683d6-Paper.pdf} {\emph {\bibinfo {booktitle} {Advances in Neural Information Processing Systems}}},\ Vol.~\bibinfo {volume} {30}\ (\bibinfo  {publisher} {Curran Associates, Inc.},\ \bibinfo {year} {2017})\ p.\ \bibinfo {pages} {5769–5779}\BibitemShut {NoStop}%
\bibitem [{\citenamefont {Arjovsky}\ and\ \citenamefont {Bottou}(2017)}]{Arjovsky2017stability}%
  \BibitemOpen
  \bibfield  {author} {\bibinfo {author} {\bibfnamefont {M.}~\bibnamefont {Arjovsky}}\ and\ \bibinfo {author} {\bibfnamefont {L.}~\bibnamefont {Bottou}},\ }\href@noop {} {\bibinfo {title} {Towards principled methods for training generative adversarial networks}} (\bibinfo {year} {2017}),\ \Eprint {https://arxiv.org/abs/1701.04862} {arXiv:1701.04862 [stat.ML]} \BibitemShut {NoStop}%
\bibitem [{\citenamefont {Mirza}\ and\ \citenamefont {Osindero}(2014)}]{Mirza2014cGAN}%
  \BibitemOpen
  \bibfield  {author} {\bibinfo {author} {\bibfnamefont {M.}~\bibnamefont {Mirza}}\ and\ \bibinfo {author} {\bibfnamefont {S.}~\bibnamefont {Osindero}},\ }\href@noop {} {\bibinfo {title} {Conditional generative adversarial nets}} (\bibinfo {year} {2014}),\ \Eprint {https://arxiv.org/abs/1411.1784} {arXiv:1411.1784 [cs.LG]} \BibitemShut {NoStop}%
\bibitem [{\citenamefont {Ho}\ \emph {et~al.}(2020)\citenamefont {Ho}, \citenamefont {Jain},\ and\ \citenamefont {Abbeel}}]{Jonathan2020DMs}%
  \BibitemOpen
  \bibfield  {author} {\bibinfo {author} {\bibfnamefont {J.}~\bibnamefont {Ho}}, \bibinfo {author} {\bibfnamefont {A.}~\bibnamefont {Jain}},\ and\ \bibinfo {author} {\bibfnamefont {P.}~\bibnamefont {Abbeel}},\ }\bibfield  {title} {\bibinfo {title} {Denoising diffusion probabilistic models},\ }in\ \href {https://proceedings.neurips.cc/paper/2020/file/4c5bcfec8584af0d967f1ab10179ca4b-Paper.pdf} {\emph {\bibinfo {booktitle} {Advances in Neural Information Processing Systems}}},\ Vol.~\bibinfo {volume} {33}\ (\bibinfo  {publisher} {Curran Associates, Inc.},\ \bibinfo {year} {2020})\ pp.\ \bibinfo {pages} {6840--6851}\BibitemShut {NoStop}%
\bibitem [{\citenamefont {Dhariwal}\ and\ \citenamefont {Nichol}(2021)}]{Dhariwal2021DM}%
  \BibitemOpen
  \bibfield  {author} {\bibinfo {author} {\bibfnamefont {P.}~\bibnamefont {Dhariwal}}\ and\ \bibinfo {author} {\bibfnamefont {A.}~\bibnamefont {Nichol}},\ }\href@noop {} {\bibinfo {title} {Diffusion models beat gans on image synthesis}} (\bibinfo {year} {2021}),\ \Eprint {https://arxiv.org/abs/2105.05233} {arXiv:2105.05233 [cs.LG]} \BibitemShut {NoStop}%
\bibitem [{\citenamefont {Radford}\ \emph {et~al.}(2016)\citenamefont {Radford}, \citenamefont {Metz},\ and\ \citenamefont {Chintala}}]{Radford2016unsupervised}%
  \BibitemOpen
  \bibfield  {author} {\bibinfo {author} {\bibfnamefont {A.}~\bibnamefont {Radford}}, \bibinfo {author} {\bibfnamefont {L.}~\bibnamefont {Metz}},\ and\ \bibinfo {author} {\bibfnamefont {S.}~\bibnamefont {Chintala}},\ }\href@noop {} {\bibinfo {title} {Unsupervised representation learning with deep convolutional generative adversarial networks}} (\bibinfo {year} {2016}),\ \Eprint {https://arxiv.org/abs/1511.06434} {arXiv:1511.06434 [cs.LG]} \BibitemShut {NoStop}%
\bibitem [{\citenamefont {Karras}\ \emph {et~al.}(2021)\citenamefont {Karras}, \citenamefont {Laine},\ and\ \citenamefont {Aila}}]{Karras2018StyleGAN}%
  \BibitemOpen
  \bibfield  {author} {\bibinfo {author} {\bibfnamefont {T.}~\bibnamefont {Karras}}, \bibinfo {author} {\bibfnamefont {S.}~\bibnamefont {Laine}},\ and\ \bibinfo {author} {\bibfnamefont {T.}~\bibnamefont {Aila}},\ }\bibfield  {title} {\bibinfo {title} {A style-based generator architecture for generative adversarial networks},\ }\href {https://doi.org/10.1109/TPAMI.2020.2970919} {\bibfield  {journal} {\bibinfo  {journal} {IEEE Transactions on Pattern Analysis and Machine Intelligence}\ }\textbf {\bibinfo {volume} {43}},\ \bibinfo {pages} {4217} (\bibinfo {year} {2021})}\BibitemShut {NoStop}%
\bibitem [{\citenamefont {Reed}\ \emph {et~al.}(2016)\citenamefont {Reed}, \citenamefont {Akata}, \citenamefont {Yan}, \citenamefont {Logeswaran}, \citenamefont {Schiele},\ and\ \citenamefont {Lee}}]{Scott2016}%
  \BibitemOpen
  \bibfield  {author} {\bibinfo {author} {\bibfnamefont {S.}~\bibnamefont {Reed}}, \bibinfo {author} {\bibfnamefont {Z.}~\bibnamefont {Akata}}, \bibinfo {author} {\bibfnamefont {X.}~\bibnamefont {Yan}}, \bibinfo {author} {\bibfnamefont {L.}~\bibnamefont {Logeswaran}}, \bibinfo {author} {\bibfnamefont {B.}~\bibnamefont {Schiele}},\ and\ \bibinfo {author} {\bibfnamefont {H.}~\bibnamefont {Lee}},\ }\bibfield  {title} {\bibinfo {title} {Generative adversarial text to image synthesis},\ }in\ \href {https://proceedings.mlr.press/v48/reed16.html} {\emph {\bibinfo {booktitle} {Proceedings of The 33rd International Conference on Machine Learning}}},\ \bibinfo {series} {Proceedings of Machine Learning Research}, Vol.~\bibinfo {volume} {48},\ \bibinfo {editor} {edited by\ \bibinfo {editor} {\bibfnamefont {M.~F.}\ \bibnamefont {Balcan}}\ and\ \bibinfo {editor} {\bibfnamefont {K.~Q.}\ \bibnamefont {Weinberger}}}\ (\bibinfo  {publisher} {PMLR},\ \bibinfo {address} {New York, New York, USA},\ \bibinfo {year} {2016})\ pp.\
  \bibinfo {pages} {1060--1069}\BibitemShut {NoStop}%
\bibitem [{\citenamefont {Kang}\ \emph {et~al.}(2023)\citenamefont {Kang}, \citenamefont {Zhu}, \citenamefont {Zhang}, \citenamefont {Park}, \citenamefont {Shechtman}, \citenamefont {Paris},\ and\ \citenamefont {Park}}]{Kang2023scaling}%
  \BibitemOpen
  \bibfield  {author} {\bibinfo {author} {\bibfnamefont {M.}~\bibnamefont {Kang}}, \bibinfo {author} {\bibfnamefont {J.-Y.}\ \bibnamefont {Zhu}}, \bibinfo {author} {\bibfnamefont {R.}~\bibnamefont {Zhang}}, \bibinfo {author} {\bibfnamefont {J.}~\bibnamefont {Park}}, \bibinfo {author} {\bibfnamefont {E.}~\bibnamefont {Shechtman}}, \bibinfo {author} {\bibfnamefont {S.}~\bibnamefont {Paris}},\ and\ \bibinfo {author} {\bibfnamefont {T.}~\bibnamefont {Park}},\ }\href@noop {} {\bibinfo {title} {Scaling up gans for text-to-image synthesis}} (\bibinfo {year} {2023}),\ \Eprint {https://arxiv.org/abs/2303.05511} {arXiv:2303.05511 [cs.CV]} \BibitemShut {NoStop}%
\bibitem [{\citenamefont {Zhu}\ \emph {et~al.}(2017)\citenamefont {Zhu}, \citenamefont {Park}, \citenamefont {Isola},\ and\ \citenamefont {Efros}}]{Zhu2017CycleGAN}%
  \BibitemOpen
  \bibfield  {author} {\bibinfo {author} {\bibfnamefont {J.-Y.}\ \bibnamefont {Zhu}}, \bibinfo {author} {\bibfnamefont {T.}~\bibnamefont {Park}}, \bibinfo {author} {\bibfnamefont {P.}~\bibnamefont {Isola}},\ and\ \bibinfo {author} {\bibfnamefont {A.~A.}\ \bibnamefont {Efros}},\ }\bibfield  {title} {\bibinfo {title} {Unpaired image-to-image translation using cycle-consistent adversarial networks},\ }in\ \href {https://doi.org/10.1109/ICCV.2017.244} {\emph {\bibinfo {booktitle} {2017 IEEE International Conference on Computer Vision (ICCV)}}}\ (\bibinfo {year} {2017})\ pp.\ \bibinfo {pages} {2242--2251}\BibitemShut {NoStop}%
\bibitem [{\citenamefont {Yi}\ \emph {et~al.}(2017)\citenamefont {Yi}, \citenamefont {Zhang}, \citenamefont {Tan},\ and\ \citenamefont {Gong}}]{Yi2017DualGAN}%
  \BibitemOpen
  \bibfield  {author} {\bibinfo {author} {\bibfnamefont {Z.}~\bibnamefont {Yi}}, \bibinfo {author} {\bibfnamefont {H.}~\bibnamefont {Zhang}}, \bibinfo {author} {\bibfnamefont {P.}~\bibnamefont {Tan}},\ and\ \bibinfo {author} {\bibfnamefont {M.}~\bibnamefont {Gong}},\ }\bibfield  {title} {\bibinfo {title} {Dualgan: Unsupervised dual learning for image-to-image translation},\ }in\ \href {https://doi.org/10.1109/ICCV.2017.310} {\emph {\bibinfo {booktitle} {2017 IEEE International Conference on Computer Vision (ICCV)}}}\ (\bibinfo  {publisher} {IEEE Computer Society},\ \bibinfo {address} {Los Alamitos, CA, USA},\ \bibinfo {year} {2017})\ pp.\ \bibinfo {pages} {2868--2876}\BibitemShut {NoStop}%
\bibitem [{\citenamefont {de~Oliveira}\ \emph {et~al.}(2017)\citenamefont {de~Oliveira}, \citenamefont {Paganini},\ and\ \citenamefont {Nachman}}]{DeOliveira2017LAGAN}%
  \BibitemOpen
  \bibfield  {author} {\bibinfo {author} {\bibfnamefont {L.}~\bibnamefont {de~Oliveira}}, \bibinfo {author} {\bibfnamefont {M.}~\bibnamefont {Paganini}},\ and\ \bibinfo {author} {\bibfnamefont {B.}~\bibnamefont {Nachman}},\ }\bibfield  {title} {\bibinfo {title} {Learning particle physics by example: Location-aware generative adversarial networks for physics synthesis},\ }\bibfield  {journal} {\bibinfo  {journal} {Computing and Software for Big Science}\ }\textbf {\bibinfo {volume} {1}},\ \href {https://doi.org/10.1007/s41781-017-0004-6} {10.1007/s41781-017-0004-6} (\bibinfo {year} {2017})\BibitemShut {NoStop}%
\bibitem [{\citenamefont {Rombach}\ \emph {et~al.}(2022)\citenamefont {Rombach}, \citenamefont {Blattmann}, \citenamefont {Lorenz}, \citenamefont {Esser},\ and\ \citenamefont {Ommer}}]{Rombach_2022_LDM_CVPR}%
  \BibitemOpen
  \bibfield  {author} {\bibinfo {author} {\bibfnamefont {R.}~\bibnamefont {Rombach}}, \bibinfo {author} {\bibfnamefont {A.}~\bibnamefont {Blattmann}}, \bibinfo {author} {\bibfnamefont {D.}~\bibnamefont {Lorenz}}, \bibinfo {author} {\bibfnamefont {P.}~\bibnamefont {Esser}},\ and\ \bibinfo {author} {\bibfnamefont {B.}~\bibnamefont {Ommer}},\ }\bibfield  {title} {\bibinfo {title} {High-resolution image synthesis with latent diffusion models},\ }in\ \href {https://doi.org/10.1109/CVPR52688.2022.01042} {\emph {\bibinfo {booktitle} {2022 IEEE/CVF Conference on Computer Vision and Pattern Recognition (CVPR)}}}\ (\bibinfo  {publisher} {IEEE Computer Society},\ \bibinfo {year} {2022})\ pp.\ \bibinfo {pages} {10674--10685}\BibitemShut {NoStop}%
\bibitem [{\citenamefont {Biamonte}\ \emph {et~al.}(2017)\citenamefont {Biamonte}, \citenamefont {Wittek}, \citenamefont {Pancotti}, \citenamefont {Rebentrost}, \citenamefont {Wiebe},\ and\ \citenamefont {Lloyd}}]{biamonte2017quantum}%
  \BibitemOpen
  \bibfield  {author} {\bibinfo {author} {\bibfnamefont {J.}~\bibnamefont {Biamonte}}, \bibinfo {author} {\bibfnamefont {P.}~\bibnamefont {Wittek}}, \bibinfo {author} {\bibfnamefont {N.}~\bibnamefont {Pancotti}}, \bibinfo {author} {\bibfnamefont {P.}~\bibnamefont {Rebentrost}}, \bibinfo {author} {\bibfnamefont {N.}~\bibnamefont {Wiebe}},\ and\ \bibinfo {author} {\bibfnamefont {S.}~\bibnamefont {Lloyd}},\ }\bibfield  {title} {\bibinfo {title} {Quantum machine learning},\ }\href {https://doi.org/10.1038/nature23474} {\bibfield  {journal} {\bibinfo  {journal} {Nature}\ }\textbf {\bibinfo {volume} {549}},\ \bibinfo {pages} {195} (\bibinfo {year} {2017})}\BibitemShut {NoStop}%
\bibitem [{\citenamefont {Cerezo}\ \emph {et~al.}(2021{\natexlab{a}})\citenamefont {Cerezo}, \citenamefont {Arrasmith}, \citenamefont {Babbush}, \citenamefont {Benjamin}, \citenamefont {Endo}, \citenamefont {Fujii}, \citenamefont {McClean}, \citenamefont {Mitarai}, \citenamefont {Yuan}, \citenamefont {Cincio},\ and\ \citenamefont {Coles}}]{cerezo2020variationalreview}%
  \BibitemOpen
  \bibfield  {author} {\bibinfo {author} {\bibfnamefont {M.}~\bibnamefont {Cerezo}}, \bibinfo {author} {\bibfnamefont {A.}~\bibnamefont {Arrasmith}}, \bibinfo {author} {\bibfnamefont {R.}~\bibnamefont {Babbush}}, \bibinfo {author} {\bibfnamefont {S.~C.}\ \bibnamefont {Benjamin}}, \bibinfo {author} {\bibfnamefont {S.}~\bibnamefont {Endo}}, \bibinfo {author} {\bibfnamefont {K.}~\bibnamefont {Fujii}}, \bibinfo {author} {\bibfnamefont {J.~R.}\ \bibnamefont {McClean}}, \bibinfo {author} {\bibfnamefont {K.}~\bibnamefont {Mitarai}}, \bibinfo {author} {\bibfnamefont {X.}~\bibnamefont {Yuan}}, \bibinfo {author} {\bibfnamefont {L.}~\bibnamefont {Cincio}},\ and\ \bibinfo {author} {\bibfnamefont {P.~J.}\ \bibnamefont {Coles}},\ }\bibfield  {title} {\bibinfo {title} {Variational quantum algorithms},\ }\href {https://doi.org/10.1038/s42254-021-00348-9} {\bibfield  {journal} {\bibinfo  {journal} {Nature Reviews Physics}\ }\textbf {\bibinfo {volume} {3}},\ \bibinfo {pages} {625–644} (\bibinfo {year}
  {2021}{\natexlab{a}})}\BibitemShut {NoStop}%
\bibitem [{\citenamefont {{Huang}}\ \emph {et~al.}(2022)\citenamefont {{Huang}}, \citenamefont {{Broughton}}, \citenamefont {{Cotler}}, \citenamefont {{Chen}}, \citenamefont {{Li}}, \citenamefont {{Mohseni}}, \citenamefont {{Neven}}, \citenamefont {{Babbush}}, \citenamefont {{Kueng}}, \citenamefont {{Preskill}},\ and\ \citenamefont {{McClean}}}]{huang2021quantum}%
  \BibitemOpen
  \bibfield  {author} {\bibinfo {author} {\bibfnamefont {H.-Y.}\ \bibnamefont {{Huang}}}, \bibinfo {author} {\bibfnamefont {M.}~\bibnamefont {{Broughton}}}, \bibinfo {author} {\bibfnamefont {J.}~\bibnamefont {{Cotler}}}, \bibinfo {author} {\bibfnamefont {S.}~\bibnamefont {{Chen}}}, \bibinfo {author} {\bibfnamefont {J.}~\bibnamefont {{Li}}}, \bibinfo {author} {\bibfnamefont {M.}~\bibnamefont {{Mohseni}}}, \bibinfo {author} {\bibfnamefont {H.}~\bibnamefont {{Neven}}}, \bibinfo {author} {\bibfnamefont {R.}~\bibnamefont {{Babbush}}}, \bibinfo {author} {\bibfnamefont {R.}~\bibnamefont {{Kueng}}}, \bibinfo {author} {\bibfnamefont {J.}~\bibnamefont {{Preskill}}},\ and\ \bibinfo {author} {\bibfnamefont {J.~R.}\ \bibnamefont {{McClean}}},\ }\bibfield  {title} {\bibinfo {title} {Quantum advantage in learning from experiments},\ }\href {https://doi.org/10.1126/science.abn7293} {\bibfield  {journal} {\bibinfo  {journal} {Science}\ }\textbf {\bibinfo {volume} {376}},\ \bibinfo {pages} {1182} (\bibinfo {year}
  {2022})}\BibitemShut {NoStop}%
\bibitem [{\citenamefont {Liu}\ \emph {et~al.}(2021)\citenamefont {Liu}, \citenamefont {Arunachalam},\ and\ \citenamefont {Temme}}]{Liu2021Speedup}%
  \BibitemOpen
  \bibfield  {author} {\bibinfo {author} {\bibfnamefont {Y.}~\bibnamefont {Liu}}, \bibinfo {author} {\bibfnamefont {S.}~\bibnamefont {Arunachalam}},\ and\ \bibinfo {author} {\bibfnamefont {K.}~\bibnamefont {Temme}},\ }\bibfield  {title} {\bibinfo {title} {A rigorous and robust quantum speed-up in supervised machine learning},\ }\href {https://doi.org/10.1038/s41567-021-01287-z} {\bibfield  {journal} {\bibinfo  {journal} {Nature Physics}\ }\textbf {\bibinfo {volume} {17}},\ \bibinfo {pages} {1013} (\bibinfo {year} {2021})}\BibitemShut {NoStop}%
\bibitem [{\citenamefont {Aharonov}\ \emph {et~al.}(2022)\citenamefont {Aharonov}, \citenamefont {Cotler},\ and\ \citenamefont {Qi}}]{aharonov2021quantum}%
  \BibitemOpen
  \bibfield  {author} {\bibinfo {author} {\bibfnamefont {D.}~\bibnamefont {Aharonov}}, \bibinfo {author} {\bibfnamefont {J.}~\bibnamefont {Cotler}},\ and\ \bibinfo {author} {\bibfnamefont {X.-L.}\ \bibnamefont {Qi}},\ }\bibfield  {title} {\bibinfo {title} {Quantum algorithmic measurement},\ }\href {https://doi.org/10.1038/s41467-021-27922-0} {\bibfield  {journal} {\bibinfo  {journal} {Nature {C}ommunications}\ }\textbf {\bibinfo {volume} {13}},\ \bibinfo {pages} {1} (\bibinfo {year} {2022})}\BibitemShut {NoStop}%
\bibitem [{\citenamefont {Gao}\ \emph {et~al.}(2022)\citenamefont {Gao}, \citenamefont {Anschuetz}, \citenamefont {Wang}, \citenamefont {Cirac},\ and\ \citenamefont {Lukin}}]{gao2022enhancing}%
  \BibitemOpen
  \bibfield  {author} {\bibinfo {author} {\bibfnamefont {X.}~\bibnamefont {Gao}}, \bibinfo {author} {\bibfnamefont {E.~R.}\ \bibnamefont {Anschuetz}}, \bibinfo {author} {\bibfnamefont {S.-T.}\ \bibnamefont {Wang}}, \bibinfo {author} {\bibfnamefont {J.~I.}\ \bibnamefont {Cirac}},\ and\ \bibinfo {author} {\bibfnamefont {M.~D.}\ \bibnamefont {Lukin}},\ }\bibfield  {title} {\bibinfo {title} {Enhancing generative models via quantum correlations},\ }\href {https://doi.org/10.1103/PhysRevX.12.021037} {\bibfield  {journal} {\bibinfo  {journal} {Phys. Rev. X}\ }\textbf {\bibinfo {volume} {12}},\ \bibinfo {pages} {021037} (\bibinfo {year} {2022})}\BibitemShut {NoStop}%
\bibitem [{\citenamefont {Huang}\ \emph {et~al.}(2021{\natexlab{a}})\citenamefont {Huang}, \citenamefont {Broughton}, \citenamefont {Mohseni}, \citenamefont {Babbush}, \citenamefont {Boixo}, \citenamefont {Neven},\ and\ \citenamefont {McClean}}]{huang2021power}%
  \BibitemOpen
  \bibfield  {author} {\bibinfo {author} {\bibfnamefont {H.-Y.}\ \bibnamefont {Huang}}, \bibinfo {author} {\bibfnamefont {M.}~\bibnamefont {Broughton}}, \bibinfo {author} {\bibfnamefont {M.}~\bibnamefont {Mohseni}}, \bibinfo {author} {\bibfnamefont {R.}~\bibnamefont {Babbush}}, \bibinfo {author} {\bibfnamefont {S.}~\bibnamefont {Boixo}}, \bibinfo {author} {\bibfnamefont {H.}~\bibnamefont {Neven}},\ and\ \bibinfo {author} {\bibfnamefont {J.~R.}\ \bibnamefont {McClean}},\ }\bibfield  {title} {\bibinfo {title} {Power of data in quantum machine learning},\ }\href {https://doi.org/10.1038/s41467-021-22539-9} {\bibfield  {journal} {\bibinfo  {journal} {Nature Communications}\ }\textbf {\bibinfo {volume} {12}},\ \bibinfo {pages} {2631} (\bibinfo {year} {2021}{\natexlab{a}})}\BibitemShut {NoStop}%
\bibitem [{\citenamefont {Wu}\ \emph {et~al.}(2023)\citenamefont {Wu}, \citenamefont {Wu}, \citenamefont {Wang},\ and\ \citenamefont {Yuan}}]{wu2023quantum}%
  \BibitemOpen
  \bibfield  {author} {\bibinfo {author} {\bibfnamefont {Y.}~\bibnamefont {Wu}}, \bibinfo {author} {\bibfnamefont {B.}~\bibnamefont {Wu}}, \bibinfo {author} {\bibfnamefont {J.}~\bibnamefont {Wang}},\ and\ \bibinfo {author} {\bibfnamefont {X.}~\bibnamefont {Yuan}},\ }\bibfield  {title} {\bibinfo {title} {Quantum phase recognition via quantum kernel methods},\ }\href@noop {} {\bibfield  {journal} {\bibinfo  {journal} {Quantum}\ }\textbf {\bibinfo {volume} {7}},\ \bibinfo {pages} {981} (\bibinfo {year} {2023})}\BibitemShut {NoStop}%
\bibitem [{\citenamefont {Nietner}\ \emph {et~al.}(2023)\citenamefont {Nietner}, \citenamefont {Ioannou}, \citenamefont {Sweke}, \citenamefont {Kueng}, \citenamefont {Eisert}, \citenamefont {Hinsche},\ and\ \citenamefont {Haferkamp}}]{nietner2023average}%
  \BibitemOpen
  \bibfield  {author} {\bibinfo {author} {\bibfnamefont {A.}~\bibnamefont {Nietner}}, \bibinfo {author} {\bibfnamefont {M.}~\bibnamefont {Ioannou}}, \bibinfo {author} {\bibfnamefont {R.}~\bibnamefont {Sweke}}, \bibinfo {author} {\bibfnamefont {R.}~\bibnamefont {Kueng}}, \bibinfo {author} {\bibfnamefont {J.}~\bibnamefont {Eisert}}, \bibinfo {author} {\bibfnamefont {M.}~\bibnamefont {Hinsche}},\ and\ \bibinfo {author} {\bibfnamefont {J.}~\bibnamefont {Haferkamp}},\ }\bibfield  {title} {\bibinfo {title} {On the average-case complexity of learning output distributions of quantum circuits},\ }\href@noop {} {\bibfield  {journal} {\bibinfo  {journal} {arXiv preprint arXiv:2305.05765}\ } (\bibinfo {year} {2023})}\BibitemShut {NoStop}%
\bibitem [{\citenamefont {Tangpanitanon}\ \emph {et~al.}(2020)\citenamefont {Tangpanitanon}, \citenamefont {Thanasilp}, \citenamefont {Dangniam}, \citenamefont {Lemonde},\ and\ \citenamefont {Angelakis}}]{tangpanitanon2020expressibility}%
  \BibitemOpen
  \bibfield  {author} {\bibinfo {author} {\bibfnamefont {J.}~\bibnamefont {Tangpanitanon}}, \bibinfo {author} {\bibfnamefont {S.}~\bibnamefont {Thanasilp}}, \bibinfo {author} {\bibfnamefont {N.}~\bibnamefont {Dangniam}}, \bibinfo {author} {\bibfnamefont {M.-A.}\ \bibnamefont {Lemonde}},\ and\ \bibinfo {author} {\bibfnamefont {D.~G.}\ \bibnamefont {Angelakis}},\ }\bibfield  {title} {\bibinfo {title} {Expressibility and trainability of parametrized analog quantum systems for machine learning applications},\ }\href {https://doi.org/10.1103/PhysRevResearch.2.043364} {\bibfield  {journal} {\bibinfo  {journal} {Physical Review Research}\ }\textbf {\bibinfo {volume} {2}},\ \bibinfo {pages} {043364} (\bibinfo {year} {2020})}\BibitemShut {NoStop}%
\bibitem [{\citenamefont {Benedetti}\ \emph {et~al.}(2019)\citenamefont {Benedetti}, \citenamefont {Garcia-Pintos}, \citenamefont {Perdomo}, \citenamefont {Leyton-Ortega}, \citenamefont {Nam},\ and\ \citenamefont {Perdomo-Ortiz}}]{Benedetti2019QCBM}%
  \BibitemOpen
  \bibfield  {author} {\bibinfo {author} {\bibfnamefont {M.}~\bibnamefont {Benedetti}}, \bibinfo {author} {\bibfnamefont {D.}~\bibnamefont {Garcia-Pintos}}, \bibinfo {author} {\bibfnamefont {O.}~\bibnamefont {Perdomo}}, \bibinfo {author} {\bibfnamefont {V.}~\bibnamefont {Leyton-Ortega}}, \bibinfo {author} {\bibfnamefont {Y.}~\bibnamefont {Nam}},\ and\ \bibinfo {author} {\bibfnamefont {A.}~\bibnamefont {Perdomo-Ortiz}},\ }\bibfield  {title} {\bibinfo {title} {A generative modeling approach for benchmarking and training shallow quantum circuits},\ }\href {https://doi.org/10.1038/s41534-019-0157-8} {\bibfield  {journal} {\bibinfo  {journal} {npj Quantum Information}\ }\textbf {\bibinfo {volume} {5}},\ \bibinfo {pages} {45} (\bibinfo {year} {2019})}\BibitemShut {NoStop}%
\bibitem [{\citenamefont {Rudolph}\ \emph {et~al.}(2023{\natexlab{a}})\citenamefont {Rudolph}, \citenamefont {Lerch}, \citenamefont {Thanasilp}, \citenamefont {Kiss}, \citenamefont {Vallecorsa}, \citenamefont {Grossi},\ and\ \citenamefont {Holmes}}]{rudolph2023trainability}%
  \BibitemOpen
  \bibfield  {author} {\bibinfo {author} {\bibfnamefont {M.~S.}\ \bibnamefont {Rudolph}}, \bibinfo {author} {\bibfnamefont {S.}~\bibnamefont {Lerch}}, \bibinfo {author} {\bibfnamefont {S.}~\bibnamefont {Thanasilp}}, \bibinfo {author} {\bibfnamefont {O.}~\bibnamefont {Kiss}}, \bibinfo {author} {\bibfnamefont {S.}~\bibnamefont {Vallecorsa}}, \bibinfo {author} {\bibfnamefont {M.}~\bibnamefont {Grossi}},\ and\ \bibinfo {author} {\bibfnamefont {Z.}~\bibnamefont {Holmes}},\ }\href@noop {} {\bibinfo {title} {Trainability barriers and opportunities in quantum generative modeling}} (\bibinfo {year} {2023}{\natexlab{a}}),\ \Eprint {https://arxiv.org/abs/2305.02881} {arXiv:2305.02881 [quant-ph]} \BibitemShut {NoStop}%
\bibitem [{\citenamefont {Coyle}\ \emph {et~al.}(2020)\citenamefont {Coyle}, \citenamefont {Mills}, \citenamefont {Danos},\ and\ \citenamefont {Kashefi}}]{coyle2020born}%
  \BibitemOpen
  \bibfield  {author} {\bibinfo {author} {\bibfnamefont {B.}~\bibnamefont {Coyle}}, \bibinfo {author} {\bibfnamefont {D.}~\bibnamefont {Mills}}, \bibinfo {author} {\bibfnamefont {V.}~\bibnamefont {Danos}},\ and\ \bibinfo {author} {\bibfnamefont {E.}~\bibnamefont {Kashefi}},\ }\bibfield  {title} {\bibinfo {title} {The born supremacy: quantum advantage and training of an ising born machine},\ }\href {https://doi.org/10.1038/s41534-020-00288-9} {\bibfield  {journal} {\bibinfo  {journal} {npj Quantum Information}\ }\textbf {\bibinfo {volume} {6}},\ \bibinfo {pages} {60} (\bibinfo {year} {2020})}\BibitemShut {NoStop}%
\bibitem [{\citenamefont {Liu}\ and\ \citenamefont {Wang}(2018)}]{liu2018differentiable}%
  \BibitemOpen
  \bibfield  {author} {\bibinfo {author} {\bibfnamefont {J.-G.}\ \bibnamefont {Liu}}\ and\ \bibinfo {author} {\bibfnamefont {L.}~\bibnamefont {Wang}},\ }\bibfield  {title} {\bibinfo {title} {Differentiable learning of quantum circuit born machines},\ }\href {https://doi.org/10.1103/PhysRevA.98.062324} {\bibfield  {journal} {\bibinfo  {journal} {Phys. Rev. A}\ }\textbf {\bibinfo {volume} {98}},\ \bibinfo {pages} {062324} (\bibinfo {year} {2018})}\BibitemShut {NoStop}%
\bibitem [{\citenamefont {Rudolph}\ \emph {et~al.}(2022)\citenamefont {Rudolph}, \citenamefont {Toussaint}, \citenamefont {Katabarwa}, \citenamefont {Johri}, \citenamefont {Peropadre},\ and\ \citenamefont {Perdomo-Ortiz}}]{Rudolph2022}%
  \BibitemOpen
  \bibfield  {author} {\bibinfo {author} {\bibfnamefont {M.~S.}\ \bibnamefont {Rudolph}}, \bibinfo {author} {\bibfnamefont {N.~B.}\ \bibnamefont {Toussaint}}, \bibinfo {author} {\bibfnamefont {A.}~\bibnamefont {Katabarwa}}, \bibinfo {author} {\bibfnamefont {S.}~\bibnamefont {Johri}}, \bibinfo {author} {\bibfnamefont {B.}~\bibnamefont {Peropadre}},\ and\ \bibinfo {author} {\bibfnamefont {A.}~\bibnamefont {Perdomo-Ortiz}},\ }\bibfield  {title} {\bibinfo {title} {Generation of high-resolution handwritten digits with an ion-trap quantum computer},\ }\href {https://doi.org/10.1103/PhysRevX.12.031010} {\bibfield  {journal} {\bibinfo  {journal} {Phys. Rev. X}\ }\textbf {\bibinfo {volume} {12}},\ \bibinfo {pages} {031010} (\bibinfo {year} {2022})}\BibitemShut {NoStop}%
\bibitem [{\citenamefont {Kiss}\ \emph {et~al.}(2022)\citenamefont {Kiss}, \citenamefont {Grossi}, \citenamefont {Kajomovitz},\ and\ \citenamefont {Vallecorsa}}]{Kiss2022QCBM}%
  \BibitemOpen
  \bibfield  {author} {\bibinfo {author} {\bibfnamefont {O.}~\bibnamefont {Kiss}}, \bibinfo {author} {\bibfnamefont {M.}~\bibnamefont {Grossi}}, \bibinfo {author} {\bibfnamefont {E.}~\bibnamefont {Kajomovitz}},\ and\ \bibinfo {author} {\bibfnamefont {S.}~\bibnamefont {Vallecorsa}},\ }\bibfield  {title} {\bibinfo {title} {Conditional born machine for monte carlo event generation},\ }\href {https://doi.org/10.1103/PhysRevA.106.022612} {\bibfield  {journal} {\bibinfo  {journal} {Phys. Rev. A}\ }\textbf {\bibinfo {volume} {106}},\ \bibinfo {pages} {022612} (\bibinfo {year} {2022})}\BibitemShut {NoStop}%
\bibitem [{\citenamefont {Kyriienko}\ \emph {et~al.}(2022)\citenamefont {Kyriienko}, \citenamefont {Paine},\ and\ \citenamefont {Elfving}}]{kyriienko2022protocols}%
  \BibitemOpen
  \bibfield  {author} {\bibinfo {author} {\bibfnamefont {O.}~\bibnamefont {Kyriienko}}, \bibinfo {author} {\bibfnamefont {A.~E.}\ \bibnamefont {Paine}},\ and\ \bibinfo {author} {\bibfnamefont {V.~E.}\ \bibnamefont {Elfving}},\ }\href@noop {} {\bibinfo {title} {Protocols for trainable and differentiable quantum generative modelling}} (\bibinfo {year} {2022}),\ \Eprint {https://arxiv.org/abs/2202.08253} {arXiv:2202.08253 [quant-ph]} \BibitemShut {NoStop}%
\bibitem [{\citenamefont {Zoufal}\ \emph {et~al.}(2019)\citenamefont {Zoufal}, \citenamefont {Lucchi},\ and\ \citenamefont {Woerner}}]{Zoufal2019}%
  \BibitemOpen
  \bibfield  {author} {\bibinfo {author} {\bibfnamefont {C.}~\bibnamefont {Zoufal}}, \bibinfo {author} {\bibfnamefont {A.}~\bibnamefont {Lucchi}},\ and\ \bibinfo {author} {\bibfnamefont {S.}~\bibnamefont {Woerner}},\ }\bibfield  {title} {\bibinfo {title} {Quantum generative adversarial networks for learning and loading random distributions},\ }\href {https://doi.org/10.1038/s41534-019-0223-2} {\bibfield  {journal} {\bibinfo  {journal} {npj Quantum Information}\ }\textbf {\bibinfo {volume} {5}},\ \bibinfo {pages} {103} (\bibinfo {year} {2019})}\BibitemShut {NoStop}%
\bibitem [{\citenamefont {Chang}\ \emph {et~al.}(2021)\citenamefont {Chang}, \citenamefont {Vallecorsa}, \citenamefont {Combarro},\ and\ \citenamefont {Carminati}}]{chang2021cvqgan}%
  \BibitemOpen
  \bibfield  {author} {\bibinfo {author} {\bibfnamefont {S.~Y.}\ \bibnamefont {Chang}}, \bibinfo {author} {\bibfnamefont {S.}~\bibnamefont {Vallecorsa}}, \bibinfo {author} {\bibfnamefont {E.~F.}\ \bibnamefont {Combarro}},\ and\ \bibinfo {author} {\bibfnamefont {F.}~\bibnamefont {Carminati}},\ }\href@noop {} {\bibinfo {title} {Quantum generative adversarial networks in a continuous-variable architecture to simulate high energy physics detectors}} (\bibinfo {year} {2021}),\ \Eprint {https://arxiv.org/abs/2101.11132} {arXiv:2101.11132 [quant-ph]} \BibitemShut {NoStop}%
\bibitem [{\citenamefont {{Chang, Su Yeon}}\ \emph {et~al.}(2021)\citenamefont {{Chang, Su Yeon}}, \citenamefont {{Herbert, Steven}}, \citenamefont {{Vallecorsa, Sofia}}, \citenamefont {{Combarro, Elías F.}},\ and\ \citenamefont {{Duncan, Ross}}}]{chang2021dual}%
  \BibitemOpen
  \bibfield  {author} {\bibinfo {author} {\bibnamefont {{Chang, Su Yeon}}}, \bibinfo {author} {\bibnamefont {{Herbert, Steven}}}, \bibinfo {author} {\bibnamefont {{Vallecorsa, Sofia}}}, \bibinfo {author} {\bibnamefont {{Combarro, Elías F.}}},\ and\ \bibinfo {author} {\bibnamefont {{Duncan, Ross}}},\ }\bibfield  {title} {\bibinfo {title} {Dual-parameterized quantum circuit gan model in high energy physics},\ }\href {https://doi.org/10.1051/epjconf/202125103050} {\bibfield  {journal} {\bibinfo  {journal} {EPJ Web Conf.}\ }\textbf {\bibinfo {volume} {251}},\ \bibinfo {pages} {03050} (\bibinfo {year} {2021})}\BibitemShut {NoStop}%
\bibitem [{\citenamefont {Huang}\ \emph {et~al.}(2021{\natexlab{b}})\citenamefont {Huang}, \citenamefont {Du}, \citenamefont {Gong}, \citenamefont {Zhao}, \citenamefont {Wu}, \citenamefont {Wang}, \citenamefont {Li}, \citenamefont {Liang}, \citenamefont {Lin}, \citenamefont {Xu}, \citenamefont {Yang}, \citenamefont {Liu}, \citenamefont {Hsieh}, \citenamefont {Deng}, \citenamefont {Rong}, \citenamefont {Peng}, \citenamefont {Lu}, \citenamefont {Chen}, \citenamefont {Tao}, \citenamefont {Zhu},\ and\ \citenamefont {Pan}}]{Huang2021qGAN}%
  \BibitemOpen
  \bibfield  {author} {\bibinfo {author} {\bibfnamefont {H.-L.}\ \bibnamefont {Huang}}, \bibinfo {author} {\bibfnamefont {Y.}~\bibnamefont {Du}}, \bibinfo {author} {\bibfnamefont {M.}~\bibnamefont {Gong}}, \bibinfo {author} {\bibfnamefont {Y.}~\bibnamefont {Zhao}}, \bibinfo {author} {\bibfnamefont {Y.}~\bibnamefont {Wu}}, \bibinfo {author} {\bibfnamefont {C.}~\bibnamefont {Wang}}, \bibinfo {author} {\bibfnamefont {S.}~\bibnamefont {Li}}, \bibinfo {author} {\bibfnamefont {F.}~\bibnamefont {Liang}}, \bibinfo {author} {\bibfnamefont {J.}~\bibnamefont {Lin}}, \bibinfo {author} {\bibfnamefont {Y.}~\bibnamefont {Xu}}, \bibinfo {author} {\bibfnamefont {R.}~\bibnamefont {Yang}}, \bibinfo {author} {\bibfnamefont {T.}~\bibnamefont {Liu}}, \bibinfo {author} {\bibfnamefont {M.-H.}\ \bibnamefont {Hsieh}}, \bibinfo {author} {\bibfnamefont {H.}~\bibnamefont {Deng}}, \bibinfo {author} {\bibfnamefont {H.}~\bibnamefont {Rong}}, \bibinfo {author} {\bibfnamefont {C.-Z.}\ \bibnamefont {Peng}}, \bibinfo {author} {\bibfnamefont
  {C.-Y.}\ \bibnamefont {Lu}}, \bibinfo {author} {\bibfnamefont {Y.-A.}\ \bibnamefont {Chen}}, \bibinfo {author} {\bibfnamefont {D.}~\bibnamefont {Tao}}, \bibinfo {author} {\bibfnamefont {X.}~\bibnamefont {Zhu}},\ and\ \bibinfo {author} {\bibfnamefont {J.-W.}\ \bibnamefont {Pan}},\ }\bibfield  {title} {\bibinfo {title} {Experimental quantum generative adversarial networks for image generation},\ }\href {https://doi.org/10.1103/PhysRevApplied.16.024051} {\bibfield  {journal} {\bibinfo  {journal} {Phys. Rev. Appl.}\ }\textbf {\bibinfo {volume} {16}},\ \bibinfo {pages} {024051} (\bibinfo {year} {2021}{\natexlab{b}})}\BibitemShut {NoStop}%
\bibitem [{\citenamefont {Letcher}\ \emph {et~al.}(2023)\citenamefont {Letcher}, \citenamefont {Woerner},\ and\ \citenamefont {Zoufal}}]{letcher2023qgan}%
  \BibitemOpen
  \bibfield  {author} {\bibinfo {author} {\bibfnamefont {A.}~\bibnamefont {Letcher}}, \bibinfo {author} {\bibfnamefont {S.}~\bibnamefont {Woerner}},\ and\ \bibinfo {author} {\bibfnamefont {C.}~\bibnamefont {Zoufal}},\ }\href@noop {} {\bibinfo {title} {From tight gradient bounds for parameterized quantum circuits to the absence of barren plateaus in qgans}} (\bibinfo {year} {2023}),\ \Eprint {https://arxiv.org/abs/2309.12681} {arXiv:2309.12681 [quant-ph]} \BibitemShut {NoStop}%
\bibitem [{\citenamefont {Amin}\ \emph {et~al.}(2018)\citenamefont {Amin}, \citenamefont {Andriyash}, \citenamefont {Rolfe}, \citenamefont {Kulchytskyy},\ and\ \citenamefont {Melko}}]{Amin2018QBM}%
  \BibitemOpen
  \bibfield  {author} {\bibinfo {author} {\bibfnamefont {M.~H.}\ \bibnamefont {Amin}}, \bibinfo {author} {\bibfnamefont {E.}~\bibnamefont {Andriyash}}, \bibinfo {author} {\bibfnamefont {J.}~\bibnamefont {Rolfe}}, \bibinfo {author} {\bibfnamefont {B.}~\bibnamefont {Kulchytskyy}},\ and\ \bibinfo {author} {\bibfnamefont {R.}~\bibnamefont {Melko}},\ }\bibfield  {title} {\bibinfo {title} {Quantum boltzmann machine},\ }\href {https://doi.org/10.1103/PhysRevX.8.021050} {\bibfield  {journal} {\bibinfo  {journal} {Phys. Rev. X}\ }\textbf {\bibinfo {volume} {8}},\ \bibinfo {pages} {021050} (\bibinfo {year} {2018})}\BibitemShut {NoStop}%
\bibitem [{\citenamefont {Coopmans}\ and\ \citenamefont {Benedetti}(2023)}]{coopmans2023sample}%
  \BibitemOpen
  \bibfield  {author} {\bibinfo {author} {\bibfnamefont {L.}~\bibnamefont {Coopmans}}\ and\ \bibinfo {author} {\bibfnamefont {M.}~\bibnamefont {Benedetti}},\ }\bibfield  {title} {\bibinfo {title} {On the sample complexity of quantum boltzmann machine learning},\ }\href {https://arxiv.org/abs/2306.14969} {\bibfield  {journal} {\bibinfo  {journal} {arXiv preprint arXiv:2306.14969}\ } (\bibinfo {year} {2023})}\BibitemShut {NoStop}%
\bibitem [{\citenamefont {Romero}\ and\ \citenamefont {Aspuru-Guzik}(2021)}]{Romero2019}%
  \BibitemOpen
  \bibfield  {author} {\bibinfo {author} {\bibfnamefont {J.}~\bibnamefont {Romero}}\ and\ \bibinfo {author} {\bibfnamefont {A.}~\bibnamefont {Aspuru-Guzik}},\ }\bibfield  {title} {\bibinfo {title} {Variational quantum generators: Generative adversarial quantum machine learning for continuous distributions},\ }\href {https://doi.org/https://doi.org/10.1002/qute.202000003} {\bibfield  {journal} {\bibinfo  {journal} {Advanced Quantum Technologies}\ }\textbf {\bibinfo {volume} {4}},\ \bibinfo {pages} {2000003} (\bibinfo {year} {2021})}\BibitemShut {NoStop}%
\bibitem [{\citenamefont {Bravo-Prieto}\ \emph {et~al.}(2022)\citenamefont {Bravo-Prieto}, \citenamefont {Baglio}, \citenamefont {C{\`{e}}}, \citenamefont {Francis}, \citenamefont {Grabowska},\ and\ \citenamefont {Carrazza}}]{BravoPrieto2022}%
  \BibitemOpen
  \bibfield  {author} {\bibinfo {author} {\bibfnamefont {C.}~\bibnamefont {Bravo-Prieto}}, \bibinfo {author} {\bibfnamefont {J.}~\bibnamefont {Baglio}}, \bibinfo {author} {\bibfnamefont {M.}~\bibnamefont {C{\`{e}}}}, \bibinfo {author} {\bibfnamefont {A.}~\bibnamefont {Francis}}, \bibinfo {author} {\bibfnamefont {D.~M.}\ \bibnamefont {Grabowska}},\ and\ \bibinfo {author} {\bibfnamefont {S.}~\bibnamefont {Carrazza}},\ }\bibfield  {title} {\bibinfo {title} {Style-based quantum generative adversarial networks for {M}onte {C}arlo events},\ }\href {https://doi.org/10.22331/q-2022-08-17-777} {\bibfield  {journal} {\bibinfo  {journal} {{Quantum}}\ }\textbf {\bibinfo {volume} {6}},\ \bibinfo {pages} {777} (\bibinfo {year} {2022})}\BibitemShut {NoStop}%
\bibitem [{\citenamefont {Barthe}\ \emph {et~al.}(2024)\citenamefont {Barthe}, \citenamefont {Grossi}, \citenamefont {Vallecorsa}, \citenamefont {Tura},\ and\ \citenamefont {Dunjko}}]{barthe2024expressivity}%
  \BibitemOpen
  \bibfield  {author} {\bibinfo {author} {\bibfnamefont {A.}~\bibnamefont {Barthe}}, \bibinfo {author} {\bibfnamefont {M.}~\bibnamefont {Grossi}}, \bibinfo {author} {\bibfnamefont {S.}~\bibnamefont {Vallecorsa}}, \bibinfo {author} {\bibfnamefont {J.}~\bibnamefont {Tura}},\ and\ \bibinfo {author} {\bibfnamefont {V.}~\bibnamefont {Dunjko}},\ }\href@noop {} {\bibinfo {title} {Expressivity of parameterized quantum circuits for generative modeling of continuous multivariate distributions}} (\bibinfo {year} {2024}),\ \Eprint {https://arxiv.org/abs/2402.09848} {arXiv:2402.09848 [quant-ph]} \BibitemShut {NoStop}%
\bibitem [{\citenamefont {Gili}\ \emph {et~al.}(2023{\natexlab{a}})\citenamefont {Gili}, \citenamefont {Hibat-Allah}, \citenamefont {Mauri}, \citenamefont {Ballance},\ and\ \citenamefont {Perdomo-Ortiz}}]{Gili2023}%
  \BibitemOpen
  \bibfield  {author} {\bibinfo {author} {\bibfnamefont {K.}~\bibnamefont {Gili}}, \bibinfo {author} {\bibfnamefont {M.}~\bibnamefont {Hibat-Allah}}, \bibinfo {author} {\bibfnamefont {M.}~\bibnamefont {Mauri}}, \bibinfo {author} {\bibfnamefont {C.}~\bibnamefont {Ballance}},\ and\ \bibinfo {author} {\bibfnamefont {A.}~\bibnamefont {Perdomo-Ortiz}},\ }\bibfield  {title} {\bibinfo {title} {Do quantum circuit born machines generalize?},\ }\href {https://doi.org/10.1088/2058-9565/acd578} {\bibfield  {journal} {\bibinfo  {journal} {Quantum Science and Technology}\ }\textbf {\bibinfo {volume} {8}},\ \bibinfo {pages} {035021} (\bibinfo {year} {2023}{\natexlab{a}})}\BibitemShut {NoStop}%
\bibitem [{\citenamefont {Basu}\ \emph {et~al.}(2015)\citenamefont {Basu}, \citenamefont {Ganguly}, \citenamefont {Mukhopadhyay}, \citenamefont {DiBiano}, \citenamefont {Karki},\ and\ \citenamefont {Nemani}}]{SAT4}%
  \BibitemOpen
  \bibfield  {author} {\bibinfo {author} {\bibfnamefont {S.}~\bibnamefont {Basu}}, \bibinfo {author} {\bibfnamefont {S.}~\bibnamefont {Ganguly}}, \bibinfo {author} {\bibfnamefont {S.}~\bibnamefont {Mukhopadhyay}}, \bibinfo {author} {\bibfnamefont {R.}~\bibnamefont {DiBiano}}, \bibinfo {author} {\bibfnamefont {M.}~\bibnamefont {Karki}},\ and\ \bibinfo {author} {\bibfnamefont {R.}~\bibnamefont {Nemani}},\ }\bibfield  {title} {\bibinfo {title} {Deepsat: a learning framework for satellite imagery},\ }in\ \href {https://doi.org/10.1145/2820783.2820816} {\emph {\bibinfo {booktitle} {Proceedings of the 23rd SIGSPATIAL International Conference on Advances in Geographic Information Systems}}},\ \bibinfo {series and number} {SIGSPATIAL '15}\ (\bibinfo  {publisher} {Association for Computing Machinery},\ \bibinfo {year} {2015})\BibitemShut {NoStop}%
\bibitem [{\citenamefont {Arjovsky}\ \emph {et~al.}(2017)\citenamefont {Arjovsky}, \citenamefont {Chintala},\ and\ \citenamefont {Bottou}}]{Arjovsky2017wGAN}%
  \BibitemOpen
  \bibfield  {author} {\bibinfo {author} {\bibfnamefont {M.}~\bibnamefont {Arjovsky}}, \bibinfo {author} {\bibfnamefont {S.}~\bibnamefont {Chintala}},\ and\ \bibinfo {author} {\bibfnamefont {L.}~\bibnamefont {Bottou}},\ }\href@noop {} {\bibinfo {title} {Wasserstein gan}} (\bibinfo {year} {2017}),\ \Eprint {https://arxiv.org/abs/1701.07875} {arXiv:1701.07875 [stat.ML]} \BibitemShut {NoStop}%
\bibitem [{\citenamefont {MacCormack}\ \emph {et~al.}(2022)\citenamefont {MacCormack}, \citenamefont {Delaney}, \citenamefont {Galda}, \citenamefont {Aggarwal},\ and\ \citenamefont {Narang}}]{MacCormack2020SU4}%
  \BibitemOpen
  \bibfield  {author} {\bibinfo {author} {\bibfnamefont {I.}~\bibnamefont {MacCormack}}, \bibinfo {author} {\bibfnamefont {C.}~\bibnamefont {Delaney}}, \bibinfo {author} {\bibfnamefont {A.}~\bibnamefont {Galda}}, \bibinfo {author} {\bibfnamefont {N.}~\bibnamefont {Aggarwal}},\ and\ \bibinfo {author} {\bibfnamefont {P.}~\bibnamefont {Narang}},\ }\bibfield  {title} {\bibinfo {title} {Branching quantum convolutional neural networks},\ }\href {https://doi.org/10.1103/PhysRevResearch.4.013117} {\bibfield  {journal} {\bibinfo  {journal} {Phys. Rev. Res.}\ }\textbf {\bibinfo {volume} {4}},\ \bibinfo {pages} {013117} (\bibinfo {year} {2022})}\BibitemShut {NoStop}%
\bibitem [{\citenamefont {P{\'{e}}rez-Salinas}\ \emph {et~al.}(2020)\citenamefont {P{\'{e}}rez-Salinas}, \citenamefont {Cervera-Lierta}, \citenamefont {Gil-Fuster},\ and\ \citenamefont {Latorre}}]{PerezSalinas2020datareuploading}%
  \BibitemOpen
  \bibfield  {author} {\bibinfo {author} {\bibfnamefont {A.}~\bibnamefont {P{\'{e}}rez-Salinas}}, \bibinfo {author} {\bibfnamefont {A.}~\bibnamefont {Cervera-Lierta}}, \bibinfo {author} {\bibfnamefont {E.}~\bibnamefont {Gil-Fuster}},\ and\ \bibinfo {author} {\bibfnamefont {J.~I.}\ \bibnamefont {Latorre}},\ }\bibfield  {title} {\bibinfo {title} {Data re-uploading for a universal quantum classifier},\ }\href {https://doi.org/10.22331/q-2020-02-06-226} {\bibfield  {journal} {\bibinfo  {journal} {{Quantum}}\ }\textbf {\bibinfo {volume} {4}},\ \bibinfo {pages} {226} (\bibinfo {year} {2020})}\BibitemShut {NoStop}%
\bibitem [{\citenamefont {Easom-Mccaldin}\ \emph {et~al.}(2021)\citenamefont {Easom-Mccaldin}, \citenamefont {Bouridane}, \citenamefont {Belatreche},\ and\ \citenamefont {Jiang}}]{Easom2021datareuploding}%
  \BibitemOpen
  \bibfield  {author} {\bibinfo {author} {\bibfnamefont {P.}~\bibnamefont {Easom-Mccaldin}}, \bibinfo {author} {\bibfnamefont {A.}~\bibnamefont {Bouridane}}, \bibinfo {author} {\bibfnamefont {A.}~\bibnamefont {Belatreche}},\ and\ \bibinfo {author} {\bibfnamefont {R.}~\bibnamefont {Jiang}},\ }\bibfield  {title} {\bibinfo {title} {On depth, robustness and performance using the data re-uploading single-qubit classifier},\ }\href {https://doi.org/10.1109/ACCESS.2021.3075492} {\bibfield  {journal} {\bibinfo  {journal} {IEEE Access}\ }\textbf {\bibinfo {volume} {9}},\ \bibinfo {pages} {65127} (\bibinfo {year} {2021})}\BibitemShut {NoStop}%
\bibitem [{\citenamefont {Zeng}\ \emph {et~al.}(2022)\citenamefont {Zeng}, \citenamefont {Wang}, \citenamefont {He}, \citenamefont {Huang},\ and\ \citenamefont {Chang}}]{Zeng2021}%
  \BibitemOpen
  \bibfield  {author} {\bibinfo {author} {\bibfnamefont {Y.}~\bibnamefont {Zeng}}, \bibinfo {author} {\bibfnamefont {H.}~\bibnamefont {Wang}}, \bibinfo {author} {\bibfnamefont {J.}~\bibnamefont {He}}, \bibinfo {author} {\bibfnamefont {Q.}~\bibnamefont {Huang}},\ and\ \bibinfo {author} {\bibfnamefont {S.}~\bibnamefont {Chang}},\ }\bibfield  {title} {\bibinfo {title} {A multi-classification hybrid quantum neural network using an all-qubit multi-observable measurement strategy},\ }\href {https://www.mdpi.com/1099-4300/24/3/394} {\bibfield  {journal} {\bibinfo  {journal} {Entropy}\ }\textbf {\bibinfo {volume} {24}} (\bibinfo {year} {2022})}\BibitemShut {NoStop}%
\bibitem [{\citenamefont {He}\ \emph {et~al.}(2016)\citenamefont {He}, \citenamefont {Zhang}, \citenamefont {Ren},\ and\ \citenamefont {Sun}}]{He2015resnet}%
  \BibitemOpen
  \bibfield  {author} {\bibinfo {author} {\bibfnamefont {K.}~\bibnamefont {He}}, \bibinfo {author} {\bibfnamefont {X.}~\bibnamefont {Zhang}}, \bibinfo {author} {\bibfnamefont {S.}~\bibnamefont {Ren}},\ and\ \bibinfo {author} {\bibfnamefont {J.}~\bibnamefont {Sun}},\ }\bibfield  {title} {\bibinfo {title} {Deep residual learning for image recognition},\ }in\ \href {https://doi.org/10.1109/CVPR.2016.90} {\emph {\bibinfo {booktitle} {2016 IEEE Conference on Computer Vision and Pattern Recognition (CVPR)}}}\ (\bibinfo {year} {2016})\ pp.\ \bibinfo {pages} {770--778}\BibitemShut {NoStop}%
\bibitem [{\citenamefont {Gili}\ \emph {et~al.}(2023{\natexlab{b}})\citenamefont {Gili}, \citenamefont {Mauri},\ and\ \citenamefont {Perdomo-Ortiz}}]{gili2023generalization}%
  \BibitemOpen
  \bibfield  {author} {\bibinfo {author} {\bibfnamefont {K.}~\bibnamefont {Gili}}, \bibinfo {author} {\bibfnamefont {M.}~\bibnamefont {Mauri}},\ and\ \bibinfo {author} {\bibfnamefont {A.}~\bibnamefont {Perdomo-Ortiz}},\ }\href@noop {} {\bibinfo {title} {Generalization metrics for practical quantum advantage in generative models}} (\bibinfo {year} {2023}{\natexlab{b}}),\ \Eprint {https://arxiv.org/abs/2201.08770} {arXiv:2201.08770 [cs.LG]} \BibitemShut {NoStop}%
\bibitem [{\citenamefont {Riofrío}\ \emph {et~al.}(2023)\citenamefont {Riofrío}, \citenamefont {Mitevski}, \citenamefont {Jones}, \citenamefont {Krellner}, \citenamefont {Vučković}, \citenamefont {Doetsch}, \citenamefont {Klepsch}, \citenamefont {Ehmer},\ and\ \citenamefont {Luckow}}]{riofrío2023performance}%
  \BibitemOpen
  \bibfield  {author} {\bibinfo {author} {\bibfnamefont {C.~A.}\ \bibnamefont {Riofrío}}, \bibinfo {author} {\bibfnamefont {O.}~\bibnamefont {Mitevski}}, \bibinfo {author} {\bibfnamefont {C.}~\bibnamefont {Jones}}, \bibinfo {author} {\bibfnamefont {F.}~\bibnamefont {Krellner}}, \bibinfo {author} {\bibfnamefont {A.}~\bibnamefont {Vučković}}, \bibinfo {author} {\bibfnamefont {J.}~\bibnamefont {Doetsch}}, \bibinfo {author} {\bibfnamefont {J.}~\bibnamefont {Klepsch}}, \bibinfo {author} {\bibfnamefont {T.}~\bibnamefont {Ehmer}},\ and\ \bibinfo {author} {\bibfnamefont {A.}~\bibnamefont {Luckow}},\ }\href@noop {} {\bibinfo {title} {A performance characterization of quantum generative models}} (\bibinfo {year} {2023}),\ \Eprint {https://arxiv.org/abs/2301.09363} {arXiv:2301.09363 [quant-ph]} \BibitemShut {NoStop}%
\bibitem [{\citenamefont {Salimans}\ \emph {et~al.}(2016)\citenamefont {Salimans}, \citenamefont {Goodfellow}, \citenamefont {Zaremba}, \citenamefont {Cheung}, \citenamefont {Radford}, \citenamefont {Chen},\ and\ \citenamefont {Chen}}]{Salimans2016gans}%
  \BibitemOpen
  \bibfield  {author} {\bibinfo {author} {\bibfnamefont {T.}~\bibnamefont {Salimans}}, \bibinfo {author} {\bibfnamefont {I.}~\bibnamefont {Goodfellow}}, \bibinfo {author} {\bibfnamefont {W.}~\bibnamefont {Zaremba}}, \bibinfo {author} {\bibfnamefont {V.}~\bibnamefont {Cheung}}, \bibinfo {author} {\bibfnamefont {A.}~\bibnamefont {Radford}}, \bibinfo {author} {\bibfnamefont {X.}~\bibnamefont {Chen}},\ and\ \bibinfo {author} {\bibfnamefont {X.}~\bibnamefont {Chen}},\ }\bibfield  {title} {\bibinfo {title} {Improved techniques for training gans},\ }in\ \href {https://proceedings.neurips.cc/paper/2016/file/8a3363abe792db2d8761d6403605aeb7-Paper.pdf} {\emph {\bibinfo {booktitle} {Advances in Neural Information Processing Systems}}},\ Vol.~\bibinfo {volume} {29}\ (\bibinfo  {publisher} {Curran Associates, Inc.},\ \bibinfo {year} {2016})\BibitemShut {NoStop}%
\bibitem [{\citenamefont {Barratt}\ and\ \citenamefont {Sharma}(2018)}]{barratt2018IS}%
  \BibitemOpen
  \bibfield  {author} {\bibinfo {author} {\bibfnamefont {S.}~\bibnamefont {Barratt}}\ and\ \bibinfo {author} {\bibfnamefont {R.}~\bibnamefont {Sharma}},\ }\href@noop {} {\bibinfo {title} {A note on the inception score}} (\bibinfo {year} {2018}),\ \Eprint {https://arxiv.org/abs/1801.01973} {arXiv:1801.01973 [stat.ML]} \BibitemShut {NoStop}%
\bibitem [{\citenamefont {Heusel}\ \emph {et~al.}(2017)\citenamefont {Heusel}, \citenamefont {Ramsauer}, \citenamefont {Unterthiner}, \citenamefont {Nessler},\ and\ \citenamefont {Hochreiter}}]{Heusel2017gans}%
  \BibitemOpen
  \bibfield  {author} {\bibinfo {author} {\bibfnamefont {M.}~\bibnamefont {Heusel}}, \bibinfo {author} {\bibfnamefont {H.}~\bibnamefont {Ramsauer}}, \bibinfo {author} {\bibfnamefont {T.}~\bibnamefont {Unterthiner}}, \bibinfo {author} {\bibfnamefont {B.}~\bibnamefont {Nessler}},\ and\ \bibinfo {author} {\bibfnamefont {S.}~\bibnamefont {Hochreiter}},\ }\bibfield  {title} {\bibinfo {title} {Gans trained by a two time-scale update rule converge to a local nash equilibrium},\ }in\ \href {https://proceedings.neurips.cc/paper_files/paper/2017/file/8a1d694707eb0fefe65871369074926d-Paper.pdf} {\emph {\bibinfo {booktitle} {Advances in Neural Information Processing Systems}}},\ Vol.~\bibinfo {volume} {30}\ (\bibinfo  {publisher} {Curran Associates, Inc.},\ \bibinfo {year} {2017})\BibitemShut {NoStop}%
\bibitem [{\citenamefont {Liu}\ and\ \citenamefont {Li}(2021)}]{metrics_gan}%
  \BibitemOpen
  \bibfield  {author} {\bibinfo {author} {\bibfnamefont {Y.}~\bibnamefont {Liu}}\ and\ \bibinfo {author} {\bibfnamefont {Y.}~\bibnamefont {Li}},\ }\href@noop {} {\bibinfo {title} {{Metrics of GANs}}},\ \bibinfo {howpublished} {\url{https://github.com/yhlleo/GAN-Metrics}} (\bibinfo {year} {2021}),\ \bibinfo {note} {[Online; accessed Nov-23-2022]}\BibitemShut {NoStop}%
\bibitem [{\citenamefont {Szegedy}\ \emph {et~al.}(2016)\citenamefont {Szegedy}, \citenamefont {Vanhoucke}, \citenamefont {Ioffe}, \citenamefont {Shlens},\ and\ \citenamefont {Wojna}}]{Szegedy2016inceptionv3}%
  \BibitemOpen
  \bibfield  {author} {\bibinfo {author} {\bibfnamefont {C.}~\bibnamefont {Szegedy}}, \bibinfo {author} {\bibfnamefont {V.}~\bibnamefont {Vanhoucke}}, \bibinfo {author} {\bibfnamefont {S.}~\bibnamefont {Ioffe}}, \bibinfo {author} {\bibfnamefont {J.}~\bibnamefont {Shlens}},\ and\ \bibinfo {author} {\bibfnamefont {Z.}~\bibnamefont {Wojna}},\ }\bibfield  {title} {\bibinfo {title} {Rethinking the inception architecture for computer vision},\ }in\ \href {https://doi.org/10.1109/CVPR.2016.308} {\emph {\bibinfo {booktitle} {2016 IEEE Conference on Computer Vision and Pattern Recognition (CVPR)}}}\ (\bibinfo {year} {2016})\ pp.\ \bibinfo {pages} {2818--2826}\BibitemShut {NoStop}%
\bibitem [{\citenamefont {Lecun}\ \emph {et~al.}(1998)\citenamefont {Lecun}, \citenamefont {Bottou}, \citenamefont {Bengio},\ and\ \citenamefont {Haffner}}]{lecun2010mnist}%
  \BibitemOpen
  \bibfield  {author} {\bibinfo {author} {\bibfnamefont {Y.}~\bibnamefont {Lecun}}, \bibinfo {author} {\bibfnamefont {L.}~\bibnamefont {Bottou}}, \bibinfo {author} {\bibfnamefont {Y.}~\bibnamefont {Bengio}},\ and\ \bibinfo {author} {\bibfnamefont {P.}~\bibnamefont {Haffner}},\ }\bibfield  {title} {\bibinfo {title} {Gradient-based learning applied to document recognition},\ }\href {https://doi.org/10.1109/5.726791} {\bibfield  {journal} {\bibinfo  {journal} {Proceedings of the IEEE}\ }\textbf {\bibinfo {volume} {86}},\ \bibinfo {pages} {2278} (\bibinfo {year} {1998})}\BibitemShut {NoStop}%
\bibitem [{\citenamefont {Xiao}\ \emph {et~al.}(2017)\citenamefont {Xiao}, \citenamefont {Rasul},\ and\ \citenamefont {Vollgraf}}]{xiao2017fashionMNIST}%
  \BibitemOpen
  \bibfield  {author} {\bibinfo {author} {\bibfnamefont {H.}~\bibnamefont {Xiao}}, \bibinfo {author} {\bibfnamefont {K.}~\bibnamefont {Rasul}},\ and\ \bibinfo {author} {\bibfnamefont {R.}~\bibnamefont {Vollgraf}},\ }\href@noop {} {\bibinfo {title} {Fashion-mnist: a novel image dataset for benchmarking machine learning algorithms}} (\bibinfo {year} {2017}),\ \Eprint {https://arxiv.org/abs/1708.07747} {arXiv:1708.07747 [cs.LG]} \BibitemShut {NoStop}%
\bibitem [{\citenamefont {Wei}\ \emph {et~al.}(2022)\citenamefont {Wei}, \citenamefont {Liu}, \citenamefont {Luo}, \citenamefont {Zhu}, \citenamefont {Davis},\ and\ \citenamefont {Liu}}]{Wei2021duelGAN}%
  \BibitemOpen
  \bibfield  {author} {\bibinfo {author} {\bibfnamefont {J.}~\bibnamefont {Wei}}, \bibinfo {author} {\bibfnamefont {M.}~\bibnamefont {Liu}}, \bibinfo {author} {\bibfnamefont {J.}~\bibnamefont {Luo}}, \bibinfo {author} {\bibfnamefont {A.}~\bibnamefont {Zhu}}, \bibinfo {author} {\bibfnamefont {J.}~\bibnamefont {Davis}},\ and\ \bibinfo {author} {\bibfnamefont {Y.}~\bibnamefont {Liu}},\ }\bibfield  {title} {\bibinfo {title} {Duelgan: A duel between two discriminators stabilizes the gan training},\ }in\ \href {https://link.springer.com/chapter/10.1007/978-3-031-20050-2_18} {\emph {\bibinfo {booktitle} {Computer Vision -- ECCV 2022}}}\ (\bibinfo  {publisher} {Springer Nature Switzerland},\ \bibinfo {year} {2022})\ pp.\ \bibinfo {pages} {290--317}\BibitemShut {NoStop}%
\bibitem [{\citenamefont {Lazcano}\ \emph {et~al.}(2021)\citenamefont {Lazcano}, \citenamefont {Franco},\ and\ \citenamefont {Creixell}}]{Lazcano2021hypGAN}%
  \BibitemOpen
  \bibfield  {author} {\bibinfo {author} {\bibfnamefont {D.}~\bibnamefont {Lazcano}}, \bibinfo {author} {\bibfnamefont {N.~F.}\ \bibnamefont {Franco}},\ and\ \bibinfo {author} {\bibfnamefont {W.}~\bibnamefont {Creixell}},\ }\bibfield  {title} {\bibinfo {title} {Hgan: Hyperbolic generative adversarial network},\ }\href {https://doi.org/10.1109/ACCESS.2021.3094723} {\bibfield  {journal} {\bibinfo  {journal} {IEEE Access}\ }\textbf {\bibinfo {volume} {9}},\ \bibinfo {pages} {96309} (\bibinfo {year} {2021})}\BibitemShut {NoStop}%
\bibitem [{\citenamefont {Bradbury}\ \emph {et~al.}(2018)\citenamefont {Bradbury}, \citenamefont {Frostig}, \citenamefont {Hawkins}, \citenamefont {Johnson}, \citenamefont {Leary}, \citenamefont {Maclaurin}, \citenamefont {Necula}, \citenamefont {Paszke}, \citenamefont {Vander{P}las}, \citenamefont {Wanderman-{M}ilne},\ and\ \citenamefont {Zhang}}]{jax2018github}%
  \BibitemOpen
  \bibfield  {author} {\bibinfo {author} {\bibfnamefont {J.}~\bibnamefont {Bradbury}}, \bibinfo {author} {\bibfnamefont {R.}~\bibnamefont {Frostig}}, \bibinfo {author} {\bibfnamefont {P.}~\bibnamefont {Hawkins}}, \bibinfo {author} {\bibfnamefont {M.~J.}\ \bibnamefont {Johnson}}, \bibinfo {author} {\bibfnamefont {C.}~\bibnamefont {Leary}}, \bibinfo {author} {\bibfnamefont {D.}~\bibnamefont {Maclaurin}}, \bibinfo {author} {\bibfnamefont {G.}~\bibnamefont {Necula}}, \bibinfo {author} {\bibfnamefont {A.}~\bibnamefont {Paszke}}, \bibinfo {author} {\bibfnamefont {J.}~\bibnamefont {Vander{P}las}}, \bibinfo {author} {\bibfnamefont {S.}~\bibnamefont {Wanderman-{M}ilne}},\ and\ \bibinfo {author} {\bibfnamefont {Q.}~\bibnamefont {Zhang}},\ }\href {http://github.com/google/jax} {\bibinfo {title} {{JAX}: composable transformations of {P}ython+{N}um{P}y programs}} (\bibinfo {year} {2018})\BibitemShut {NoStop}%
\bibitem [{\citenamefont {Heek}\ \emph {et~al.}(2020)\citenamefont {Heek}, \citenamefont {Levskaya}, \citenamefont {Oliver}, \citenamefont {Ritter}, \citenamefont {Rondepierre}, \citenamefont {Steiner},\ and\ \citenamefont {van {Z}ee}}]{flax2020github}%
  \BibitemOpen
  \bibfield  {author} {\bibinfo {author} {\bibfnamefont {J.}~\bibnamefont {Heek}}, \bibinfo {author} {\bibfnamefont {A.}~\bibnamefont {Levskaya}}, \bibinfo {author} {\bibfnamefont {A.}~\bibnamefont {Oliver}}, \bibinfo {author} {\bibfnamefont {M.}~\bibnamefont {Ritter}}, \bibinfo {author} {\bibfnamefont {B.}~\bibnamefont {Rondepierre}}, \bibinfo {author} {\bibfnamefont {A.}~\bibnamefont {Steiner}},\ and\ \bibinfo {author} {\bibfnamefont {M.}~\bibnamefont {van {Z}ee}},\ }\href {http://github.com/google/flax} {\bibinfo {title} {{F}lax: A neural network library and ecosystem for {JAX}}} (\bibinfo {year} {2020})\BibitemShut {NoStop}%
\bibitem [{\citenamefont {Bergholm}\ \emph {et~al.}(2022)\citenamefont {Bergholm}, \citenamefont {Izaac}, \citenamefont {Schuld}, \citenamefont {Gogolin}, \citenamefont {Ahmed}, \citenamefont {Ajith}, \citenamefont {Alam}, \citenamefont {Alonso-Linaje}, \citenamefont {AkashNarayanan}, \citenamefont {Asadi}, \citenamefont {Arrazola}, \citenamefont {Azad}, \citenamefont {Banning}, \citenamefont {Blank}, \citenamefont {Bromley}, \citenamefont {Cordier}, \citenamefont {Ceroni}, \citenamefont {Delgado}, \citenamefont {Matteo}, \citenamefont {Dusko}, \citenamefont {Garg}, \citenamefont {Guala}, \citenamefont {Hayes}, \citenamefont {Hill}, \citenamefont {Ijaz}, \citenamefont {Isacsson}, \citenamefont {Ittah}, \citenamefont {Jahangiri}, \citenamefont {Jain}, \citenamefont {Jiang}, \citenamefont {Khandelwal}, \citenamefont {Kottmann}, \citenamefont {Lang}, \citenamefont {Lee}, \citenamefont {Loke}, \citenamefont {Lowe}, \citenamefont {McKiernan}, \citenamefont {Meyer}, \citenamefont {Montañez-Barrera}, \citenamefont
  {Moyard}, \citenamefont {Niu}, \citenamefont {O'Riordan}, \citenamefont {Oud}, \citenamefont {Panigrahi}, \citenamefont {Park}, \citenamefont {Polatajko}, \citenamefont {Quesada}, \citenamefont {Roberts}, \citenamefont {Sá}, \citenamefont {Schoch}, \citenamefont {Shi}, \citenamefont {Shu}, \citenamefont {Sim}, \citenamefont {Singh}, \citenamefont {Strandberg}, \citenamefont {Soni}, \citenamefont {Száva}, \citenamefont {Thabet}, \citenamefont {Vargas-Hernández}, \citenamefont {Vincent}, \citenamefont {Vitucci}, \citenamefont {Weber}, \citenamefont {Wierichs}, \citenamefont {Wiersema}, \citenamefont {Willmann}, \citenamefont {Wong}, \citenamefont {Zhang},\ and\ \citenamefont {Killoran}}]{pennylane2018}%
  \BibitemOpen
  \bibfield  {author} {\bibinfo {author} {\bibfnamefont {V.}~\bibnamefont {Bergholm}}, \bibinfo {author} {\bibfnamefont {J.}~\bibnamefont {Izaac}}, \bibinfo {author} {\bibfnamefont {M.}~\bibnamefont {Schuld}}, \bibinfo {author} {\bibfnamefont {C.}~\bibnamefont {Gogolin}}, \bibinfo {author} {\bibfnamefont {S.}~\bibnamefont {Ahmed}}, \bibinfo {author} {\bibfnamefont {V.}~\bibnamefont {Ajith}}, \bibinfo {author} {\bibfnamefont {M.~S.}\ \bibnamefont {Alam}}, \bibinfo {author} {\bibfnamefont {G.}~\bibnamefont {Alonso-Linaje}}, \bibinfo {author} {\bibfnamefont {B.}~\bibnamefont {AkashNarayanan}}, \bibinfo {author} {\bibfnamefont {A.}~\bibnamefont {Asadi}}, \bibinfo {author} {\bibfnamefont {J.~M.}\ \bibnamefont {Arrazola}}, \bibinfo {author} {\bibfnamefont {U.}~\bibnamefont {Azad}}, \bibinfo {author} {\bibfnamefont {S.}~\bibnamefont {Banning}}, \bibinfo {author} {\bibfnamefont {C.}~\bibnamefont {Blank}}, \bibinfo {author} {\bibfnamefont {T.~R.}\ \bibnamefont {Bromley}}, \bibinfo {author} {\bibfnamefont {B.~A.}\
  \bibnamefont {Cordier}}, \bibinfo {author} {\bibfnamefont {J.}~\bibnamefont {Ceroni}}, \bibinfo {author} {\bibfnamefont {A.}~\bibnamefont {Delgado}}, \bibinfo {author} {\bibfnamefont {O.~D.}\ \bibnamefont {Matteo}}, \bibinfo {author} {\bibfnamefont {A.}~\bibnamefont {Dusko}}, \bibinfo {author} {\bibfnamefont {T.}~\bibnamefont {Garg}}, \bibinfo {author} {\bibfnamefont {D.}~\bibnamefont {Guala}}, \bibinfo {author} {\bibfnamefont {A.}~\bibnamefont {Hayes}}, \bibinfo {author} {\bibfnamefont {R.}~\bibnamefont {Hill}}, \bibinfo {author} {\bibfnamefont {A.}~\bibnamefont {Ijaz}}, \bibinfo {author} {\bibfnamefont {T.}~\bibnamefont {Isacsson}}, \bibinfo {author} {\bibfnamefont {D.}~\bibnamefont {Ittah}}, \bibinfo {author} {\bibfnamefont {S.}~\bibnamefont {Jahangiri}}, \bibinfo {author} {\bibfnamefont {P.}~\bibnamefont {Jain}}, \bibinfo {author} {\bibfnamefont {E.}~\bibnamefont {Jiang}}, \bibinfo {author} {\bibfnamefont {A.}~\bibnamefont {Khandelwal}}, \bibinfo {author} {\bibfnamefont {K.}~\bibnamefont {Kottmann}},
  \bibinfo {author} {\bibfnamefont {R.~A.}\ \bibnamefont {Lang}}, \bibinfo {author} {\bibfnamefont {C.}~\bibnamefont {Lee}}, \bibinfo {author} {\bibfnamefont {T.}~\bibnamefont {Loke}}, \bibinfo {author} {\bibfnamefont {A.}~\bibnamefont {Lowe}}, \bibinfo {author} {\bibfnamefont {K.}~\bibnamefont {McKiernan}}, \bibinfo {author} {\bibfnamefont {J.~J.}\ \bibnamefont {Meyer}}, \bibinfo {author} {\bibfnamefont {J.~A.}\ \bibnamefont {Montañez-Barrera}}, \bibinfo {author} {\bibfnamefont {R.}~\bibnamefont {Moyard}}, \bibinfo {author} {\bibfnamefont {Z.}~\bibnamefont {Niu}}, \bibinfo {author} {\bibfnamefont {L.~J.}\ \bibnamefont {O'Riordan}}, \bibinfo {author} {\bibfnamefont {S.}~\bibnamefont {Oud}}, \bibinfo {author} {\bibfnamefont {A.}~\bibnamefont {Panigrahi}}, \bibinfo {author} {\bibfnamefont {C.-Y.}\ \bibnamefont {Park}}, \bibinfo {author} {\bibfnamefont {D.}~\bibnamefont {Polatajko}}, \bibinfo {author} {\bibfnamefont {N.}~\bibnamefont {Quesada}}, \bibinfo {author} {\bibfnamefont {C.}~\bibnamefont {Roberts}},
  \bibinfo {author} {\bibfnamefont {N.}~\bibnamefont {Sá}}, \bibinfo {author} {\bibfnamefont {I.}~\bibnamefont {Schoch}}, \bibinfo {author} {\bibfnamefont {B.}~\bibnamefont {Shi}}, \bibinfo {author} {\bibfnamefont {S.}~\bibnamefont {Shu}}, \bibinfo {author} {\bibfnamefont {S.}~\bibnamefont {Sim}}, \bibinfo {author} {\bibfnamefont {A.}~\bibnamefont {Singh}}, \bibinfo {author} {\bibfnamefont {I.}~\bibnamefont {Strandberg}}, \bibinfo {author} {\bibfnamefont {J.}~\bibnamefont {Soni}}, \bibinfo {author} {\bibfnamefont {A.}~\bibnamefont {Száva}}, \bibinfo {author} {\bibfnamefont {S.}~\bibnamefont {Thabet}}, \bibinfo {author} {\bibfnamefont {R.~A.}\ \bibnamefont {Vargas-Hernández}}, \bibinfo {author} {\bibfnamefont {T.}~\bibnamefont {Vincent}}, \bibinfo {author} {\bibfnamefont {N.}~\bibnamefont {Vitucci}}, \bibinfo {author} {\bibfnamefont {M.}~\bibnamefont {Weber}}, \bibinfo {author} {\bibfnamefont {D.}~\bibnamefont {Wierichs}}, \bibinfo {author} {\bibfnamefont {R.}~\bibnamefont {Wiersema}}, \bibinfo {author}
  {\bibfnamefont {M.}~\bibnamefont {Willmann}}, \bibinfo {author} {\bibfnamefont {V.}~\bibnamefont {Wong}}, \bibinfo {author} {\bibfnamefont {S.}~\bibnamefont {Zhang}},\ and\ \bibinfo {author} {\bibfnamefont {N.}~\bibnamefont {Killoran}},\ }\href@noop {} {\bibinfo {title} {Pennylane: Automatic differentiation of hybrid quantum-classical computations}} (\bibinfo {year} {2022}),\ \Eprint {https://arxiv.org/abs/1811.04968} {arXiv:1811.04968 [quant-ph]} \BibitemShut {NoStop}%
\bibitem [{Fas()}]{FashionMNIST_FID}%
  \BibitemOpen
  \href@noop {} {\bibinfo {title} {{Image Generation on Fashion-MNIST}}},\ \bibinfo {howpublished} {\url{https://paperswithcode.com/sota/image-generation-on-fashion-mnist}},\ \bibinfo {note} {[Online; accessed Apr-26-2024]}\BibitemShut {NoStop}%
\bibitem [{\citenamefont {Böhm}\ and\ \citenamefont {Seljak}(2022)}]{Bohm2022PAE}%
  \BibitemOpen
  \bibfield  {author} {\bibinfo {author} {\bibfnamefont {V.}~\bibnamefont {Böhm}}\ and\ \bibinfo {author} {\bibfnamefont {U.}~\bibnamefont {Seljak}},\ }\href@noop {} {\bibinfo {title} {Probabilistic autoencoder}} (\bibinfo {year} {2022}),\ \Eprint {https://arxiv.org/abs/2006.05479} {arXiv:2006.05479 [cs.LG]} \BibitemShut {NoStop}%
\bibitem [{\citenamefont {van~der Maaten}\ and\ \citenamefont {Hinton}(2008)}]{Matten2008tsne}%
  \BibitemOpen
  \bibfield  {author} {\bibinfo {author} {\bibfnamefont {L.}~\bibnamefont {van~der Maaten}}\ and\ \bibinfo {author} {\bibfnamefont {G.}~\bibnamefont {Hinton}},\ }\bibfield  {title} {\bibinfo {title} {Visualizing data using t-sne},\ }\href {http://jmlr.org/papers/v9/vandermaaten08a.html} {\bibfield  {journal} {\bibinfo  {journal} {Journal of Machine Learning Research}\ }\textbf {\bibinfo {volume} {9}},\ \bibinfo {pages} {2579} (\bibinfo {year} {2008})}\BibitemShut {NoStop}%
\bibitem [{\citenamefont {Pedregosa}\ \emph {et~al.}(2011)\citenamefont {Pedregosa}, \citenamefont {Varoquaux}, \citenamefont {Gramfort}, \citenamefont {Michel}, \citenamefont {Thirion}, \citenamefont {Grisel}, \citenamefont {Blondel}, \citenamefont {Prettenhofer}, \citenamefont {Weiss}, \citenamefont {Dubourg}, \citenamefont {Vanderplas}, \citenamefont {Passos}, \citenamefont {Cournapeau}, \citenamefont {Brucher}, \citenamefont {Perrot},\ and\ \citenamefont {{{\'E}}douard Duchesnay}}]{scikit-learn}%
  \BibitemOpen
  \bibfield  {author} {\bibinfo {author} {\bibfnamefont {F.}~\bibnamefont {Pedregosa}}, \bibinfo {author} {\bibfnamefont {G.}~\bibnamefont {Varoquaux}}, \bibinfo {author} {\bibfnamefont {A.}~\bibnamefont {Gramfort}}, \bibinfo {author} {\bibfnamefont {V.}~\bibnamefont {Michel}}, \bibinfo {author} {\bibfnamefont {B.}~\bibnamefont {Thirion}}, \bibinfo {author} {\bibfnamefont {O.}~\bibnamefont {Grisel}}, \bibinfo {author} {\bibfnamefont {M.}~\bibnamefont {Blondel}}, \bibinfo {author} {\bibfnamefont {P.}~\bibnamefont {Prettenhofer}}, \bibinfo {author} {\bibfnamefont {R.}~\bibnamefont {Weiss}}, \bibinfo {author} {\bibfnamefont {V.}~\bibnamefont {Dubourg}}, \bibinfo {author} {\bibfnamefont {J.}~\bibnamefont {Vanderplas}}, \bibinfo {author} {\bibfnamefont {A.}~\bibnamefont {Passos}}, \bibinfo {author} {\bibfnamefont {D.}~\bibnamefont {Cournapeau}}, \bibinfo {author} {\bibfnamefont {M.}~\bibnamefont {Brucher}}, \bibinfo {author} {\bibfnamefont {M.}~\bibnamefont {Perrot}},\ and\ \bibinfo {author} {\bibnamefont
  {{{\'E}}douard Duchesnay}},\ }\bibfield  {title} {\bibinfo {title} {Scikit-learn: Machine learning in python},\ }\href {http://jmlr.org/papers/v12/pedregosa11a.html} {\bibfield  {journal} {\bibinfo  {journal} {Journal of Machine Learning Research}\ }\textbf {\bibinfo {volume} {12}},\ \bibinfo {pages} {2825} (\bibinfo {year} {2011})}\BibitemShut {NoStop}%
\bibitem [{\citenamefont {Abbas}\ \emph {et~al.}(2021)\citenamefont {Abbas}, \citenamefont {Sutter}, \citenamefont {Zoufal}, \citenamefont {Lucchi}, \citenamefont {Figalli},\ and\ \citenamefont {Woerner}}]{Abbas2021}%
  \BibitemOpen
  \bibfield  {author} {\bibinfo {author} {\bibfnamefont {A.}~\bibnamefont {Abbas}}, \bibinfo {author} {\bibfnamefont {D.}~\bibnamefont {Sutter}}, \bibinfo {author} {\bibfnamefont {C.}~\bibnamefont {Zoufal}}, \bibinfo {author} {\bibfnamefont {A.}~\bibnamefont {Lucchi}}, \bibinfo {author} {\bibfnamefont {A.}~\bibnamefont {Figalli}},\ and\ \bibinfo {author} {\bibfnamefont {S.}~\bibnamefont {Woerner}},\ }\bibfield  {title} {\bibinfo {title} {The power of quantum neural networks},\ }\href {https://doi.org/10.1038/s43588-021-00084-1} {\bibfield  {journal} {\bibinfo  {journal} {Nature Computational Science}\ }\textbf {\bibinfo {volume} {1}},\ \bibinfo {pages} {403} (\bibinfo {year} {2021})}\BibitemShut {NoStop}%
\bibitem [{\citenamefont {McClean}\ \emph {et~al.}(2018)\citenamefont {McClean}, \citenamefont {Boixo}, \citenamefont {Smelyanskiy}, \citenamefont {Babbush},\ and\ \citenamefont {Neven}}]{McClean2018BP}%
  \BibitemOpen
  \bibfield  {author} {\bibinfo {author} {\bibfnamefont {J.~R.}\ \bibnamefont {McClean}}, \bibinfo {author} {\bibfnamefont {S.}~\bibnamefont {Boixo}}, \bibinfo {author} {\bibfnamefont {V.~N.}\ \bibnamefont {Smelyanskiy}}, \bibinfo {author} {\bibfnamefont {R.}~\bibnamefont {Babbush}},\ and\ \bibinfo {author} {\bibfnamefont {H.}~\bibnamefont {Neven}},\ }\bibfield  {title} {\bibinfo {title} {Barren plateaus in quantum neural network training landscapes},\ }\href {https://doi.org/10.1038/s41467-018-07090-4} {\bibfield  {journal} {\bibinfo  {journal} {Nature Communications}\ }\textbf {\bibinfo {volume} {9}},\ \bibinfo {pages} {4812} (\bibinfo {year} {2018})}\BibitemShut {NoStop}%
\bibitem [{\citenamefont {Larocca}\ \emph {et~al.}(2024)\citenamefont {Larocca}, \citenamefont {Thanasilp}, \citenamefont {Wang}, \citenamefont {Sharma}, \citenamefont {Biamonte}, \citenamefont {Coles}, \citenamefont {Cincio}, \citenamefont {McClean}, \citenamefont {Holmes},\ and\ \citenamefont {Cerezo}}]{larocca2024review}%
  \BibitemOpen
  \bibfield  {author} {\bibinfo {author} {\bibfnamefont {M.}~\bibnamefont {Larocca}}, \bibinfo {author} {\bibfnamefont {S.}~\bibnamefont {Thanasilp}}, \bibinfo {author} {\bibfnamefont {S.}~\bibnamefont {Wang}}, \bibinfo {author} {\bibfnamefont {K.}~\bibnamefont {Sharma}}, \bibinfo {author} {\bibfnamefont {J.}~\bibnamefont {Biamonte}}, \bibinfo {author} {\bibfnamefont {P.~J.}\ \bibnamefont {Coles}}, \bibinfo {author} {\bibfnamefont {L.}~\bibnamefont {Cincio}}, \bibinfo {author} {\bibfnamefont {J.~R.}\ \bibnamefont {McClean}}, \bibinfo {author} {\bibfnamefont {Z.}~\bibnamefont {Holmes}},\ and\ \bibinfo {author} {\bibfnamefont {M.}~\bibnamefont {Cerezo}},\ }\href@noop {} {\bibinfo {title} {A review of barren plateaus in variational quantum computing}} (\bibinfo {year} {2024}),\ \Eprint {https://arxiv.org/abs/2405.00781} {arXiv:2405.00781 [quant-ph]} \BibitemShut {NoStop}%
\bibitem [{\citenamefont {Arrasmith}\ \emph {et~al.}(2022)\citenamefont {Arrasmith}, \citenamefont {Holmes}, \citenamefont {Cerezo},\ and\ \citenamefont {Coles}}]{arrasmith2021equivalence}%
  \BibitemOpen
  \bibfield  {author} {\bibinfo {author} {\bibfnamefont {A.}~\bibnamefont {Arrasmith}}, \bibinfo {author} {\bibfnamefont {Z.}~\bibnamefont {Holmes}}, \bibinfo {author} {\bibfnamefont {M.}~\bibnamefont {Cerezo}},\ and\ \bibinfo {author} {\bibfnamefont {P.~J.}\ \bibnamefont {Coles}},\ }\bibfield  {title} {\bibinfo {title} {Equivalence of quantum barren plateaus to cost concentration and narrow gorges},\ }\href {https://doi.org/10.1088/2058-9565/ac7d06} {\bibfield  {journal} {\bibinfo  {journal} {Quantum Science and Technology}\ }\textbf {\bibinfo {volume} {7}},\ \bibinfo {pages} {045015} (\bibinfo {year} {2022})}\BibitemShut {NoStop}%
\bibitem [{\citenamefont {Holmes}\ \emph {et~al.}(2022)\citenamefont {Holmes}, \citenamefont {Sharma}, \citenamefont {Cerezo},\ and\ \citenamefont {Coles}}]{Holmes2022ExprBP}%
  \BibitemOpen
  \bibfield  {author} {\bibinfo {author} {\bibfnamefont {Z.}~\bibnamefont {Holmes}}, \bibinfo {author} {\bibfnamefont {K.}~\bibnamefont {Sharma}}, \bibinfo {author} {\bibfnamefont {M.}~\bibnamefont {Cerezo}},\ and\ \bibinfo {author} {\bibfnamefont {P.~J.}\ \bibnamefont {Coles}},\ }\bibfield  {title} {\bibinfo {title} {Connecting ansatz expressibility to gradient magnitudes and barren plateaus},\ }\href {https://doi.org/10.1103/PRXQuantum.3.010313} {\bibfield  {journal} {\bibinfo  {journal} {PRX Quantum}\ }\textbf {\bibinfo {volume} {3}},\ \bibinfo {pages} {010313} (\bibinfo {year} {2022})}\BibitemShut {NoStop}%
\bibitem [{\citenamefont {Cerezo}\ \emph {et~al.}(2021{\natexlab{b}})\citenamefont {Cerezo}, \citenamefont {Sone}, \citenamefont {Volkoff}, \citenamefont {Cincio},\ and\ \citenamefont {Coles}}]{Cerezo2021CostBP}%
  \BibitemOpen
  \bibfield  {author} {\bibinfo {author} {\bibfnamefont {M.}~\bibnamefont {Cerezo}}, \bibinfo {author} {\bibfnamefont {A.}~\bibnamefont {Sone}}, \bibinfo {author} {\bibfnamefont {T.}~\bibnamefont {Volkoff}}, \bibinfo {author} {\bibfnamefont {L.}~\bibnamefont {Cincio}},\ and\ \bibinfo {author} {\bibfnamefont {P.~J.}\ \bibnamefont {Coles}},\ }\bibfield  {title} {\bibinfo {title} {Cost function dependent barren plateaus in shallow parametrized quantum circuits},\ }\href {https://doi.org/10.1038/s41467-021-21728-w} {\bibfield  {journal} {\bibinfo  {journal} {Nature Communications}\ }\textbf {\bibinfo {volume} {12}},\ \bibinfo {pages} {1791} (\bibinfo {year} {2021}{\natexlab{b}})}\BibitemShut {NoStop}%
\bibitem [{\citenamefont {Wang}\ \emph {et~al.}(2021)\citenamefont {Wang}, \citenamefont {Fontana}, \citenamefont {Cerezo}, \citenamefont {Sharma}, \citenamefont {Sone}, \citenamefont {Cincio},\ and\ \citenamefont {Coles}}]{Wang2021NoiseBP}%
  \BibitemOpen
  \bibfield  {author} {\bibinfo {author} {\bibfnamefont {S.}~\bibnamefont {Wang}}, \bibinfo {author} {\bibfnamefont {E.}~\bibnamefont {Fontana}}, \bibinfo {author} {\bibfnamefont {M.}~\bibnamefont {Cerezo}}, \bibinfo {author} {\bibfnamefont {K.}~\bibnamefont {Sharma}}, \bibinfo {author} {\bibfnamefont {A.}~\bibnamefont {Sone}}, \bibinfo {author} {\bibfnamefont {L.}~\bibnamefont {Cincio}},\ and\ \bibinfo {author} {\bibfnamefont {P.~J.}\ \bibnamefont {Coles}},\ }\bibfield  {title} {\bibinfo {title} {Noise-induced barren plateaus in variational quantum algorithms},\ }\href {https://doi.org/10.1038/s41467-021-27045-6} {\bibfield  {journal} {\bibinfo  {journal} {Nature Communications}\ }\textbf {\bibinfo {volume} {12}},\ \bibinfo {pages} {6961} (\bibinfo {year} {2021})}\BibitemShut {NoStop}%
\bibitem [{\citenamefont {Marrero}\ \emph {et~al.}(2021)\citenamefont {Marrero}, \citenamefont {Kieferov{\'a}},\ and\ \citenamefont {Wiebe}}]{marrero2020entanglement}%
  \BibitemOpen
  \bibfield  {author} {\bibinfo {author} {\bibfnamefont {C.~O.}\ \bibnamefont {Marrero}}, \bibinfo {author} {\bibfnamefont {M.}~\bibnamefont {Kieferov{\'a}}},\ and\ \bibinfo {author} {\bibfnamefont {N.}~\bibnamefont {Wiebe}},\ }\bibfield  {title} {\bibinfo {title} {Entanglement-induced barren plateaus},\ }\href {https://doi.org/10.1103/PRXQuantum.2.040316} {\bibfield  {journal} {\bibinfo  {journal} {PRX Quantum}\ }\textbf {\bibinfo {volume} {2}},\ \bibinfo {pages} {040316} (\bibinfo {year} {2021})}\BibitemShut {NoStop}%
\bibitem [{\citenamefont {Patti}\ \emph {et~al.}(2021)\citenamefont {Patti}, \citenamefont {Najafi}, \citenamefont {Gao},\ and\ \citenamefont {Yelin}}]{patti2020entanglement}%
  \BibitemOpen
  \bibfield  {author} {\bibinfo {author} {\bibfnamefont {T.~L.}\ \bibnamefont {Patti}}, \bibinfo {author} {\bibfnamefont {K.}~\bibnamefont {Najafi}}, \bibinfo {author} {\bibfnamefont {X.}~\bibnamefont {Gao}},\ and\ \bibinfo {author} {\bibfnamefont {S.~F.}\ \bibnamefont {Yelin}},\ }\bibfield  {title} {\bibinfo {title} {Entanglement devised barren plateau mitigation},\ }\href {https://doi.org/10.1103/PhysRevResearch.3.033090} {\bibfield  {journal} {\bibinfo  {journal} {Physical Review Research}\ }\textbf {\bibinfo {volume} {3}},\ \bibinfo {pages} {033090} (\bibinfo {year} {2021})}\BibitemShut {NoStop}%
\bibitem [{\citenamefont {Larocca}\ \emph {et~al.}(2022)\citenamefont {Larocca}, \citenamefont {Czarnik}, \citenamefont {Sharma}, \citenamefont {Muraleedharan}, \citenamefont {Coles},\ and\ \citenamefont {Cerezo}}]{larocca2021diagnosing}%
  \BibitemOpen
  \bibfield  {author} {\bibinfo {author} {\bibfnamefont {M.}~\bibnamefont {Larocca}}, \bibinfo {author} {\bibfnamefont {P.}~\bibnamefont {Czarnik}}, \bibinfo {author} {\bibfnamefont {K.}~\bibnamefont {Sharma}}, \bibinfo {author} {\bibfnamefont {G.}~\bibnamefont {Muraleedharan}}, \bibinfo {author} {\bibfnamefont {P.~J.}\ \bibnamefont {Coles}},\ and\ \bibinfo {author} {\bibfnamefont {M.}~\bibnamefont {Cerezo}},\ }\bibfield  {title} {\bibinfo {title} {Diagnosing {B}arren {P}lateaus with {T}ools from {Q}uantum {O}ptimal {C}ontrol},\ }\href {https://doi.org/10.22331/q-2022-09-29-824} {\bibfield  {journal} {\bibinfo  {journal} {{Quantum}}\ }\textbf {\bibinfo {volume} {6}},\ \bibinfo {pages} {824} (\bibinfo {year} {2022})}\BibitemShut {NoStop}%
\bibitem [{\citenamefont {Holmes}\ \emph {et~al.}(2021)\citenamefont {Holmes}, \citenamefont {Arrasmith}, \citenamefont {Yan}, \citenamefont {Coles}, \citenamefont {Albrecht},\ and\ \citenamefont {Sornborger}}]{holmes2020barren}%
  \BibitemOpen
  \bibfield  {author} {\bibinfo {author} {\bibfnamefont {Z.}~\bibnamefont {Holmes}}, \bibinfo {author} {\bibfnamefont {A.}~\bibnamefont {Arrasmith}}, \bibinfo {author} {\bibfnamefont {B.}~\bibnamefont {Yan}}, \bibinfo {author} {\bibfnamefont {P.~J.}\ \bibnamefont {Coles}}, \bibinfo {author} {\bibfnamefont {A.}~\bibnamefont {Albrecht}},\ and\ \bibinfo {author} {\bibfnamefont {A.~T.}\ \bibnamefont {Sornborger}},\ }\bibfield  {title} {\bibinfo {title} {Barren plateaus preclude learning scramblers},\ }\href {https://doi.org/10.1103/PhysRevLett.126.190501} {\bibfield  {journal} {\bibinfo  {journal} {Physical Review Letters}\ }\textbf {\bibinfo {volume} {126}},\ \bibinfo {pages} {190501} (\bibinfo {year} {2021})}\BibitemShut {NoStop}%
\bibitem [{\citenamefont {Thanasilp}\ \emph {et~al.}(2023)\citenamefont {Thanasilp}, \citenamefont {Wang}, \citenamefont {Nghiem}, \citenamefont {Coles},\ and\ \citenamefont {Cerezo}}]{Thanasilp2023QMLBP}%
  \BibitemOpen
  \bibfield  {author} {\bibinfo {author} {\bibfnamefont {S.}~\bibnamefont {Thanasilp}}, \bibinfo {author} {\bibfnamefont {S.}~\bibnamefont {Wang}}, \bibinfo {author} {\bibfnamefont {N.~A.}\ \bibnamefont {Nghiem}}, \bibinfo {author} {\bibfnamefont {P.}~\bibnamefont {Coles}},\ and\ \bibinfo {author} {\bibfnamefont {M.}~\bibnamefont {Cerezo}},\ }\bibfield  {title} {\bibinfo {title} {Subtleties in the trainability of quantum machine learning models},\ }\href {https://doi.org/10.1007/s42484-023-00103-6} {\bibfield  {journal} {\bibinfo  {journal} {Quantum Machine Intelligence}\ }\textbf {\bibinfo {volume} {5}},\ \bibinfo {pages} {21} (\bibinfo {year} {2023})}\BibitemShut {NoStop}%
\bibitem [{\citenamefont {Cerezo}\ \emph {et~al.}(2023)\citenamefont {Cerezo}, \citenamefont {Larocca}, \citenamefont {García-Martín}, \citenamefont {Diaz}, \citenamefont {Braccia}, \citenamefont {Fontana}, \citenamefont {Rudolph}, \citenamefont {Bermejo}, \citenamefont {Ijaz}, \citenamefont {Thanasilp}, \citenamefont {Anschuetz},\ and\ \citenamefont {Holmes}}]{Cerezo2023BP}%
  \BibitemOpen
  \bibfield  {author} {\bibinfo {author} {\bibfnamefont {M.}~\bibnamefont {Cerezo}}, \bibinfo {author} {\bibfnamefont {M.}~\bibnamefont {Larocca}}, \bibinfo {author} {\bibfnamefont {D.}~\bibnamefont {García-Martín}}, \bibinfo {author} {\bibfnamefont {N.~L.}\ \bibnamefont {Diaz}}, \bibinfo {author} {\bibfnamefont {P.}~\bibnamefont {Braccia}}, \bibinfo {author} {\bibfnamefont {E.}~\bibnamefont {Fontana}}, \bibinfo {author} {\bibfnamefont {M.~S.}\ \bibnamefont {Rudolph}}, \bibinfo {author} {\bibfnamefont {P.}~\bibnamefont {Bermejo}}, \bibinfo {author} {\bibfnamefont {A.}~\bibnamefont {Ijaz}}, \bibinfo {author} {\bibfnamefont {S.}~\bibnamefont {Thanasilp}}, \bibinfo {author} {\bibfnamefont {E.~R.}\ \bibnamefont {Anschuetz}},\ and\ \bibinfo {author} {\bibfnamefont {Z.}~\bibnamefont {Holmes}},\ }\href@noop {} {\bibinfo {title} {Does provable absence of barren plateaus imply classical simulability? or, why we need to rethink variational quantum computing}} (\bibinfo {year} {2023}),\ \Eprint
  {https://arxiv.org/abs/2312.09121} {arXiv:2312.09121 [quant-ph]} \BibitemShut {NoStop}%
\bibitem [{\citenamefont {Ragone}\ \emph {et~al.}(2023)\citenamefont {Ragone}, \citenamefont {Bakalov}, \citenamefont {Sauvage}, \citenamefont {Kemper}, \citenamefont {Marrero}, \citenamefont {Larocca},\ and\ \citenamefont {Cerezo}}]{Ragone2023UnifiedBP}%
  \BibitemOpen
  \bibfield  {author} {\bibinfo {author} {\bibfnamefont {M.}~\bibnamefont {Ragone}}, \bibinfo {author} {\bibfnamefont {B.~N.}\ \bibnamefont {Bakalov}}, \bibinfo {author} {\bibfnamefont {F.}~\bibnamefont {Sauvage}}, \bibinfo {author} {\bibfnamefont {A.~F.}\ \bibnamefont {Kemper}}, \bibinfo {author} {\bibfnamefont {C.~O.}\ \bibnamefont {Marrero}}, \bibinfo {author} {\bibfnamefont {M.}~\bibnamefont {Larocca}},\ and\ \bibinfo {author} {\bibfnamefont {M.}~\bibnamefont {Cerezo}},\ }\href@noop {} {\bibinfo {title} {A unified theory of barren plateaus for deep parametrized quantum circuits}} (\bibinfo {year} {2023}),\ \Eprint {https://arxiv.org/abs/2309.09342} {arXiv:2309.09342 [quant-ph]} \BibitemShut {NoStop}%
\bibitem [{\citenamefont {Fontana}\ \emph {et~al.}(2023)\citenamefont {Fontana}, \citenamefont {Herman}, \citenamefont {Chakrabarti}, \citenamefont {Kumar}, \citenamefont {Yalovetzky}, \citenamefont {Heredge}, \citenamefont {Hari~Sureshbabu},\ and\ \citenamefont {Pistoia}}]{fontana2023theadjoint}%
  \BibitemOpen
  \bibfield  {author} {\bibinfo {author} {\bibfnamefont {E.}~\bibnamefont {Fontana}}, \bibinfo {author} {\bibfnamefont {D.}~\bibnamefont {Herman}}, \bibinfo {author} {\bibfnamefont {S.}~\bibnamefont {Chakrabarti}}, \bibinfo {author} {\bibfnamefont {N.}~\bibnamefont {Kumar}}, \bibinfo {author} {\bibfnamefont {R.}~\bibnamefont {Yalovetzky}}, \bibinfo {author} {\bibfnamefont {J.}~\bibnamefont {Heredge}}, \bibinfo {author} {\bibfnamefont {S.}~\bibnamefont {Hari~Sureshbabu}},\ and\ \bibinfo {author} {\bibfnamefont {M.}~\bibnamefont {Pistoia}},\ }\bibfield  {title} {\bibinfo {title} {The adjoint is all you need: Characterizing barren plateaus in quantum ans\"atze},\ }\href {https://arxiv.org/abs/2309.07902} {\bibfield  {journal} {\bibinfo  {journal} {arXiv preprint arXiv:2309.07902}\ } (\bibinfo {year} {2023})}\BibitemShut {NoStop}%
\bibitem [{\citenamefont {Diaz}\ \emph {et~al.}(2023)\citenamefont {Diaz}, \citenamefont {Garc{\'\i}a-Mart{\'\i}n}, \citenamefont {Kazi}, \citenamefont {Larocca},\ and\ \citenamefont {Cerezo}}]{diaz2023showcasing}%
  \BibitemOpen
  \bibfield  {author} {\bibinfo {author} {\bibfnamefont {N.~L.}\ \bibnamefont {Diaz}}, \bibinfo {author} {\bibfnamefont {D.}~\bibnamefont {Garc{\'\i}a-Mart{\'\i}n}}, \bibinfo {author} {\bibfnamefont {S.}~\bibnamefont {Kazi}}, \bibinfo {author} {\bibfnamefont {M.}~\bibnamefont {Larocca}},\ and\ \bibinfo {author} {\bibfnamefont {M.}~\bibnamefont {Cerezo}},\ }\bibfield  {title} {\bibinfo {title} {Showcasing a barren plateau theory beyond the dynamical lie algebra},\ }\href {https://arxiv.org/abs/2310.11505} {\bibfield  {journal} {\bibinfo  {journal} {arXiv preprint arXiv:2310.11505}\ } (\bibinfo {year} {2023})}\BibitemShut {NoStop}%
\bibitem [{\citenamefont {Leone}\ \emph {et~al.}(2022)\citenamefont {Leone}, \citenamefont {Oliviero}, \citenamefont {Cincio},\ and\ \citenamefont {Cerezo}}]{leone2022practical}%
  \BibitemOpen
  \bibfield  {author} {\bibinfo {author} {\bibfnamefont {L.}~\bibnamefont {Leone}}, \bibinfo {author} {\bibfnamefont {S.~F.}\ \bibnamefont {Oliviero}}, \bibinfo {author} {\bibfnamefont {L.}~\bibnamefont {Cincio}},\ and\ \bibinfo {author} {\bibfnamefont {M.}~\bibnamefont {Cerezo}},\ }\bibfield  {title} {\bibinfo {title} {On the practical usefulness of the hardware efficient ansatz},\ }\href {https://arxiv.org/abs/2211.01477} {\bibfield  {journal} {\bibinfo  {journal} {arXiv preprint arXiv:2211.01477}\ } (\bibinfo {year} {2022})}\BibitemShut {NoStop}%
\bibitem [{\citenamefont {Barthe}\ and\ \citenamefont {P{\'e}rez-Salinas}(2023)}]{barthe2023gradients}%
  \BibitemOpen
  \bibfield  {author} {\bibinfo {author} {\bibfnamefont {A.}~\bibnamefont {Barthe}}\ and\ \bibinfo {author} {\bibfnamefont {A.}~\bibnamefont {P{\'e}rez-Salinas}},\ }\bibfield  {title} {\bibinfo {title} {Gradients and frequency profiles of quantum re-uploading models},\ }\href@noop {} {\bibfield  {journal} {\bibinfo  {journal} {arXiv preprint arXiv:2311.10822}\ } (\bibinfo {year} {2023})}\BibitemShut {NoStop}%
\bibitem [{\citenamefont {Thanasilp}\ \emph {et~al.}(2022)\citenamefont {Thanasilp}, \citenamefont {Wang}, \citenamefont {Cerezo},\ and\ \citenamefont {Holmes}}]{thanasilp2022exponential}%
  \BibitemOpen
  \bibfield  {author} {\bibinfo {author} {\bibfnamefont {S.}~\bibnamefont {Thanasilp}}, \bibinfo {author} {\bibfnamefont {S.}~\bibnamefont {Wang}}, \bibinfo {author} {\bibfnamefont {M.}~\bibnamefont {Cerezo}},\ and\ \bibinfo {author} {\bibfnamefont {Z.}~\bibnamefont {Holmes}},\ }\bibfield  {title} {\bibinfo {title} {Exponential concentration in quantum kernel methods},\ }\href {https://arxiv.org/abs/2208.11060} {\bibfield  {journal} {\bibinfo  {journal} {arXiv preprint arXiv:2208.11060}\ } (\bibinfo {year} {2022})}\BibitemShut {NoStop}%
\bibitem [{\citenamefont {Xiong}\ \emph {et~al.}(2023)\citenamefont {Xiong}, \citenamefont {Facelli}, \citenamefont {Sahebi}, \citenamefont {Agnel}, \citenamefont {Chotibut}, \citenamefont {Thanasilp},\ and\ \citenamefont {Holmes}}]{xiong2023fundamental}%
  \BibitemOpen
  \bibfield  {author} {\bibinfo {author} {\bibfnamefont {W.}~\bibnamefont {Xiong}}, \bibinfo {author} {\bibfnamefont {G.}~\bibnamefont {Facelli}}, \bibinfo {author} {\bibfnamefont {M.}~\bibnamefont {Sahebi}}, \bibinfo {author} {\bibfnamefont {O.}~\bibnamefont {Agnel}}, \bibinfo {author} {\bibfnamefont {T.}~\bibnamefont {Chotibut}}, \bibinfo {author} {\bibfnamefont {S.}~\bibnamefont {Thanasilp}},\ and\ \bibinfo {author} {\bibfnamefont {Z.}~\bibnamefont {Holmes}},\ }\bibfield  {title} {\bibinfo {title} {On fundamental aspects of quantum extreme learning machines},\ }\href {https://arxiv.org/abs/2312.15124} {\bibfield  {journal} {\bibinfo  {journal} {arXiv preprint arXiv:2312.15124}\ } (\bibinfo {year} {2023})}\BibitemShut {NoStop}%
\bibitem [{\citenamefont {Suzuki}\ \emph {et~al.}(2022)\citenamefont {Suzuki}, \citenamefont {Kawaguchi},\ and\ \citenamefont {Yamamoto}}]{suzuki2022quantumfisher}%
  \BibitemOpen
  \bibfield  {author} {\bibinfo {author} {\bibfnamefont {Y.}~\bibnamefont {Suzuki}}, \bibinfo {author} {\bibfnamefont {H.}~\bibnamefont {Kawaguchi}},\ and\ \bibinfo {author} {\bibfnamefont {N.}~\bibnamefont {Yamamoto}},\ }\bibfield  {title} {\bibinfo {title} {Quantum fisher kernel for mitigating the vanishing similarity issue},\ }\href {https://arxiv.org/abs/2210.16581} {\bibfield  {journal} {\bibinfo  {journal} {arXiv preprint arXiv:2210.16581}\ } (\bibinfo {year} {2022})}\BibitemShut {NoStop}%
\bibitem [{\citenamefont {Suzuki}\ and\ \citenamefont {Li}(2023)}]{suzuki2023effect}%
  \BibitemOpen
  \bibfield  {author} {\bibinfo {author} {\bibfnamefont {Y.}~\bibnamefont {Suzuki}}\ and\ \bibinfo {author} {\bibfnamefont {M.}~\bibnamefont {Li}},\ }\bibfield  {title} {\bibinfo {title} {Effect of alternating layered ansatzes on trainability of projected quantum kernel},\ }\href {https://arxiv.org/abs/2310.00361} {\bibfield  {journal} {\bibinfo  {journal} {arXiv preprint arXiv:2310.00361}\ } (\bibinfo {year} {2023})}\BibitemShut {NoStop}%
\bibitem [{\citenamefont {Rychlik}(1976)}]{rychlik1976central}%
  \BibitemOpen
  \bibfield  {author} {\bibinfo {author} {\bibfnamefont {Z.}~\bibnamefont {Rychlik}},\ }\bibfield  {title} {\bibinfo {title} {A central limit theorem for sums of a random number of independent random variables},\ }in\ \href {https://doi.org/10.1016/0167-6377(84)90008-7} {\emph {\bibinfo {booktitle} {Colloquium Mathematicum}}},\ Vol.~\bibinfo {volume} {1}\ (\bibinfo {year} {1976})\ pp.\ \bibinfo {pages} {147--158}\BibitemShut {NoStop}%
\bibitem [{\citenamefont {Grant}\ \emph {et~al.}(2019)\citenamefont {Grant}, \citenamefont {Wossnig}, \citenamefont {Ostaszewski},\ and\ \citenamefont {Benedetti}}]{Grant2019initialization}%
  \BibitemOpen
  \bibfield  {author} {\bibinfo {author} {\bibfnamefont {E.}~\bibnamefont {Grant}}, \bibinfo {author} {\bibfnamefont {L.}~\bibnamefont {Wossnig}}, \bibinfo {author} {\bibfnamefont {M.}~\bibnamefont {Ostaszewski}},\ and\ \bibinfo {author} {\bibfnamefont {M.}~\bibnamefont {Benedetti}},\ }\bibfield  {title} {\bibinfo {title} {An initialization strategy for addressing barren plateaus in parametrized quantum circuits},\ }\href {https://doi.org/10.22331/q-2019-12-09-214} {\bibfield  {journal} {\bibinfo  {journal} {{Quantum}}\ }\textbf {\bibinfo {volume} {3}},\ \bibinfo {pages} {214} (\bibinfo {year} {2019})}\BibitemShut {NoStop}%
\bibitem [{\citenamefont {Zhang}\ \emph {et~al.}(2022)\citenamefont {Zhang}, \citenamefont {Liu}, \citenamefont {Hsieh},\ and\ \citenamefont {Tao}}]{zhang2022escaping}%
  \BibitemOpen
  \bibfield  {author} {\bibinfo {author} {\bibfnamefont {K.}~\bibnamefont {Zhang}}, \bibinfo {author} {\bibfnamefont {L.}~\bibnamefont {Liu}}, \bibinfo {author} {\bibfnamefont {M.-H.}\ \bibnamefont {Hsieh}},\ and\ \bibinfo {author} {\bibfnamefont {D.}~\bibnamefont {Tao}},\ }\href@noop {} {\bibinfo {title} {Escaping from the barren plateau via gaussian initializations in deep variational quantum circuits}} (\bibinfo {year} {2022}),\ \Eprint {https://arxiv.org/abs/2203.09376} {arXiv:2203.09376 [quant-ph]} \BibitemShut {NoStop}%
\bibitem [{IBM()}]{IBMansatz}%
  \BibitemOpen
  \href@noop {} {\bibinfo {title} {{IBM Quantum EfficientSU2}}},\ \bibinfo {howpublished} {\url{https://docs.quantum.ibm.com/api/qiskit/qiskit.circuit.library.EfficientSU2}},\ \bibinfo {note} {[Online; accessed Mar-15-2024]}\BibitemShut {NoStop}%
\bibitem [{\citenamefont {i~Valls}\ \emph {et~al.}(2024)\citenamefont {i~Valls}, \citenamefont {Drudis}, \citenamefont {Thanasilp},\ and\ \citenamefont {Holmes}}]{valls2024warmstart}%
  \BibitemOpen
  \bibfield  {author} {\bibinfo {author} {\bibfnamefont {R.~P.}\ \bibnamefont {i~Valls}}, \bibinfo {author} {\bibfnamefont {M.}~\bibnamefont {Drudis}}, \bibinfo {author} {\bibfnamefont {S.}~\bibnamefont {Thanasilp}},\ and\ \bibinfo {author} {\bibfnamefont {Z.}~\bibnamefont {Holmes}},\ }\href@noop {} {\bibinfo {title} {Variational quantum simulation: a case study for understanding warm starts}} (\bibinfo {year} {2024}),\ \Eprint {https://arxiv.org/abs/2404.10044} {arXiv:2404.10044 [quant-ph]} \BibitemShut {NoStop}%
\bibitem [{\citenamefont {Rudolph}\ \emph {et~al.}(2023{\natexlab{b}})\citenamefont {Rudolph}, \citenamefont {Miller}, \citenamefont {Motlagh}, \citenamefont {Chen}, \citenamefont {Acharya},\ and\ \citenamefont {Perdomo-Ortiz}}]{rudolph2022synergy}%
  \BibitemOpen
  \bibfield  {author} {\bibinfo {author} {\bibfnamefont {M.~S.}\ \bibnamefont {Rudolph}}, \bibinfo {author} {\bibfnamefont {J.}~\bibnamefont {Miller}}, \bibinfo {author} {\bibfnamefont {D.}~\bibnamefont {Motlagh}}, \bibinfo {author} {\bibfnamefont {J.}~\bibnamefont {Chen}}, \bibinfo {author} {\bibfnamefont {A.}~\bibnamefont {Acharya}},\ and\ \bibinfo {author} {\bibfnamefont {A.}~\bibnamefont {Perdomo-Ortiz}},\ }\bibfield  {title} {\bibinfo {title} {Synergistic pretraining of parametrized quantum circuits via tensor networks},\ }\href {https://doi.org/10.1038/s41467-023-43908-6} {\bibfield  {journal} {\bibinfo  {journal} {Nature Communications}\ }\textbf {\bibinfo {volume} {14}},\ \bibinfo {pages} {8367} (\bibinfo {year} {2023}{\natexlab{b}})}\BibitemShut {NoStop}%
\bibitem [{\citenamefont {Mele}\ \emph {et~al.}(2022)\citenamefont {Mele}, \citenamefont {Mbeng}, \citenamefont {Santoro}, \citenamefont {Collura},\ and\ \citenamefont {Torta}}]{mele2022avoiding}%
  \BibitemOpen
  \bibfield  {author} {\bibinfo {author} {\bibfnamefont {A.~A.}\ \bibnamefont {Mele}}, \bibinfo {author} {\bibfnamefont {G.~B.}\ \bibnamefont {Mbeng}}, \bibinfo {author} {\bibfnamefont {G.~E.}\ \bibnamefont {Santoro}}, \bibinfo {author} {\bibfnamefont {M.}~\bibnamefont {Collura}},\ and\ \bibinfo {author} {\bibfnamefont {P.}~\bibnamefont {Torta}},\ }\bibfield  {title} {\bibinfo {title} {Avoiding barren plateaus via transferability of smooth solutions in a hamiltonian variational ansatz},\ }\href {https://doi.org/10.1103/PhysRevA.106.L060401} {\bibfield  {journal} {\bibinfo  {journal} {Physical Review A}\ }\textbf {\bibinfo {volume} {106}},\ \bibinfo {pages} {L060401} (\bibinfo {year} {2022})}\BibitemShut {NoStop}%
\bibitem [{\citenamefont {Landman}\ \emph {et~al.}(2022)\citenamefont {Landman}, \citenamefont {Thabet}, \citenamefont {Dalyac}, \citenamefont {Mhiri},\ and\ \citenamefont {Kashefi}}]{landman2022classically}%
  \BibitemOpen
  \bibfield  {author} {\bibinfo {author} {\bibfnamefont {J.}~\bibnamefont {Landman}}, \bibinfo {author} {\bibfnamefont {S.}~\bibnamefont {Thabet}}, \bibinfo {author} {\bibfnamefont {C.}~\bibnamefont {Dalyac}}, \bibinfo {author} {\bibfnamefont {H.}~\bibnamefont {Mhiri}},\ and\ \bibinfo {author} {\bibfnamefont {E.}~\bibnamefont {Kashefi}},\ }\href@noop {} {\bibinfo {title} {Classically approximating variational quantum machine learning with random fourier features}} (\bibinfo {year} {2022}),\ \Eprint {https://arxiv.org/abs/2210.13200} {arXiv:2210.13200 [quant-ph]} \BibitemShut {NoStop}%
\bibitem [{\citenamefont {Karras}\ \emph {et~al.}(2018)\citenamefont {Karras}, \citenamefont {Aila}, \citenamefont {Laine},\ and\ \citenamefont {Lehtinen}}]{Karrs2017Progan}%
  \BibitemOpen
  \bibfield  {author} {\bibinfo {author} {\bibfnamefont {T.}~\bibnamefont {Karras}}, \bibinfo {author} {\bibfnamefont {T.}~\bibnamefont {Aila}}, \bibinfo {author} {\bibfnamefont {S.}~\bibnamefont {Laine}},\ and\ \bibinfo {author} {\bibfnamefont {J.}~\bibnamefont {Lehtinen}},\ }\href@noop {} {\bibinfo {title} {Progressive growing of gans for improved quality, stability, and variation}} (\bibinfo {year} {2018}),\ \Eprint {https://arxiv.org/abs/1710.10196} {arXiv:1710.10196 [cs.NE]} \BibitemShut {NoStop}%
\bibitem [{\citenamefont {Lloyd}\ and\ \citenamefont {Weedbrook}(2018)}]{Lloyd2018}%
  \BibitemOpen
  \bibfield  {author} {\bibinfo {author} {\bibfnamefont {S.}~\bibnamefont {Lloyd}}\ and\ \bibinfo {author} {\bibfnamefont {C.}~\bibnamefont {Weedbrook}},\ }\bibfield  {title} {\bibinfo {title} {Quantum generative adversarial learning},\ }\href {https://doi.org/10.1103/PhysRevLett.121.040502} {\bibfield  {journal} {\bibinfo  {journal} {Phys. Rev. Lett.}\ }\textbf {\bibinfo {volume} {121}},\ \bibinfo {pages} {040502} (\bibinfo {year} {2018})}\BibitemShut {NoStop}%
\bibitem [{\citenamefont {Zeng}\ \emph {et~al.}(2019)\citenamefont {Zeng}, \citenamefont {Wu}, \citenamefont {Liu}, \citenamefont {Wang},\ and\ \citenamefont {Hu}}]{Zeng2019GAN}%
  \BibitemOpen
  \bibfield  {author} {\bibinfo {author} {\bibfnamefont {J.}~\bibnamefont {Zeng}}, \bibinfo {author} {\bibfnamefont {Y.}~\bibnamefont {Wu}}, \bibinfo {author} {\bibfnamefont {J.-G.}\ \bibnamefont {Liu}}, \bibinfo {author} {\bibfnamefont {L.}~\bibnamefont {Wang}},\ and\ \bibinfo {author} {\bibfnamefont {J.}~\bibnamefont {Hu}},\ }\bibfield  {title} {\bibinfo {title} {Learning and inference on generative adversarial quantum circuits},\ }\href {https://doi.org/10.1103/PhysRevA.99.052306} {\bibfield  {journal} {\bibinfo  {journal} {Phys. Rev. A}\ }\textbf {\bibinfo {volume} {99}},\ \bibinfo {pages} {052306} (\bibinfo {year} {2019})}\BibitemShut {NoStop}%
\bibitem [{\citenamefont {Assouel}\ \emph {et~al.}(2022)\citenamefont {Assouel}, \citenamefont {Jacquier},\ and\ \citenamefont {Kondratyev}}]{Assouel2022}%
  \BibitemOpen
  \bibfield  {author} {\bibinfo {author} {\bibfnamefont {A.}~\bibnamefont {Assouel}}, \bibinfo {author} {\bibfnamefont {A.}~\bibnamefont {Jacquier}},\ and\ \bibinfo {author} {\bibfnamefont {A.}~\bibnamefont {Kondratyev}},\ }\bibfield  {title} {\bibinfo {title} {A quantum generative adversarial network for distributions},\ }\href {https://doi.org/10.1007/s42484-022-00083-z} {\bibfield  {journal} {\bibinfo  {journal} {Quantum Machine Intelligence}\ }\textbf {\bibinfo {volume} {4}},\ \bibinfo {pages} {28} (\bibinfo {year} {2022})}\BibitemShut {NoStop}%
\bibitem [{\citenamefont {Schuld}\ and\ \citenamefont {Killoran}(2019)}]{Schuld2019featuremap}%
  \BibitemOpen
  \bibfield  {author} {\bibinfo {author} {\bibfnamefont {M.}~\bibnamefont {Schuld}}\ and\ \bibinfo {author} {\bibfnamefont {N.}~\bibnamefont {Killoran}},\ }\bibfield  {title} {\bibinfo {title} {Quantum machine learning in feature hilbert spaces},\ }\href {https://doi.org/10.1103/PhysRevLett.122.040504} {\bibfield  {journal} {\bibinfo  {journal} {Phys. Rev. Lett.}\ }\textbf {\bibinfo {volume} {122}},\ \bibinfo {pages} {040504} (\bibinfo {year} {2019})}\BibitemShut {NoStop}%
\bibitem [{\citenamefont {Li}\ \emph {et~al.}(2021)\citenamefont {Li}, \citenamefont {Topaloglu},\ and\ \citenamefont {Ghosh}}]{Li2021drug}%
  \BibitemOpen
  \bibfield  {author} {\bibinfo {author} {\bibfnamefont {J.}~\bibnamefont {Li}}, \bibinfo {author} {\bibfnamefont {R.~O.}\ \bibnamefont {Topaloglu}},\ and\ \bibinfo {author} {\bibfnamefont {S.}~\bibnamefont {Ghosh}},\ }\bibfield  {title} {\bibinfo {title} {Quantum generative models for small molecule drug discovery},\ }\href {https://doi.org/10.1109/TQE.2021.3104804} {\bibfield  {journal} {\bibinfo  {journal} {IEEE Transactions on Quantum Engineering}\ }\textbf {\bibinfo {volume} {2}},\ \bibinfo {pages} {1} (\bibinfo {year} {2021})}\BibitemShut {NoStop}%
\bibitem [{\citenamefont {Tsang}\ \emph {et~al.}(2023)\citenamefont {Tsang}, \citenamefont {West}, \citenamefont {Erfani},\ and\ \citenamefont {Usman}}]{Tsang2022patch}%
  \BibitemOpen
  \bibfield  {author} {\bibinfo {author} {\bibfnamefont {S.}~\bibnamefont {Tsang}}, \bibinfo {author} {\bibfnamefont {M.~T.}\ \bibnamefont {West}}, \bibinfo {author} {\bibfnamefont {S.~M.}\ \bibnamefont {Erfani}},\ and\ \bibinfo {author} {\bibfnamefont {M.}~\bibnamefont {Usman}},\ }\bibfield  {title} {\bibinfo {title} {Hybrid quantum–classical generative adversarial network for high-resolution image generation},\ }\href {https://doi.org/10.1109/TQE.2023.3319319} {\bibfield  {journal} {\bibinfo  {journal} {IEEE Transactions on Quantum Engineering}\ }\textbf {\bibinfo {volume} {4}},\ \bibinfo {pages} {1} (\bibinfo {year} {2023})}\BibitemShut {NoStop}%
\end{thebibliography}
\end{document}